\documentclass[bibyear]{aa}  

\usepackage{graphicx}
\usepackage{txfonts}
\usepackage{natbib}
\bibpunct[]{(}{)}{;}{a}{}{,}
\begin{document}

   \title{Rotational spectroscopy, tentative interstellar detection,\\ and
    chemical modelling of N-methylformamide}


   \author{A. Belloche \inst{1}
          \and A.~A. Meshcheryakov \inst{2}
          \and R.~T. Garrod \inst{3}
          \and V.~V. Ilyushin \inst{2}
          \and E.~A. Alekseev \inst{2,4}
          \and\\ R.~A. Motiyenko \inst{5}
          \and L. Margul{\`e}s \inst{5}
          \and H.~S.~P.~M{\"u}ller\inst{6}
          \and K.~M. Menten \inst{1}
          }

\institute{Max-Planck-Institut f\"{u}r Radioastronomie, Auf dem H\"{u}gel 69, 53121 Bonn, Germany\\ \email{belloche@mpifr-bonn.mpg.de}
         \and
           Microwave spectroscopy department, Institute of Radio Astronomy of NASU, Chervonopraporna Str. 4, 61002 Kharkiv, Ukraine
         \and  
         Departments of Chemistry and Astronomy, University of Virginia, Charlottesville, VA 22904, USA        
         \and 
           Quantum Radiophysics Department of V. N. Karazin Kharkiv National 
           University, Svobody Square 4, 61022 Kharkiv, Ukraine
         \and
          Laboratoire de Physique des Lasers, Atomes, et Mol{\'e}cules, UMR 8523, CNRS - Universit{\'e} de Lille 1, 59655 Villeneuve d'Ascq Cedex, France
         \and I. Physikalisches Institut, Universit{\"a}t zu K{\"o}ln, Z{\"u}lpicher Str. 77, 50937 K{\"o}ln, Germany
             }

   \date{Received 16 September 2016; accepted 17 January 2017}

 
  \abstract
   {N-methylformamide,  CH$_3$NHCHO, may be an important molecule for 
interstellar pre-biotic chemistry because it contains a peptide bond, which in 
terrestrial chemistry is responsible for linking amino acids in proteins.
The rotational spectrum of the most stable \textit{trans} conformer of 
N-methylformamide is complicated by strong torsion-rotation interaction due 
to the low barrier of the methyl torsion. For this reason, the theoretical 
description of the rotational spectrum of the \textit{trans} conformer has
up to now not been accurate enough to provide a firm basis for its 
interstellar detection. }
   {In this context, as a prerequisite for a successful interstellar 
detection, our goal is to improve the characterization of the rotational 
spectrum of N-methylformamide.}
   {We use two absorption spectrometers in Kharkiv and Lille to measure the
rotational spectra over the frequency range 45--630~GHz. The analysis is 
carried out using the Rho-axis method and the RAM36 code. We search for 
N-methylformamide toward the hot molecular core Sagittarius (Sgr) B2(N2) using 
a spectral line survey carried out with the Atacama Large 
Millimeter/submillimeter Array (ALMA). The 
astronomical spectra are analyzed under the assumption of local thermodynamic 
equilibrium. The astronomical results are put into a broader astrochemical 
context with the help of a gas-grain chemical kinetics model.}
   {The new laboratory data set for the \textit{trans} conformer of 
N-methylformamide consists of 9469 distinct line frequencies with $J \leq 62$, 
including the first assignment of the rotational spectra of the first and 
second excited torsional states. All these lines are fitted within 
experimental accuracy for the first time. Based on the reliable frequency 
predictions obtained in this study, we report the tentative detection of 
N-methylformamide towards Sgr~B2(N2). We find N-methylformamide to be more than
one order of magnitude less abundant than formamide (NH$_2$CHO), a factor of two
less abundant than the unsaturated molecule methyl isocyanate (CH$_3$NCO), but
only slightly less abundant than acetamide (CH$_3$CONH$_2$). We also report the 
tentative detection of the $^{15}$N isotopologue of formamide 
($^{15}$NH$_2$CHO) toward Sgr~B2(N2). The chemical models indicate that the 
efficient formation of HNCO via NH + CO on grains is a necessary step in the 
achievement of the observed gas-phase abundance of CH$_3$NCO. Production of 
CH$_3$NHCHO may plausibly occur on grains either through the direct addition 
of functional-group radicals or through the hydrogenation of CH$_3$NCO.
}
   {Provided the detection of N-methylformamide is confirmed, the only slight
underabundance of this molecule compared to its more stable structural isomer 
acetamide and the sensitivity of the model abundances to the chemical kinetics 
parameters suggest that the formation of these two molecules is controlled by 
kinetics rather than thermal equilibrium.}


   \maketitle
%

\section{Introduction}
\label{s:introduction}

The peptide linkage is a fundamental building block of life on Earth 
\citep{kaiser2013formation}. Therefore peptide molecules have for a long 
time attracted much attention. The simplest molecule containing a peptide 
bond, formamide (NH$_2$CHO), was detected in the interstellar medium (ISM) 
back in the 1970s \citep{rubin1971microwave}. The relatively high abundance of 
formamide also permitted the detection of rotational lines of its first excited 
vibrational state $\varv_{12}=1$ in Orion~KL \citep{motiyenko2012rotational} 
and in Sagittarius (Sgr) B2(N) \citep{Belloche13}. N-methylformamide, 
CH$_3$NHCHO, is one of the simplest derivatives of formamide and also a peptide
molecule.
It is of interest as a candidate for interstellar detection because its 
structural isomer acetamide (CH$_3$CONH$_2$) has already been detected in the 
ISM \citep{hollis2006detection, Halfen11}. CH$_3$NHCHO is the second most 
stable C$_2$H$_5$NO isomer after acetamide \citep{Lattelais10}.

\begin{figure}
\centering
\resizebox{\hsize}{!}{\includegraphics{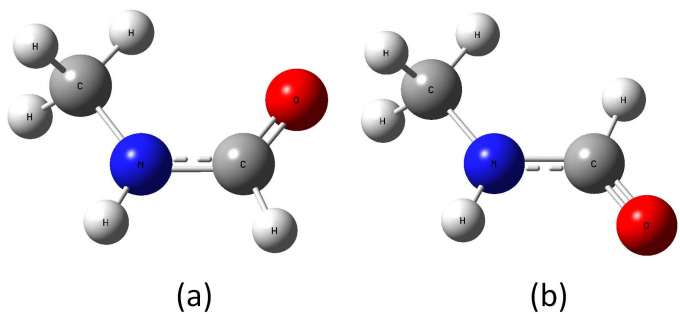}}
\caption{Structure of \textit{trans} (a) and \textit{cis} (b) conformations of 
N-methylformamide.}
\label{fig:struct}
\end{figure}

N-methylformamide exists in two stable rotameric forms, \textit{trans} and 
\textit{cis}. Their structures are shown in Fig.~\ref{fig:struct}. According to 
quantum chemical calculations, the \textit{trans} conformer is more stable than 
\textit{cis} by 466~cm$^{-1}$ \citep[666~K,][]{kawashima2010dynamical}. The 
\textit{trans} conformer is also characterized by a very low barrier to 
internal rotation of the methyl top. The coupling between the overall rotation 
of the molecule and the almost free rotation of the methyl top significantly 
complicates the description of the spectrum. For this reason, 
N-methylformamide has been the subject of extensive spectroscopic 
investigations, but the analysis of its microwave rotational spectrum for a 
long time did not yield satisfactory results. 
\citet{fantoni1996rotational} were the first to succeed in
assigning rotational spectral lines belonging to the \textit{trans} 
conformer in the CH$_3$ internal rotation ground state of $A$ symmetry. They 
performed measurements between 18 and 40 GHz, however they could not identify 
any spectral lines of the $E$ species. Later, \citet{fantoni2002very} 
published results of measurements and analysis for $E$ lines. The $V_3$ 
barrier of the methyl group internal rotation was determined to be 
$55.17 \pm 0.84$~cm$^{-1}$.

Recently, \citet{kawashima2010dynamical} carried out a new spectroscopic 
investigation of N-methylformamide. In that study, rotational spectra of both 
\textit{trans} and \textit{cis} conformers of the normal as well as deuterated 
CH$_3$NHCHO isotopologues were measured in the frequency range of 5--118 GHz. 
Molecular parameters including rotational constants and $V_3$ barriers to 
methyl-group internal rotation were determined for all investigated species 
and conformers. Owing to the relatively high barrier to internal rotation for 
the \textit{cis} conformer, a good description was obtained for its 108 
measured transitions. However, the low barrier and the limitations in the 
model used for the theoretical description of the rotational 
spectrum allowed to fit only low $J$ quantum number transitions ($J < 11$) for 
the \textit{trans} conformer. In addition, 60 out of 467 assigned transitions 
of the \textit{trans} conformer were excluded from the final fit as their 
residuals ranged from 1 to 67~MHz and were much higher than the 
experimental accuracy estimated to be 0.004--0.05~MHz. 

One should note that in previous publications two different schemes were used 
for naming the conformations of N-methylformamide. 
\citet{fantoni1996rotational} and \citet{fantoni2002very} used the dihedral 
angle $D (\mathrm{H-N-C-H}')$ where $\mathrm{H}'$ is the carbonyl group 
hydrogen. The conformer with $D = 0^\circ$ was named \textit{cis} and the 
conformer with $D = 180^\circ$ was named \textit{trans}. 
\citet{kawashima2010dynamical} used another convention widely accepted for 
molecules with a peptide bond. According to this convention applied for 
N-methylformamide, one should use the $D^*(\mathrm{Y-N-C-X})$ dihedral angle, 
where X is the carbonyl hydrogen and Y is the methyl group. \textit{Cis} and 
\textit{trans} conformers named using this convention are thus the opposite of 
the \textit{cis} and \textit{trans} conformers in 
\citet{fantoni1996rotational} and \citet{fantoni2002very}. Here we use the 
naming adopted by \citet{kawashima2010dynamical}, i.e. using the
$D^*$ dihedral angle.

In the present study, we extend the measurement and analysis of the rotational 
spectrum of \textit{trans} N-methylformamide which is of higher interest 
for an astrophysical detection. 
We use the accurate frequency predictions obtained in this study to search for
N-methylformamide in the ISM. We target the high-mass star 
forming region Sgr~B2(N), one of the most prolific sources for the detection of 
complex organic molecules in the ISM \citep[e.g.,][]{Belloche13}. For this, we
use a spectral survey of Sgr~B2(N) conducted with the Atacama Large 
Millimeter/submillimeter Array (ALMA) in its Cycle 0 and 1. This survey aims 
at exploring molecular complexity with ALMA \citep[EMoCA, see][]{Belloche16}.

The experimental setup is presented in Sect.~\ref{s:experiment}. The analysis
of the rotational spectrum of N-methylformamide and the results that follow 
are described in Sect.~\ref{s:spectro}. A tentative detection of 
N-methylformamide in Sgr~B2(N) is presented in Sect.~\ref{s:obsresults} along 
with the derivation of column densities of other related molecules of interest. 
Chemical modelling is performed in Sect.~\ref{s:chemistry} to put the 
observational results into a broader astrochemical context. The results are 
discussed in Sect.~\ref{s:discussion} and the conclusions are presented in 
Sect.~\ref{s:conclusions}.

\section{Experimental setup}
\label{s:experiment}

A sample of N-methylformamide was purchased from Sigma-Aldrich and used without 
further purification. The experimental investigation of the absorption 
spectrum of N-methylformamide was carried out over the frequency range 
45--630~GHz using 
two microwave spectrometers. The first one is the automated millimeter wave 
spectrometer of the Institute of Radio Astronomy of NASU in Kharkiv, Ukraine 
\citep{alekseev2012millimeter}. The second one is the terahertz spectrometer 
of the Laboratory of Physics of Lasers, Atoms, and Molecules in Lille, France 
\citep{zakharenko2015terahertz}.

The spectrometer in Kharkiv is built according to the classical scheme of 
absorption spectrometers, and its detailed description can be found in
\citet{alekseev2012millimeter}. The spectrometer was slightly upgraded with the 
aim of expanding the operating frequency range: a new backward-wave oscillator 
(BWO) unit covering frequencies from 34 to 52 GHz has been put into operation. 
Thus at present this spectrometer can record spectra between 
34 and 250~GHz. In order to improve the sensitivity below 50~GHz, a new 
waveguide absorbing cell (a copper waveguide of $10 \times 72$~mm$^2$ internal 
cross section and 295~cm length) was used instead of the commonly employed 
quasi-optic absorption cell. The measurements of N-methylformamide with the 
Kharkiv spectrometer were done in the frequency range 45 to 150~GHz. All 
measurements were performed at room temperature and with sample pressures 
(about 10~mTorr) that provided close to Doppler-limited spectral resolution. 
The frequency determination errors were estimated to be 10, 30, and 100~kHz 
depending on the measured signal-to-noise ratio. 

The measurements in Lille were performed between 150 and 630~GHz at typical 
pressures of 10~Pa and at room temperature. The frequency determination errors 
were estimated to be 30~kHz and 50~kHz below and above 330~GHz, respectively. 
The frequencies of the lines with poor signal-to-noise ratio or distorted 
lineshape were measured with 50~kHz or 100~kHz accuracy.

\section{Spectroscopic analysis and results}
\label{s:spectro}

We performed the analysis using the Rho-axis method (RAM), which was 
already applied successfully to a number of molecules with large-amplitude 
torsional motion. The method uses the axis system obtained by rotation of the 
principal axis system to make the new $z$ axis parallel to the $\rho$ vector. 
The coordinates of the $\rho$ vector are calculated using the following 
expression:
\[ \rho_g = \frac{\lambda_g I_\alpha}{I_g}, \; (g = x, y, z) \]
where $\lambda_g$ are the direction cosines of the internal rotation axis of 
the top in the principal axis system, $I_g$ are the principal inertia moments, 
and $I_\alpha$ is the inertia moment of the methyl top. The RAM Hamiltonian may 
be written as \citep{kleinerreview2010}:
\begin{equation} 
\label{hram}
H_{RAM} = H_T + H_R + H_{cd} + H_{int} 
\end{equation}
$H_T$ represents the torsional Hamiltonian defined as:
\begin{equation}
\label{HT}
H_T = F(p_\alpha - \rho J_z)^2+V(\alpha)
\end{equation}
where $F$ is the internal rotation constant, $p_\alpha$ is the internal 
rotation angular momentum conjugate to the torsion angle $\alpha$, and 
$V(\alpha)$ the internal rotation potential function:
\begin{equation}
\label{VA}
V(\alpha) = \frac{1}{2} V_3(1-\cos 3\alpha)+\frac{1}{2}V_6(1-\cos 6\alpha) + ...
\end{equation}
$H_R$ represents the rigid rotor rotational Hamiltonian in the rho-axis 
system. In 
addition to usual $A$, $B$, and $C$ terms, for the molecules with ($xz$) plane 
of symmetry (as is appropriate for N-methylformamide), it contains a 
non-diagonal term $D_{xz}$. In the $I^r$ coordinate representation $H_R$ has 
the following form:
\begin{equation}
\label{HR}
H_R = A_{RAM}J_z^2+ B_{RAM} J_x^2 + C_{RAM} J_y^2 + D_{xz}(J_xJ_z+J_zJ_x)
\end{equation}
The last two terms in Eq. \ref{hram}, $H_{cd}$ and $H_{int}$, are respectively 
the usual centrifugal distortion and  higher-order torsional-rotational
interaction Hamiltonians.

To fit and predict the rotational spectra we used the RAM36 
(Rho-axis method for 3 and 6-fold barriers) program that allows to include in 
the model almost any symmetry-allowed torsion-rotation Hamiltonian term up to 
the twelfth order \citep{ilyushin2010new, ilyushin2013submillimeter}. The RAM 
Hamiltonian in Eq. \ref{hram} may be expressed in the following form used 
in the RAM36 program :
\begin{align}
\label{ram36}
\begin{split}
H = & \frac{1}{2} \sum \limits_{knpqrs} B_{knpqrs0}\left[J^{2k} J_z^n J_x^p J_y^q p_\alpha^r \cos (3s\alpha) \right.\\
 & \left. + \cos (3s\alpha)p_\alpha^r J_y^q J_x^p J_z^n J^{2k} \right] \\
 & \frac{1}{2} \sum \limits_{knpqrt} B_{knpqr0t}\left[J^{2k} J_z^n J_x^p J_y^q p_\alpha^r \sin (3t\alpha) \right.\\
 & \left. + \sin (3t\alpha)p_\alpha^r J_y^q J_x^p J_z^n J^{2k} \right]
\end{split} 
\end{align}
where the $B_{knpqrst}$ are fitting parameters. In the case of a $C_{3v}$ top 
and $C_s$ frame (as is appropriate for N-methylformamide), the allowed terms 
in the torsion-rotation Hamiltonian must be totally symmetric in the group 
$G_6$ (and also must be Hermitian and invariant to the time reversal 
operation). Since all individual operators $p_\alpha$, $J_x$, $J_y$, $J_z$, 
$J^2$, $\cos (3s\alpha)$, and $\sin (3t\alpha)$ used in Eq. \ref{ram36} are 
Hermitian, all possible terms provided by Eq. \ref{ram36} will automatically 
be Hermitian. The particular term to be fitted is represented in the input file 
with a set of $k, n, p, q, r, s, t$ integer indices that are checked by the 
program for conformity with time reversal and symmetry requirements, to 
prevent accidental introduction of symmetry-forbidden terms into the 
Hamiltonian. For example, $B_{0000200}$ corresponds to $F$ in Eq. \ref{HT}, 
$B_{0200000}$ to $A_{RAM}$ in Eq. \ref{HR} etc.
In Table~\ref{tabpar} that presents the final set of molecular parameters, we 
give, instead of $B_{knpqrst}$, more conventional 
names for the parameters whose nomenclature is based on the subscript 
procedures of \citet{Xu2008305}.

The RAM36 program uses the free-rotor quantum number $m$ to label the 
torsional energy levels. It is well known that the internal rotation of a 
methyl top attached to a molecular frame should be treated as an anharmonic 
vibrational motion well below the top of the barrier to internal rotation, and 
as a nearly free internal rotation motion well above the top of the barrier. 
The $v_t$ labeling assumes that the spacings between degenerate and 
nondegenerate levels of the torsional Hamiltonian associated with a given $v_t$ 
are much smaller than those between levels with different $v_t$ 
\citep{linswalen1959}. The \textit{trans} conformation of N-methylformamide 
represents an intermediate case because of the relatively low barrier to 
internal 
rotation \citep{linswalen1959}. Owing to the low barrier, the $A-E$ splitting 
in the first excited torsional state $v_t = 1$ of the \textit{trans} 
conformation is comparable with the energy difference between the $v_t = 0$ 
and $v_t = 1$ states, 
whereas the $v_t = 2$ state lies well above the barrier. Because of this 
intermediate situation we decided to keep in Tables \ref{t:fit} and 
\ref{t:pred} the quantum number labeling of torsional states $m$ which is used 
inside the RAM36 program.

The RAM36 code was modified to take into account the quadrupole hyperfine 
structure of the transitions which is present in the spectrum of 
N-methylformamide due to nonzero electric quadrupole moment of the nitrogen 
atom. We used the standard hyperfine energy expression:
\begin{align*}
E_{hf} & = \left[ \chi_{aa}\left\langle J_a^2\right\rangle + \chi_{bb}\left\langle J_b^2\right\rangle - \left(  \chi_{aa}+\chi_{bb} \right) \left\langle J_c^2\right\rangle \right. \\
& \left. \vphantom{J_c^2} + \chi_{ab}\left\langle \vphantom{J_c^2} J_aJ_b+J_bJ_a\right\rangle \right] \frac{2f(I,J,F)}{J(J+1)}
\end{align*}
where $f(I,J,F)$ is the Casimir function. Typically a resolved pattern of the
hyperfine structure was observed as a doublet with an approximately two-to-one 
ratio in intensities. The stronger doublet component contains unresolved 
hyperfine transitions with selection rules $F = J+1 \leftarrow F = J + 1$ and 
$F = J - 1 \leftarrow F = J - 1$, whereas the weaker doublet component 
corresponds to the $F = J \leftarrow F = J$ transition. 

We started our analysis of the \textit{trans} N-methylformamide spectrum from 
refitting the data available from the literature 
\citep{fantoni1996rotational, fantoni2002very, kawashima2010dynamical} using 
the RAM36 code. Application of the RAM36 code allowed us to fit almost within 
experimental error the data available in the literature including those lines 
which were previously excluded from the fits due to rather high 
observed-minus-calculated values \citep{kawashima2010dynamical}. Thus we 
obtained a reliable basis for assigning the newly measured lines in the 
45--630~GHz range. Assigning and fitting of the new data using the RAM36 
program proceeded in a fairly conventional iterative way going up in frequency.

\longtab{
\begin{longtable}{cccc}
\caption{\label{tabpar} Molecular parameters of the trans conformer of N-methylformamide obtained with the RAM36 program.}\\
\hline\hline
$ntr$\tablefootmark{a} & Parameter\tablefootmark{b} & Operator\tablefootmark{c} & Value\tablefootmark{d} \\
\hline
\endfirsthead
\caption{continued.}\\
\hline\hline
$ntr$\tablefootmark{a} & Parameter\tablefootmark{b} & Operator\tablefootmark{c} & Value\tablefootmark{d} \\
\hline
\endhead
\hline
\endfoot
220 & $F    $                      &  $p_\alpha^2       $                             &   5.5825023(37)                 \\
220 & $V_3  $                      &  $\frac{1}{2}(1-\cos3\alpha)$                           &   51.7199088(90)                \\
211 & $\rho $                      &  $J_zp_\alpha      $                             &   0.080976579(84)               \\
202 & $A_{RAM}-0.5(B_{RAM}+C_{RAM})$ & $ J_z^2 $                                      &   0.3540036(48)                 \\
202 & $0.5(B_{RAM}+C_{RAM})$       &  $J^2                                          $ &   0.22175219(40)                \\
202 & $0.5(B_{RAM}-C_{RAM})$       &  $J_x^2-J_y^2                                  $ &   0.05717595(45)                \\
202 & $D_{zx}     $                &  $\lbrace J_z,J_x \rbrace                      $ &  $ -0.155070742(38)             $ \\
440 & $F_m        $                &  $p_\alpha^4                                   $ &  $ -0.67979(73)\times 10^{-3}   $ \\
440 & $V_6        $                &  $\frac{1}{2}(1-\cos 6\alpha)                          $ &  $ 8.02866(40)                  $ \\
431 & $\rho_m     $                &  $J_zp_\alpha^3                                $ &  $ 0.222402(74)\times 10^{-3}   $ \\
422 & $F_J        $                &  $J^2p_\alpha^2                                $ &  $ -0.1029(15)\times 10^{-6}    $ \\
422 & $F_K        $                &  $J_z^2p_\alpha^2                              $ &  $ -0.63309(14)\times 10^{-4}   $ \\
422 & $F_{xy}     $                &  $p_\alpha^2(J_x^2-J_y^2)                      $ &  $ -0.23266(15)\times 10^{-5}   $ \\
422 & $F_{zx}     $                &  $\frac{1}{2}p_\alpha^2\lbrace J_z,J_x \rbrace         $ &  $ 0.226160(59)\times 10^{-4}   $ \\
422 & $V_{3J}     $                &  $J^2(1-\cos 3\alpha)                            $ &  $ -0.15939626(49)\times 10^{-2}$ \\
422 & $V_{3K}     $                &  $J_z^2(1-\cos 3\alpha)                          $ &  $ 0.8718486(73)\times 10^{-2}  $ \\
422 & $V_{3zx}    $                &  $\frac{1}{2}(1-\cos 3\alpha)\lbrace J_z,J_x \rbrace     $ &  $ -0.9181495(26)\times 10^{-2} $ \\
422 & $V_{3xy}    $                &  $(J_x^2-J_y^2)(1-\cos 3\alpha)                  $ &  $ -0.194794(44)\times 10^{-4}  $ \\
422 & $D_{3xy}    $                &  $\frac{1}{2}\sin 3\alpha\lbrace J_x,J_y \rbrace         $ &  $ 0.235469(41) \times 10^{-3}  $ \\
413 & $\rho_J     $                &  $J^2J_zp_\alpha                               $ &  $ 0.240348(50)\times 10^{-5}   $ \\
413 & $\rho_K     $                &  $J_z^3p_\alpha                                $ &  $ 0.112304(16)\times 10^{-4}   $ \\
413 & $\rho_{zx}  $                &  $\frac{1}{2}p_\alpha \lbrace J_z^2,J_x \rbrace         $ &  $ -0.186678(14)\times 10^{-4}  $ \\
413 & $\rho_{xy}  $                &  $\frac{1}{2}p_\alpha \lbrace J_z,(J_x^2-J_y^2) \rbrace$ &  $ 0.473931(43)\times 10^{-5}   $ \\
404 & $D_{zxK}    $                &  $\lbrace J_z^3,J_x \rbrace                    $ &  $ 0.203044(10)\times 10^{-5}   $ \\
404 & $\Delta_J   $                &  $-J^4                                         $ &  $ 0.300775(22)\times 10^{-6}   $ \\
404 & $\Delta_{JK}$                &  $-J^2J_z^2                                    $ &  $ -0.83337(25)\times 10^{-6}   $ \\
404 & $\Delta_K   $                &  $-J_z^4                                       $ &  $ 0.302147(39)\times 10^{-5}   $ \\
404 & $\delta_J   $                &  $-2J^2(J_x^2-J_y^2)                           $ &  $ 0.108271(13)\times 10^{-6}   $ \\
404 & $\delta_K   $                &  $-\lbrace J_z^2,(J_x^2-J_y^2) \rbrace         $ &  $ 0.367453(69)\times 10^{-6}   $ \\
660 & $F_{mm}     $                &  $p_\alpha^6                                                                     $ & $ -0.10119(30)\times 10^{-4}  $ \\
660 & $V_9        $                &  $\frac{1}{2}(1-\cos 9\alpha)                                                            $ & $ 2.0575(23)                  $ \\
651 & $\rho_{mm}  $                &  $p_\alpha^5J_z                                                                  $ & $ 0.26713(46)\times 10^{-5}   $ \\
642 & $F_{mJ}     $                &  $J^2p_\alpha^4                                                                  $ & $ 0.3393(19)\times 10^{-7}    $ \\
642 & $F_{mK}     $                &  $J_z^2p_\alpha^4                                                                $ & $ -0.6445(16)\times 10^{-6}   $ \\
642 & $F_{mxy}    $                &  $p_\alpha^4(J_x^2-J_y^2)                                                        $ & $ 0.2157(17)\times 10^{-7}    $ \\
642 & $F_{mzx}    $                &  $\frac{1}{2}p_\alpha^4 \lbrace J_z,J_x \rbrace                                          $ & $ -0.1406(36)\times 10^{-7}   $ \\
642 & $V_{6J}     $                &  $J^2(1-\cos 6\alpha)                                                            $ & $ -0.124130(82) \times 10^{-3}$ \\
642 & $V_{6K}     $                &  $J_z^2(1-\cos 6\alpha)                                                          $ & $ 0.4820(10)\times 10^{-3}    $ \\
642 & $V_{6zx}    $                &  $\frac{1}{2}(1-\cos 6\alpha)\lbrace J_z,J_x  \rbrace                                    $ & $ -0.543394(97)\times 10^{-3} $ \\
642 & $V_{6xy}    $                &  $(1-\cos 6\alpha)(J_x^2-J_y^2)                                                  $ & $ -0.20435(93)\times 10^{-4}  $ \\
642 & $D_{6xy}    $                &  $\frac{1}{2} \sin 6\alpha \lbrace J_x,J_y \rbrace                                       $ & $ 0.6217(56)\times 10^{-4}    $ \\
642 & $D_{6zy}    $                &  $\frac{1}{2} \sin 6\alpha \lbrace J_z,J_y \rbrace                                       $ & $ -0.12563(23)\times 10^{-3}  $ \\
633 & $\rho_{mJ}  $                &  $J^2p_\alpha^3J_z                                                               $ & $ -0.3725(33)\times 10^{-8}   $ \\
633 & $\rho_{mK}  $                &  $J_z^3p_\alpha^3                                                                $ & $ 0.5362(28)\times 10^{-7}    $ \\
633 & $\rho_{mxy} $                &  $\frac{1}{2}p_\alpha^3\lbrace J_z,(J_x^2-J_y^2) \rbrace                                 $ & $ -0.819(25)\times 10^{-9}    $ \\
633 & $\rho_{mzx} $                &  $\frac{1}{2}p_\alpha^3 \lbrace J_z^2,J_x \rbrace                                        $ & $ -0.1058(18)\times 10^{-7}   $ \\
624 & $F_{JJ}     $                &  $J^4p_\alpha^2                                                                  $ & $ 0.2173(36)\times 10^{-10}   $ \\
624 & $F_{KK}     $                &  $J_z^4p_\alpha^2                                                                $ & $ -0.2049(35)\times 10^{-8}   $ \\
624 & $F_{zxK}    $                &  $\frac{1}{2}p_\alpha^2 \lbrace J_z^3,J_x \rbrace                                        $ & $ -0.1137(15) \times 10^{-8}  $ \\
624 & $V_{3JJ}    $                &  $J^4(1-\cos 3\alpha)                                                            $ & $ 0.131102(86)\times 10^{-7}  $ \\
624 & $V_{3JK}    $                &  $J^2J_z^2(1-\cos 3\alpha)                                                       $ & $ 0.15148(95)\times 10^{-7}   $ \\
624 & $V_{3KK}    $                &  $J_z^4(1-\cos3\alpha)                                                           $ & $ -0.5940(31)\times 10^{-7}   $ \\
624 & $V_{3zxJ}   $                &  $\frac{1}{2}J^2(1-\cos 3\alpha) \lbrace J_z,J_x \rbrace                                 $ & $ 0.3290(72)\times 10^{-8}    $ \\
624 & $V_{3xyJ}   $                &  $J^2(1-\cos 3\alpha)(J_x^2-J_y^2)                                               $ & $ -0.2988(10)\times 10^{-8}   $ \\
624 & $V_{3xyK}   $                &  $\frac{1}{2}(1-\cos 3\alpha)\lbrace J_z^2,(J_x^2-J_y^2) \rbrace                         $ & $ -0.1819(10)\times 10^{-7}   $ \\
624 & $V_{3zxK}   $                &  $\frac{1}{2}(1-\cos 3\alpha)\lbrace J_z^3,J_x \rbrace                                   $ & $ -0.9915(12)\times 10^{-7}   $ \\
624 & $V_{3zxx}   $                &  $\frac{1}{2}\cos 3\alpha \lbrace J_z,J_x^3 \rbrace                                      $ & $ -0.15343(15)\times 10^{-6}  $ \\
624 & $V_{3x4y4}  $                &  $\cos 3\alpha (J_x^4+J_y^4)                                                     $ & $ 0.7525(14)\times 10^{-8}    $ \\
624 & $D_{3xyJ}   $                &  $\frac{1}{2}J^2 \sin 3\alpha \lbrace J_x,J_y \rbrace                                    $ & $ -0.6130(23)\times 10^{-8}   $ \\
624 & $D_{3xyK}   $                &  $\frac{1}{2}\sin 3\alpha \lbrace J_z^2,J_x,J_y \rbrace                                  $ & $ 0.23854(49)\times 10^{-6}   $ \\
624 & $D_{3zyJ}   $                &  $\frac{1}{2} J^2\sin 3\alpha \lbrace J_z,J_y \rbrace                                    $ & $ -0.6514(19)\times 10^{-7}   $ \\
624 & $D_{3zyK}   $                &  $\frac{1}{2}\sin 3\alpha \lbrace J_z^3,J_y \rbrace                                      $ & $ -0.5861(61)\times 10^{-7}   $ \\
624 & $D_{3zyy}   $                &  $\frac{1}{2}\sin 3\alpha \lbrace J_z,J_y^3 \rbrace                                      $ & $ 0.8302(24)\times 10^{-7}    $ \\
624 & $D_{3x3yxy3}$                &  $\frac{1}{2}\sin 3\alpha \left[\lbrace J_x^3,J_y \rbrace-\lbrace J_x,J_y^3\rbrace \right]$ & $ 0.4008(24)\times 10^{-8}    $ \\
615 & $\rho_{JJ}  $                &  $J^4J_zp_\alpha                                                                 $ & $ -0.1203(18)\times 10^{-10}  $ \\
615 & $\rho_{JK}  $                &  $J^2J_z^3p_\alpha                                                               $ & $ -0.6148(28) \times 10^{-9}  $ \\
615 & $\rho_{KK}  $                &  $J_z^5p_\alpha                                                                  $ & $ 0.4502(53)\times 10^{-9}    $ \\
615 & $\rho_{zxJ} $                &  $\frac{1}{2}J^2p_\alpha \lbrace J_z^2,J_x \rbrace                                       $ & $ 0.475(11)\times 10^{-10}    $ \\
615 & $\rho_{zxK} $                &  $\frac{1}{2}p_\alpha \lbrace J_z^4,J_x \rbrace                                          $ & $ 0.13926(45)\times 10^{-8}   $ \\
615 & $\rho_{x2y2}$                &  $\frac{1}{2}p_\alpha \lbrace J_z,J_x^2,J_y^2 \rbrace                                    $ & $ 0.23934(84)\times 10^{-9}   $ \\
606 & $D_{zxKK}   $                &  $\lbrace J_z^5,J_x \rbrace                                                              $ & $ -0.7781(20)\times 10^{-10}  $ \\
606 & $\Phi_J     $                &  $J^6                                                                            $ & $ 0.4542(10)\times 10^{-12}   $ \\
606 & $\Phi_{KJ}  $                &  $J^2J_z^4                                                                       $ & $ 0.3394(14)\times 10^{-10}   $ \\
606 & $\Phi_K     $                &  $J_z^6                                                                          $ & $ 0.2619(35)\times 10^{-10}   $ \\
606 & $\phi_J     $                &  $2J^4(J_x^2-J_y^2)                                                              $ & $ 0.23367(53)\times 10^{-12}  $ \\
606 & $\phi_{JK}  $                &  $J^2\lbrace J_z^2,(J_x^2-J_y^2) \rbrace                                         $ & $ -0.6207(42)\times 10^{-12}  $ \\
606 & $\phi_K     $                &  $\lbrace J_z^4,(J_x^2-J_y^2) \rbrace                                            $ & $ 0.22553(51)\times 10^{-10}  $ \\
871 & $ \rho_{mmm} $               &  $ J_zp_\alpha^7               $&                          $ 0.7121(15)\times 10^{-7}   $          \\
862 & $ F_{mmJ}    $               &  $ J^2p_\alpha^6               $&                          $ 0.4570(73)\times 10^{-9}   $          \\
862 & $ F_{mmK}    $               &  $ J_z^2p_\alpha^6             $ &                         $ -0.19488(81)\times 10^{-7} $           \\
862 & $ F_{mmxy}   $               &  $ p_\alpha^6(J_x^2-J_y^2)     $     &                     $ 0.3816(70)\times 10^{-9}   $             \\
862 & $ V_{9J}     $               &  $ J^2(1-\cos 9\alpha)         $  &                        $ -0.15061(48)\times 10^{-3} $          \\
862 & $ V_{9K}     $               &  $ J_z^2(1-\cos 9\alpha)       $   &                       $ 0.1550(58)\times 10^{-3}   $           \\
862 & $ V_{9xy}    $               &  $ (1-\cos 9\alpha)(J_x^2-J_y^2)$ &                        $ -0.5022(55)\times 10^{-4}  $          \\
862 & $ D_{9xy}    $               &  $ \frac{1}{2}\sin 9\alpha \lbrace J_x,J_y \rbrace $       &       $ -0.1195(32)\times 10^{-3}  $         \\
853 & $ \rho_{mmK} $               &  $ J_z^3p_\alpha^5  $            &                         $ 0.2984(20)\times 10^{-8}   $         \\
853 & $ \rho_{mmzx}$               &  $ \frac{1}{2}p_\alpha^5 \lbrace J_z^2,J_x \rbrace$          &     $ -0.411(11)\times 10^{-9}   $          \\
844 & $ F_{mJK}    $               &  $ J^2J_z^2p_\alpha^4$            &                        $ 0.980(55)\times 10^{-11}   $          \\
844 & $ F_{mKK}    $               &  $ J_z^4p_\alpha^4   $           &                         $ -0.2444(30)\times 10^{-9}  $         \\
844 & $ F_{mxyK}   $               &  $ \frac{1}{2}p_\alpha^4 \lbrace J_z^2,(J_x^2-J_y^2) \rbrace $ &   $ 0.1692(29)\times 10^{-10}  $            \\
844 & $ V_{6zxx}   $               &  $ \frac{1}{2}\cos 6\alpha \lbrace J_z,J_x^3 \rbrace   $    &      $ -0.1034(16)\times 10^{-7}  $         \\
844 & $ D_{6xyJ}   $               &  $ \frac{1}{2}J^2\sin 6\alpha \lbrace J_x,J_y \rbrace  $    &      $ -0.1224(35)\times 10^{-8}  $         \\
844 & $ D_{6zyK}   $               &  $ \frac{1}{2}\sin 6\alpha \lbrace J_z^3,J_y \rbrace   $    &      $ 0.573(14)\times 10^{-7}    $         \\
835 & $ \rho_{mKK} $               &  $ J_z^5p_\alpha^3            $  &                         $ 0.1312(35)\times 10^{-10}  $        \\
826 & $ F_{KKK}    $               &  $ J_z^6p_\alpha^2            $  &                         $ -0.491(25)\times 10^{-12}  $        \\
826 & $ V_{3JJK}   $               &  $ J^4J_z^2(1-\cos 3\alpha)   $   &                        $ -0.365(13)\times 10^{-12}  $         \\
826 & $ V_{3KKK}   $               &  $ J_z^6(1-\cos 3\alpha)      $    &                       $ -0.909(34)\times 10^{-11}  $          \\
826 & $ V_{3zxxx}  $               &  $ \frac{1}{2}\cos3\alpha \lbrace J_z,J_x^5 \rbrace      $ &       $ 0.675(14)\times 10^{-12}   $        \\
826 & $ D_{3xyJK}  $               &  $ \frac{1}{2}J^2\sin 3\alpha \lbrace J_z^2,J_x,J_y \rbrace$ &     $ -0.1837(65)\times 10^{-11} $          \\
826 & $ D_{3xyKK}  $               &  $ \frac{1}{2}\sin 3\alpha \lbrace J_z^4,J_x,J_y \rbrace   $ &     $ -0.1309(33)\times 10^{-10} $          \\
\multicolumn{4}{c}{}\\
  & $ \chi_{aa} $ & $0.70093(90)\times 10^{-4}$  & \\
  & $ \chi_{bb} $ & $0.64466(92)\times 10^{-4}$  & \\
  & $ 2\chi_{ab} $ & $0.1755(15)\times 10^{-4} $  & \\
\end{longtable} 
\tablefoot{
\tablefoottext{a}{$n = t + r$, where $n$ is the total order of the operator, $t$ is the order of the torsional part and $r$ is the order of the rotational part, respectively.}
\tablefoottext{b}{Parameter nomenclature based on the subscript procedures of \citet{Xu2008305}}
\tablefoottext{c}{$\lbrace A,B,C \rbrace = ABC+CBA$, $\lbrace A,B \rbrace = AB+BA$. The product of the operator in the third column of a given row and the parameter in the second 
column of that row gives the term actually used in the torsion-rotation Hamiltonian of the program, except for $F$, $\rho$ and $A_{RAM}$, which occur in the Hamiltonian in the form 
$F(p_\alpha-\rho P_a)^2+A_{RAM}P_a^2$.}
\tablefoottext{a}{All values are in cm$^{-1}$ (except $\rho$ which is unitless). Statistical uncertainties are shown as one standard uncertainty in the units of the last two digits. }
}
}

The complete data set treated at the final stage of the current study 
includes both our new data and data from the literature 
\citep{fantoni1996rotational, fantoni2002very, kawashima2010dynamical}. The 
data set contains 12456 $A$- and $E$-type transitions with 
$J \leq 62$ and $K_a \leq 21$ for \textit{trans} N-methylformamide in 
the lowest three torsional 
states. Due to blending these 12456 transitions correspond to 9469 distinct 
line frequencies (mainly because of not fully resolved quadrupole hyperfine 
structure). The fit chosen as the "best" one uses a model consisting of 103 
parameters. The weighted root-mean-square deviation of the fit of 12456 
microwave transition frequencies with $J \leq 62$ is 0.84, indicating that 
assumed statistical uncertainties were slightly overestimated. 
The largest residual of 0.293 MHz is observed in the fit for the 
$v_t=0$ $E$-symmetry species transition $49_{9,41} \leftarrow 48_{9,40}$.
The final set of 
molecular parameters is presented in Table~\ref{tabpar}. The final 
data set of fitted transitions of the N-methylformamide \textit{trans} 
conformer is presented in Table~\ref{t:fit}, where we provide quantum numbers 
for each level, followed by observed transition frequencies, measurement 
uncertainties and residuals from the fit. 
The complete version of Table~\ref{t:fit} is available at the CDS, 
here only part of the table is presented for illustration purposes.

\begin{table*}
\centering
 \caption{Measured transitions of \textit{trans} N-methylformamide in the 
 $v_t=$ 0, 1, and 2 states.}  
 \label{t:fit}	
\begin{tabular}{ccccccccccccc}
\hline\hline                                                                                                                 
$m'$ & $F'$\tablefootmark{a} & $J'$ & $K_a'$ & $K_c'$ & $m''$ & $F''$\tablefootmark{a} & $J''$ & $K_a''$ & $K_c''$ &Obs. freq. & Uncertainty &  Obs.-calc.   \\ 
     &      &      &        &        &       &       &       &         &         &  (MHz)     &    (MHz)   &   (MHz)       \\
\hline                                                                                                                                           
 -3 & 18 & 19  & 3 & 16  & -3 & 17 & 18 &  2 & 16  &  251285.2870 & 0.1000  &    0.0203 \\
 -3 & 20 & 19  & 3 & 16  & -3 & 19 & 18 &  2 & 16  &  251285.2870 & 0.1000  &   -0.0235 \\
 -3 & 19 & 19  & 3 & 16  & -3 & 18 & 18 &  2 & 16  &  251286.1600 & 0.1000  &    0.0037 \\
  3 &    & 21  & 4 & 17  &  3 &    & 20 &  5 & 16  &  251379.8920 & 0.0500  &    0.0007 \\
 -2 &    & 23  & 3 & 21  & -2 &    & 22 &  3 & 20  &  251386.7420 & 0.0500  &   -0.0109 \\
 -2 &    & 23  & 3 & 21  & -2 &    & 22 &  3 & 20  &  251386.7430 & 0.0500  &   -0.0099 \\
  3 &    & 22  & 2 & 21  &  3 &    & 21 &  2 & 19  &  251468.4440 & 0.0500  &    0.0002 \\
  0 &    & 22  & 8 & 15  &  0 &    & 21 &  8 & 14  &  251472.2980 & 0.0500  &   -0.0016 \\
  0 &    & 22  & 8 & 14  &  0 &    & 21 &  8 & 13  &  251493.1300 & 0.0500  &    0.0136 \\
  1 &    & 22  & 8 & 15  &  1 &    & 21 &  8 & 14  &  251565.7320 & 0.0500  &    0.0034 \\
  0 & 18 & 19  & 4 & 16  &  0 & 17 & 18 &  3 & 15  &  251591.3020 & 0.0500  &    0.0035 \\
  0 & 20 & 19  & 4 & 16  &  0 & 19 & 18 &  3 & 15  &  251591.3020 & 0.0500  &   -0.0310 \\
  0 & 19 & 19  & 4 & 16  &  0 & 18 & 18 &  3 & 15  &  251591.9750 & 0.0500  &   -0.0035 \\
 \hline                                                                       
\end{tabular}
\tablefoot{The complete table is available at the CDS.
\tablefoottext{a}{The quantum number $F$ is not indicated for the transitions with unresolved hyperfine structure.}
}
\end{table*}

\begin{figure}
\centering
\resizebox{\hsize}{!}{\includegraphics{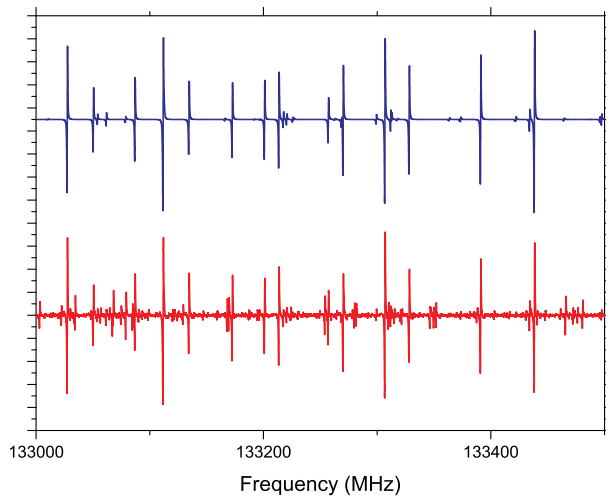}}
\vspace*{-6.5ex}
\caption{Predicted (upper panel) and measured (lower panel) rotational 
spectrum of N-methylformamide between 133 and 133.5~GHz. The observed 
lineshapes correspond to the first derivative of the actual line profile 
because frequency modulation and first harmonic 
lock-in detection is employed. The intensity axis is in arbitrary units.}
\label{fig:spec}
\end{figure}

Comparison of the low-order parameter values from Table~\ref{tabpar} with the 
corresponding  parameters determined by \citet{kawashima2010dynamical} reveals 
rather significant shifts in the values on a background of a general 
qualitative agreement (see Table~\ref{t:Scompar}). 
Part of these shifts come from the basic difference in 
the models used, although both refer to the Rho-axis method.
We follow the definition of the 
Rho-axis method given in \citet{hougen1994ram} and, in our case, 
the $A$ and $E$ species are treated together with a single set of rotational 
parameters. On the contrary, 
\citet{kawashima2010dynamical} treated the $A$ and $E$ species with distinct
sets of rotational parameters (see their Table~2).
As a result, a direct comparison of the changes in the rotational parameters 
is not possible. The same is true for the centrifugal distortion 
constants. At the same time, the change in $V_3$ value is in good agreement 
with the prediction made in \citet{kawashima2010dynamical} where a quite large 
$V_6$ term in the potential function was postulated by analogy with acetamide 
\citep{ilyushin2004acetamide}. The data set available in 
\citet{kawashima2010dynamical} was limited to the ground torsional state 
transitions only and did not give the opportunity to determine the $V_6$ value. 
Thus they examined how the $V_3$ potential barrier is changed when a $V_6$ 
term is added assuming that the coupling of the vibrational modes of the NH 
group with the CH$_3$ internal rotation has an effect similar to the one 
observed for the NH$_2$ group in acetamide \citep{hirota2010acetamide}. 
According to our 
results, the value of an index $R$ used in \citet{kawashima2010dynamical} to 
express the effect of the $V_6$ term quantitatively is estimated to be 26.5 
which is in a relatively good agreement with the value of 23.0 obtained in 
\citet{kawashima2010dynamical} on the basis of ground state data only. Thus our 
results support the general analysis of the CH$_3$ internal rotation potential 
barrier in N-methylformamide provided in \citet{kawashima2010dynamical}.

A portion of the rotational spectrum of N-methylformamide measured around 
133~GHz in the laboratory is shown in Fig.~\ref{fig:spec} and compared to the 
predicted rotational spectrum as provided by our current theoretical model. As 
can be seen from Fig.~\ref{fig:spec}, the overall correspondence between 
experimental and theoretical spectra is very good. A slight inconsistency with
intensity between predicted and observed spectra that may be visible for some 
strong lines is due to variations of source power and detector sensitivity.

\begin{table*}
\centering
 \caption{Predicted transitions of \textit{trans} N-methylformamide in the
 $v_t=$ 0, 1, and 2 states.}  
 \label{t:pred}	
\begin{tabular}{cccccccccccccc}
\hline\hline                                                                                                                 
$m'$ & $F'$ & $J'$ & $K_{\rm a}'$ & $K_{\rm c}'$ & $m''$ & $F''$ & $J''$ & $K_{\rm a}''$ & $K_{\rm c}''$ &Calc. freq. & Uncertainty &  $E_{\rm l}$    & $\mu^2S$  \\ 
     &      &      &        &        &       &       &       &         &         &  {\small (MHz)}     &    {\small (MHz)}    &  {\small (cm$^{-1}$)} &   {\small (D$^2$)} \\
\hline                                                                                                                                           
  1 & 30 &  30 & 15 & 15  &  1 & 30 &  30 & 14 & 17  & 393571.2045 &  0.0384 &    266.1079  &  20.6    \\
  1 & 31 &  30 & 15 & 15  &  1 & 31 &  30 & 14 & 17  & 393571.2531 &  0.0384 &    266.1079  &  21.3    \\
  1 & 29 &  30 & 15 & 15  &  1 & 29 &  30 & 14 & 17  & 393571.2547 &  0.0384 &    266.1079  &  20.0    \\
  0 & 24 &  24 &  5 & 19  &  0 & 24 &  24 &  3 & 22  & 393583.8360 &  0.0055 &    115.5411  &  0.0382  \\
  0 & 25 &  24 &  5 & 19  &  0 & 25 &  24 &  3 & 22  & 393585.7457 &  0.0055 &    115.5411  &  0.0398  \\
  0 & 23 &  24 &  5 & 19  &  0 & 23 &  24 &  3 & 22  & 393585.8253 &  0.0055 &    115.5411  &  0.0367  \\
  0 & 15 &  15 &  8 &  8  &  0 & 15 &  15 &  6 &  9  & 393600.5805 &  0.0054 &     62.7047  &  0.0364  \\
  0 & 16 &  15 &  8 &  8  &  0 & 16 &  15 &  6 &  9  & 393600.7174 &  0.0054 &     62.7047  &  0.0389  \\
  0 & 14 &  15 &  8 &  8  &  0 & 14 &  15 &  6 &  9  & 393600.7265 &  0.0054 &     62.7047  &  0.0342  \\
 -3 & 41 &  41 &  3 & 38  & -3 & 40 &  40 &  3 & 37  & 393611.4811 &  0.0389 &    337.7355  &  36.9    \\
 -3 & 42 &  41 &  3 & 38  & -3 & 41 &  40 &  3 & 37  & 393611.5587 &  0.0389 &    337.7355  &  37.8    \\
 -3 & 40 &  41 &  3 & 38  & -3 & 39 &  40 &  3 & 37  & 393611.5617 &  0.0389 &    337.7355  &  36.0    \\
 \hline                                                                       
\end{tabular}
\tablefoot{The complete table is available at the CDS.
}
\end{table*}

The predictions of rotational transitions of \textit{trans} N-methylformamide 
in the $v_t = $ 0, 1, and 2 torsionally excited states resulting from the fit 
are presented in Table~\ref{t:pred}. They are calculated for the frequency 
range up to 650~GHz and for the transitions with $J \leq 65$. The table 
provides quantum numbers, followed by calculated transition frequencies and 
their uncertainties, the energy of the lower state and the product $\mu^2S$, 
where $\mu$ is the dipole moment of the molecule and $S$ is the line strength 
of the transition. Owing to its significant size the complete version of 
Table~\ref{t:pred} is available at the CDS.

\begin{table}
 \begin{center}
 \caption{Partition functions of N-methylformamide and acetamide.}
 \label{t:qpart}
 \vspace*{-3.0ex}
 \begin{tabular}{rrccrc}
 \hline\hline
 \multicolumn{1}{c}{} & \multicolumn{2}{c}{CH$_3$NHCHO} && \multicolumn{2}{c}{CH$_3$CONH$_2$} \\ 
\cline{2-3} \cline{5-6} \\
 \multicolumn{1}{c}{$T$} & \multicolumn{1}{c}{$Q_{\rm tr}$$^{a}$} & \multicolumn{1}{c}{$Q_{\rm v}$$^{b}$} && \multicolumn{1}{c}{$Q_{\rm tr}$$^{a}$} & \multicolumn{1}{c}{$Q_{\rm v}$$^{b}$} \\ 
 \multicolumn{1}{c}{\scriptsize (K)} & & && & \\
 \hline
 10 &    365.16247 & 1.00000  &&    359.18058 & 1.00000  \\
 20 &   1189.23944 & 1.00000  &&   1272.66459 & 1.00000  \\
 30 &   2488.15240 & 1.00000  &&   2766.98655 & 1.00000  \\
 40 &   4317.90735 & 1.00008  &&   4871.11731 & 1.00009  \\
 50 &   6701.57116 & 1.00058  &&   7590.48345 & 1.00058  \\
 60 &   9648.68076 & 1.00224  &&  10925.10798 & 1.00205  \\
 70 &  13164.31826 & 1.00592  &&  14874.85986 & 1.00508  \\
 80 &  17251.81144 & 1.01233  &&  19439.98556 & 1.01017  \\
 90 &  21913.57312 & 1.02193  &&  24620.89439 & 1.01769  \\
100 &  27151.45535 & 1.03496  &&  30418.00805 & 1.02792  \\
110 &  32966.94532 & 1.05150  &&  36831.70759 & 1.04108  \\
120 &  39361.28161 & 1.07156  &&  43862.32239 & 1.05734  \\
130 &  46335.51849 & 1.09509  &&  51510.12815 & 1.07686  \\
140 &  53890.55428 & 1.12204  &&  59775.34167 & 1.09979  \\
150 &  62027.13560 & 1.15237  &&  68658.10939 & 1.12626  \\
160 &  70745.84645 & 1.18606  &&  78158.49050 & 1.15642  \\
170 &  80047.08923 & 1.22311  &&  88276.43644 & 1.19042  \\
180 &  89931.06287 & 1.26356  &&  99011.76908 & 1.22844  \\
190 & 100397.74181 & 1.30747  && 110364.15950 & 1.27064  \\
200 & 111446.85817 & 1.35493  && 122333.10876 & 1.31725  \\
210 & 123077.88845 & 1.40605  && 134917.93177 & 1.36849  \\
220 & 135290.04535 & 1.46097  && 148117.74490 & 1.42463  \\
230 & 148082.27451 & 1.51987  && 161931.45737 & 1.48594  \\
240 & 161453.25592 & 1.58293  && 176357.76649 & 1.55275  \\
250 & 175401.40932 & 1.65036  && 191395.15645 & 1.62543  \\
260 & 189924.90280 & 1.72242  && 207041.90022 & 1.70434  \\
270 & 205021.66401 & 1.79936  && 223296.06409 & 1.78994  \\
280 & 220689.39314 & 1.88148  && 240155.51446 & 1.88269  \\
290 & 236925.57717 & 1.96908  && 257617.92629 & 1.98311  \\
300 & 253727.50471 & 2.06251  && 275680.79287 & 2.09175  \\
310 & 271092.28113 & 2.16213  && 294341.43638 & 2.20925  \\
320 & 289016.84348 & 2.26834  && 313597.01910 & 2.33626  \\
330 & 307497.97493 & 2.38156  && 333444.55472 & 2.47352  \\
\hline
 \end{tabular}
 \end{center}
 \vspace*{-2.5ex}
 \tablefoot{
 \tablefoottext{a}{$Q_{\rm tr}$ is the torsional-rotational partition function. 
 It does not take the hyperfine splitting into account.}
 \tablefoottext{b}{$Q_{\rm v}$ is the vibrational partition function. The total
 partition function of the molecule (without hyperfine splitting) is 
$Q_{\rm tr} \times Q_{\rm v}$.}
 }
 \end{table}

We provide the torsional-rotational ($Q_{\rm tr}$) and vibrational 
($Q_{\rm v}$) partition functions of N-methylformamide in Table~\ref{t:qpart}. 
The values of $Q_{\rm tr}$ were calculated from first principles, i.e. via 
direct summation over the rotational-torsional states. The maximum value of 
the $J$ quantum number of the energy levels taken into account to calculate the 
partition function is 130, and excited torsional states up to 
$\varv_{\rm t} = 8$ were considered. The vibrational part, $Q_{\rm v}$, was 
estimated using an harmonic approximation and a simple formula that may be 
found in \citet[][, see their Eq.~3.60]{gordycook84}. The frequencies of the 
normal modes were obtained from DFT calculations of the harmonic force field 
using the B3LYP method and a 6-311++(3df,2pd) basis set. Table~\ref{t:qpart} 
also lists the partition function values of acetamide that we calculated in a 
similar way as for N-methylformamide. The torsional-rotational part was 
calculated from first principales on the basis of the results presented in the 
paper by \citet{ilyushin2004acetamide}. To compute the $Q_{\rm v}$ values, we 
used, when available, vibrational frequencies reported in the literature 
\citep{kutz62, kydd80}, but also the results of DFT calculations with the same 
method and basis set as for N-methylformamide. For both molecules, the values 
of $Q_{\rm v}$ were calculated by taking all the vibrational modes into 
account except for the torsional mode which is already considered in 
$Q_{\rm tr}$. The full partition function, $Q_{\rm tot}$, is thus the product of 
$Q_{\rm tr}$ and $Q_v$. 

\section{Astronomical results}
\label{s:obsresults}

\subsection{Observations}
\label{ss:observations}

We use the full data set of the EMoCA spectral line survey obtained 
toward Sgr~B2(N) with ALMA in its Cycles 0 and 1. The survey covers the 
frequency range between 84.1 and 114.4~GHz with a spectral resolution of 
488.3~kHz (1.7 to 1.3~km~s$^{-1}$). The median angular resolution is 
1.6$\arcsec$. A detailed description of the observations, the data reduction 
process, and the method used to identify the detected lines and derive column 
densities was presented in \citet{Belloche16}. Population diagrams are 
constructed in the same way as in our previous work. Here, we would like to 
emphasize the fact that the apparent discrepancy between the synthetic 
populations and the fit to the observed populations, the former lying below 
the latter (see, e.g., Fig~\ref{f:popdiag_ch3nhcho}b), results from the fact 
that the model is optimized to match (i.e., not overestimate) the peak 
temperatures of the detected 
transitions while the population diagram is based on integrated intensities. 
Because of the high level of line confusion, the wings of the detected lines 
are often still partially contaminated even after removing the contribution of 
the other known molecules included in our complete model. Therefore, the 
populations derived from the integrated intensities are most of the time a bit 
overestimated.

The complete model mentioned above refers to the synthetic spectrum 
that includes the emission of all the molecules that we have identified in 
Sgr~B2(N2) so far 
\citep[][]{Belloche13,Belloche14,Belloche16,Mueller16a,Mueller16b,Margules16}. 
It is overlaid in green in all figures that display observed spectra.

We focus our analysis on the secondary hot core Sgr~B2(N2), which is located
$\sim$5$\arcsec$ to the North of the main hot core Sgr~B2(N1) 
\citep[][]{Belloche16}. Sgr~B2(N2) has narrower linewidths 
($FWHM \sim 5$~km~s$^{-1}$) and thus exhibits a lower degree of line confusion 
than Sgr~B2(N1). While some of the molecules reported below are also
present on larger scales in the envelope of Sgr~B2 
\citep[see, e.g.,][]{Jones08}, our 
interferometric observations are only sensitive to the compact emission arising
from the embedded hot cores. As shown in Fig.~\ref{f:maps}, the emission
analyzed in this section is compact and the derived column densities refer to
the hot core Sgr~B2(N2) only.

\subsection{Tentative detection of N-methylformamide (CH$_3$NHCHO)}
\label{s:obs_ch3nhcho}

We searched for emission lines of CH$_3$NHCHO toward Sgr~B2(N2) using the
spectroscopic predictions obtained in Sect.~\ref{s:spectro}. We compared the 
observed ALMA spectrum of Sgr~B2(N2) to synthetic spectra of CH$_3$NHCHO 
produced under the assumption of local thermodynamic equilibrium (LTE), which 
is expected to be valid given the high densities of the hot core regions 
probed by the EMoCA survey \citep[][]{Belloche16}. 
Figures~\ref{f:spec_ch3nhcho_ve0}--\ref{f:spec_ch3nhcho_ve2} show all the
transitions of CH$_3$NHCHO in its ground state and its first and second 
torsionally excited states that are covered by our survey and are expected to
contribute significantly to the detected signal for typical hot core 
temperatures (150--200~K). In these figures, the synthetic spectrum containing
the contribution of all molecules that we have identified toward Sgr~B2(N2) so 
far, including CH$_3$NHCHO, is overlaid in green on the observed spectrum, 
while the red spectrum shows the contribution of CH$_3$NHCHO only, as derived 
from our best-fit LTE model. Most transitions of CH$_3$NHCHO are blended with 
lines emitted by other molecules and therefore cannot be unambiguously 
assigned to CH$_3$NHCHO. However, several lines are relatively free of 
contamination and match lines detected toward Sgr~B2(N2), both in terms of 
linewidths and peak temperatures. These lines are marked with a star in Col.
11 of Tables~\ref{t:list_ch3nhcho_ve0}--\ref{t:list_ch3nhcho_ve2}. In total,
five lines can be clearly assigned to CH$_3$NHCHO, four within its 
torsional ground state and one within its first torsionally excited state.
Given the small number of clearly detected lines, we consider our 
detection of CH$_3$NHCHO as tentative rather than secure.

\begin{figure}
\centerline{\resizebox{0.95\hsize}{!}{\includegraphics[angle=0]{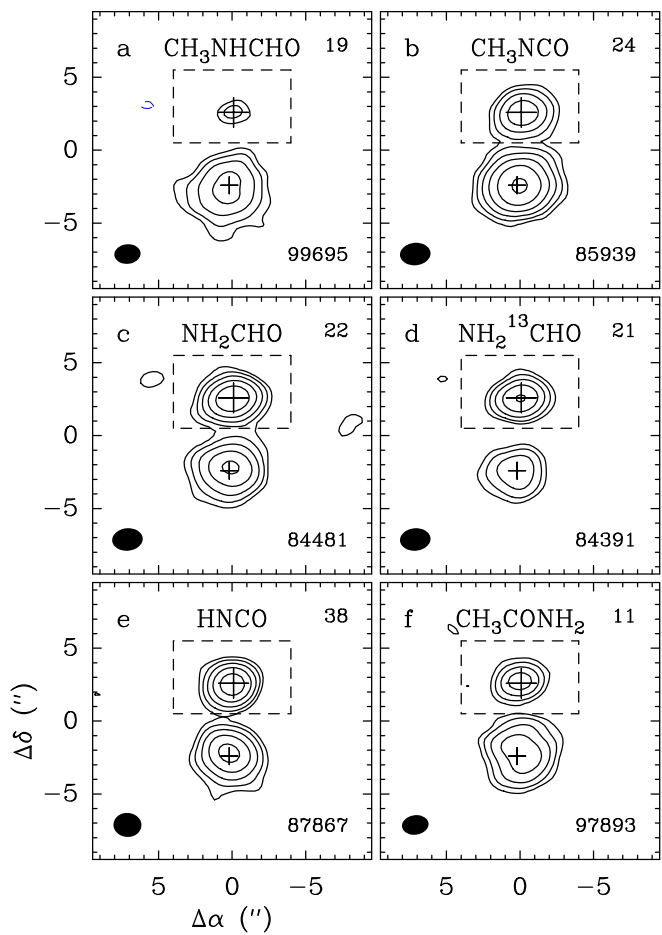}}}
\caption{Integrated intensity maps toward Sgr~B2(N2) of 
\textbf{a} CH$_3$NHCHO $9_{2,8}$--$8_{2,7}$, 
\textbf{b} CH$_3$NCO $10_{1,10}$--$9_{1,9}$,
\textbf{c} NH$_2$CHO $\varv_{12}$=1 $4_{0,4}$--$3_{0,3}$,
\textbf{d} NH$_2$$^{13}$CHO $4_{0,4}$--$3_{0,3}$, 
\textbf{e} HNCO $4_{3,1}$--$3_{3,0}$ and $4_{3,2}$--$3_{3,1}$,
and \textbf{f} CH$_3$CONH$_2$ $9_{0,9}$--$8_{1,8}$, $9_{1,9}$--$8_{1,8}$, 
$9_{0,9}$--$8_{0,8}$, and $9_{1,9}$--$8_{0,8}$. The
position of Sgr B2(N2) is marked with a big cross. The smaller cross indicates
the position of Sgr B2(N1). Because of the different systemic velocities of the
two sources, the line assignment is valid only for Sgr B2(N2), inside the 
dashed box. The contours start at 4$\sigma$ and increase by a factor of 2 at 
each step. The dashed (blue) contour is at $-$4$\sigma$. The value of the rms
noise level, $\sigma$, is given in mJy~beam$^{-1}$~km~s$^{-1}$ in the top 
right corner of each panel. The mean frequency of the integration range in 
MHz and the synthesized beam are shown in the bottom right and left corners of 
each panel, respectively.
}
\label{f:maps}
\end{figure}

Gaussian fits to the integrated intensity maps of four of the five detected
lines\footnote{We did not attempt to fit a size to the integrated intensity map 
of the line at 111.22~GHz because it is somewhat more contaminated on its 
low-frequency side.} indicate that the size 
of the emission is smaller than the beam. We obtain a size of 
$\sim$1.8$\arcsec$ for the line at 93.41~GHz observed in setup S3 but it is 
the setup with the worst angular resolution ($2.9\arcsec \times 1.5\arcsec$)
so we do not trust this size much. Indeed the other lines, at 91.89, 99.69 
(see Fig.~\ref{f:maps}a), 
and 113.61~GHz, are all unresolved, suggesting a size smaller than 1$\arcsec$.
We adopt a size of 0.9$\arcsec$ for our LTE modelling of CH$_3$NHCHO.

\begin{figure}
\centerline{\resizebox{0.95\hsize}{!}{\includegraphics[angle=0]{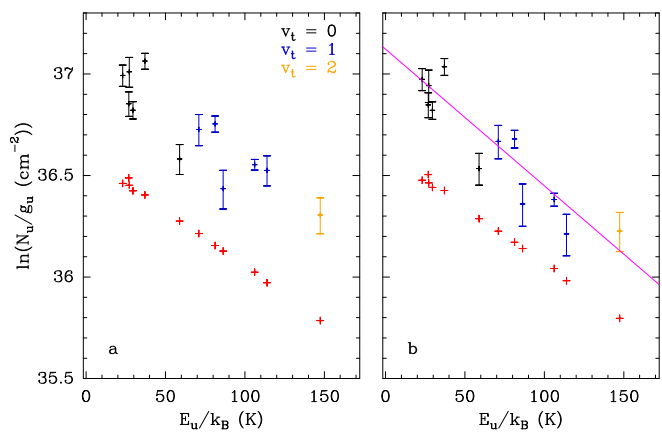}}}
\caption{Population diagram of CH$_3$NHCHO toward Sgr~B2(N2). 
The observed datapoints are 
shown in various colors (but not red) as indicated in the upper right corner 
of panel \textbf{a} while the synthetic populations are shown in red. No 
correction is applied in panel \textbf{a}. 
In panel \textbf{b}, the optical depth correction has been applied to both the 
observed and synthetic populations and the contamination by all other 
species included in the full model has been removed from the observed 
datapoints. The purple line is a linear fit to the observed populations (in 
linear-logarithmic space).
}
\label{f:popdiag_ch3nhcho}
\end{figure}

\begin{table}
 \begin{center}
 \caption{
 Rotational temperatures derived from population diagrams of selected complex organic molecules toward Sgr~B2(N2).
}
 \label{t:popfit}
 \vspace*{0.0ex}
 \begin{tabular}{lll}
 \hline\hline
 \multicolumn{1}{c}{Molecule} & \multicolumn{1}{c}{States\tablefootmark{a}} & \multicolumn{1}{c}{$T_{\rm fit}$\tablefootmark{b}} \\ 
  & & \multicolumn{1}{c}{\scriptsize (K)} \\ 
 \hline
CH$_3$NHCHO & $\varv_{\rm t}=0$, $\varv_{\rm t}=1$, $\varv_{\rm t}=2$ &   149 (20) \\ 
\hline 
CH$_3$NCO & $\varv_{\rm b}=0$, $\varv_{\rm b}=1$ & 140.9 (7.2) \\ 
\hline 
NH$_2$CHO & $\varv=0$, $\varv_{12}=1$ & 154.8 (3.3) \\ 
NH$_2$$^{13}$CHO & $\varv=0$, $\varv_{12}=1$ &   222 (37) \\ 
\hline 
HNCO & $\varv=0$, $\varv_5=1$, $\varv_6=1$, $\varv_4=1$ & 220.1 (9.6) \\ 
\hline 
CH$_3$CONH$_2$ & $\varv_{\rm t}=0$, $\varv_{\rm t}=1$, $\varv_{\rm t}=2$ &   226 (33) \\ 
\hline 
 \end{tabular}
 \end{center}
 \vspace*{-2.5ex}
 \tablefoot{
 \tablefoottext{a}{Vibrational or torsional states that were taken into account to fit the population diagram.}
 \tablefoottext{b}{The standard deviation of the fit is given in parentheses. As explained in Sect.~3 of \citet{Belloche16}, these uncertainties are purely statistical and should be viewed with caution. They may be underestimated.}
 }
 \end{table}

The population diagram of CH$_3$NHCHO is shown in 
Fig.~\ref{f:popdiag_ch3nhcho}. We used all detected lines plus a few other ones
that contribute significantly to detected lines and are contaminated by species
that we have already identified and included in our full model. All the 
transitions used for the population diagram are listed in 
Tables~\ref{t:list_ch3nhcho_ve0}--\ref{t:list_ch3nhcho_ve2}. Panel b of 
Fig.~\ref{f:popdiag_ch3nhcho} shows the population diagram after correcting
for the optical depth of the transitions, based on our best-fit LTE model of
CH$_3$NHCHO, and after removing the contamination by other species, based on 
our full model. A fit to this diagram yields a rotational temperature of 
$149 \pm 20$~K, with a significant uncertainty (Table~\ref{t:popfit}). For the 
LTE modelling of the spectrum, we adopt a temperature of 180~K. The best-fit 
parameters are listed in Table~\ref{t:coldens}.

\begin{table*}[!ht]
 \begin{center}
 \caption{
 Parameters of our best-fit LTE model of selected complex organic molecules toward Sgr~B2(N2).
}
 \label{t:coldens}
 \vspace*{-1.2ex}
 \begin{tabular}{lcrccccccr}
 \hline\hline
 \multicolumn{1}{c}{Molecule} & \multicolumn{1}{c}{Status\tablefootmark{a}} & \multicolumn{1}{c}{$N_{\rm det}$\tablefootmark{b}} & \multicolumn{1}{c}{Size\tablefootmark{c}} & \multicolumn{1}{c}{$T_{\mathrm{rot}}$\tablefootmark{d}} & \multicolumn{1}{c}{$N$\tablefootmark{e}} & \multicolumn{1}{c}{$F_{\rm vib}$\tablefootmark{f}} & \multicolumn{1}{c}{$\Delta V$\tablefootmark{g}} & \multicolumn{1}{c}{$V_{\mathrm{off}}$\tablefootmark{h}} & \multicolumn{1}{c}{$\frac{N_{\rm ref}}{N}$\tablefootmark{i}} \\ 
  & & & \multicolumn{1}{c}{\scriptsize ($''$)} & \multicolumn{1}{c}{\scriptsize (K)} & \multicolumn{1}{c}{\scriptsize (cm$^{-2}$)} & & \multicolumn{1}{c}{\scriptsize (km~s$^{-1}$)} & \multicolumn{1}{c}{\scriptsize (km~s$^{-1}$)} & \\ 
 \hline
 CH$_3$NHCHO, $\varv_{\rm t}=0$$^\star$ & t & 4 &  0.9 &  180 &  1.0 (17) & 1.26 & 5.0 & 0.5 &       1 \\ 
 \hspace*{12.5ex} $\varv_{\rm t}=1$ & t & 1 &  0.9 &  180 &  1.0 (17) & 1.26 & 5.0 & 0.5 &       1 \\ 
 \hspace*{12.5ex} $\varv_{\rm t}=2$ & t & 0 &  0.9 &  180 &  1.0 (17) & 1.26 & 5.0 & 0.5 &       1 \\ 
\hline 
 CH$_3$NCO, $\varv_{\rm b}=0$$^\star$ & d & 60 &  1.2 &  150 &  2.2 (17) & 1.00 & 5.0 & -0.6 &       1 \\ 
 \hspace*{9.3ex} $\varv_{\rm b}=1$ & d & 4 &  1.2 &  150 &  2.2 (17) & 1.00 & 5.0 & -0.6 &       1 \\ 
\hline 
 NH$_2$CHO, $\varv=0$$^\star$ & d & 30 &  0.8 &  200 &  3.5 (18) & 1.17 & 5.5 & 0.2 &       1 \\ 
 \hspace*{9.5ex} $\varv_{12}=1$ & d & 13 &  0.8 &  200 &  2.6 (18) & 1.17 & 5.5 & 0.2 &     1.4 \\ 
 NH$_2$$^{13}$CHO, $\varv=0$ & d & 11 &  0.8 &  200 &  1.3 (17) & 1.17 & 5.5 & 0.5 &      27 \\ 
 \hspace*{11.0ex} $\varv_{12}=1$ & d & 2 &  0.8 &  200 &  1.3 (17) & 1.17 & 5.5 & 0.5 &      27 \\ 
 $^{15}$NH$_2$CHO, $\varv=0$ & t & 1 &  0.8 &  200 &  1.2 (16) & 1.17 & 5.5 & 0.5 &     300 \\ 
\hline 
 HNCO, $\varv=0$$^\star$ & d & 12 &  0.9 &  240 &  2.0 (18) & 1.06 & 5.5 & 0.0 &       1 \\ 
 \hspace*{6.8ex} $\varv_5=1$ & d & 4 &  0.9 &  240 &  2.0 (18) & 1.06 & 5.5 & 0.0 &       1 \\ 
 \hspace*{6.8ex} $\varv_6=1$ & d & 1 &  0.9 &  240 &  2.0 (18) & 1.06 & 5.5 & 0.0 &       1 \\ 
 \hspace*{6.8ex} $\varv_4=1$ & t & 0 &  0.9 &  240 &  2.0 (18) & 1.06 & 5.5 & 0.0 &       1 \\ 
 HN$^{13}$CO, $\varv=0$ & t & 0 &  0.9 &  240 &  1.0 (17) & 1.06 & 5.5 & 0.0 &      20 \\ 
\hline 
 CH$_3$CONH$_2$, $\varv_{\rm t}=0$$^\star$ & d & 10 &  0.9 &  180 &  1.4 (17) & 1.23 & 5.0 & 1.5 &       1 \\ 
 \hspace*{11.8ex} $\varv_{\rm t}=1$ & d & 8 &  0.9 &  180 &  1.4 (17) & 1.23 & 5.0 & 1.5 &       1 \\ 
 \hspace*{11.8ex} $\varv_{\rm t}=2$ & d & 5 &  0.9 &  180 &  1.4 (17) & 1.23 & 5.0 & 1.5 &       1 \\ 
 \hspace*{11.8ex} $\Delta\varv_{\rm t} \neq 0$ & t & 0 &  0.9 &  180 &  1.4 (17) & 1.23 & 5.0 & 1.5 &       1 \\ 
\hline 
 \end{tabular}
 \end{center}
 \vspace*{-2.5ex}
 \tablefoot{
 \tablefoottext{a}{d: detection, t: tentative detection.}
 \tablefoottext{b}{Number of detected lines \citep[conservative estimate, see Sect.~3 of][]{Belloche16}. One line of a given species may mean a group of transitions of that species that are blended together.}
 \tablefoottext{c}{Source diameter (\textit{FWHM}).}
 \tablefoottext{d}{Rotational temperature.}
 \tablefoottext{e}{Total column density of the molecule. $X$ ($Y$) means $X \times 10^Y$. An identical value for all listed vibrational/torsional states of a molecule means that LTE is an adequate description of the vibrational/torsional excitation.}
 \tablefoottext{f}{Correction factor that was applied to the column density to account for the contribution of vibrationally excited states, in the cases where this contribution was not included in the partition function of the spectroscopic predictions.}
 \tablefoottext{g}{Linewidth (\textit{FWHM}).}
 \tablefoottext{h}{Velocity offset with respect to the assumed systemic velocity of Sgr~B2(N2), $V_{\mathrm{lsr}} = 74$ km~s$^{-1}$.}
 \tablefoottext{i}{Column density ratio, with $N_{\rm ref}$ the column density of the previous reference species marked with a $\star$.}
 }
 \end{table*}

In the following sections, we derive the column density of molecules that may
be related to N-methylformamide in order to put the tentative detection of
this molecule into a broader astrochemical context.

\subsection{Methyl isocyanate (CH$_3$NCO)}
\label{ss:ch3nco}

The first interstellar detection of methyl isocyanate was obtained toward
Sgr~B2(N) based on single-dish observations 
\citep[][; see also \citeauthor{Cernicharo16} \citeyear{Cernicharo16}]{Halfen15},
shortly after its in-situ detection in the frozen surface of comet 
67P/Churyumov-Gerasimenko \citep[][]{Goesmann15}. Here, we use the predictions
available in the Cologne database for molecular 
spectroscopy\footnote{http://www.astro.uni-koeln.de/cdms/} 
\citep[CDMS,][]{Mueller05} (tags 57\,505 and 57\,506, both version 1), which 
are based on measurements reported by \citet{Cernicharo16} and \citet{Koput86}.

Methyl isocyanate is well detected toward Sgr~B2(N2) in the EMoCA spectral 
survey: 60 lines are clearly detected in its vibrational ground state and four 
in its first vibrationally excited state (Figs.~\ref{f:spec_ch3nco_ve0} and 
\ref{f:spec_ch3nco_ve1}). Gaussian fits to the integrated intensity maps of
the detected lines indicate a median emission size of $\sim$1.2$\arcsec$ with
a rms dispersion of $\sim$0.1$\arcsec$ (see Fig.~\ref{f:maps}b). A 
fit to its population diagram shown
in Fig.~\ref{f:popdiag_ch3nco} yields a rotational temperature of 
$141 \pm 7$~K (Table~\ref{t:popfit}). We adopt a temperature of 150~K for our 
LTE modelling. The parameters of our best-fit model are given in 
Table~\ref{t:coldens}. The model overestimates the peak temperature of a few
transitions by $\sim$30--40\% (at 87.236, 96.062, 96.120, 104.793, and 
104.856~GHz) while all other transitions are well fitted; the reason for these 
discrepancies is unclear.

\subsection{Formamide (NH$_2$CHO)}
\label{ss:nh2cho}

We use the CDMS entries for formamide in its ground and first vibrationally
excited states (tags 45\,512 and 45\,516, versions 2 and 1, respectively) and 
for its $^{13}$C and $^{15}$N isotopologues in their ground state (tags 46\,512
and 46\,513, versions 2 and 1, respectively). These entries are based largely 
on \citet{motiyenko2012rotational}, but also contain additional data. 
Laboratory data in the range of our survey were published by \citet{Kryvda09}. 
The entry for $\varv_{12} = 1$ of the $^{13}$C isotopologue was prepared by one 
of us (HSPM) based on data from \citet{Stubgaard78}.

Formamide is well detected toward Sgr~B2(N2), with 30 lines in its vibrational
ground state and 13 in its vibrationally excited state $\varv_{12}=1$ 
(Figs.~\ref{f:spec_nh2cho_ve0} and \ref{f:spec_nh2cho_v12e1}). Its $^{13}$C
isotopologue is also clearly detected, with 11 and 2 lines in its $\varv=0$
and $\varv_{12}=1$ states, respectively (Figs.~\ref{f:spec_nh2cho_13c_ve0} and 
\ref{f:spec_nh2cho_13c_v12e1}). Finally, we report a tentative detection
of $^{15}$NH$_2$CHO, with one clearly detected line consistent with a 
$^{14}$N/$^{15}$N isotopic ratio of 300 (Fig.~\ref{f:spec_nh2cho_15n_ve0}).

We derive a median emission size of $\sim$0.9--1.0$\arcsec$ from Gaussian
fits to the integrated intensity maps of the main and $^{13}$C isotopologues, 
with a rms dispersion of $\sim$0.1$\arcsec$ (see Figs.~\ref{f:maps}c and d). A 
number of lines of the main isotopologue are saturated and not well fitted by 
our simple LTE model. We selected only the transitions with an optical depth 
lower than 2 to build the population diagram shown in 
Fig.~\ref{f:popdiag_nh2cho}. A fit to 
this diagram yields a rotational temperature of $155 \pm 3$~K 
(Table~\ref{t:popfit}). However, we obtain a rotational temperature of 
$200 \pm 14$~K when we limit the fit to the transitions that belong to the 
vibrational ground state. The population diagram of the $^{13}$C isotopologue 
is less populated (Fig.~\ref{f:popdiag_nh2cho_13c}) and yields a more uncertain
rotational temperature of $222 \pm 35$~K (Table~\ref{t:popfit}). We adopt a 
temperature of 200~K for our LTE model of all isotopologues of formamide. 

We initially modeled the spectra assuming a size of $0.9\arcsec$ as derived 
above but, in order to fit the $\varv_{12}=1$ transitions, we then had to 
assume a total column density of NH$_2$CHO much lower than the one needed to 
fit the ground state transitions. By reducing the size to $0.8\arcsec$ we 
could attenuate the discrepancy between $\varv=0$ and $\varv_{12}=1$ but we 
still need a total column density 1.4 times lower to fit $\varv_{12}=1$. This 
is surprising because we do not face this problem for the $^{13}$C 
isotopologue for which both states are well fitted assuming the same total 
column density. The discrepancy that affects the main isotopologue may be due 
to its higher optical depth although we do not feel that it is a satisfactory 
explanation.

\subsection{Isocyanic acid (HNCO)}
\label{ss:hnco}

We use the CDMS entry for HNCO in its ground state (tag 43\,511 version 1) by 
\citet{Lapinov07} with additional measurements in the range of our survey by 
\citet{Hocking75}, and the JPL entry for the $^{13}$C isotopologue 
\citep[tag 44\,008 version 1,][]{Hocking75}. Private entries for the 
vibrationally excited states $\varv_5 = 1$, $\varv_6 = 1$, and $\varv_4 = 1$ 
of the main isotopologue were prepared by one of us (HSPM). They are based on 
a preliminary, unpublished analysis of the ground and the four lowest excited 
states and were already used in \citet{Belloche13}. The data on the excited 
states was summarized in \citet{Niedenhoff96}. Transition frequencies in the 
range of our survey were published by \citet{Yamada77a} and \citet{Yamada77b}.

Isocyanic acid is also well detected toward Sgr~B2(N2), with twelves lines in 
its vibrational ground state, four in its vibrationally excited state 
$\varv_5=1$, and one in its state $\varv_6=1$ 
(Figs.~\ref{f:spec_hnco_ve0}--\ref{f:spec_hnco_v6e1}). Its state $\varv_4=1$
is not unambiguously detected but it contributes significantly to the signal 
detected at 87.97~GHz so we have included it in our full model 
(Fig.~\ref{f:spec_hnco_v4e1}). The $^{13}$C isotopologue is not unambiguously 
detected because all its significant transitions are located in the 
blueshifted wing of transitions of the main isotopologue, some of these 
suffering from 
absorption of the main isotopologue from the outer envelope of Sgr~B2, which we 
have not taken into account in our full model yet 
(Fig.~\ref{f:spec_hnco_13c_ve0}). Still, the $^{13}$C isotopologue contributes 
significantly to the signal detected at several frequencies, so we have 
included it in our full model, based on the model derived below for the main 
isotopologue and assuming a $^{12}$C/$^{13}$C isotopic ratio of 20.

We derive a median size of 0.9$\arcsec$ from Gaussian fits to the integrated
intensity maps of the main isotopologue, with a dispersion of 0.2$\arcsec$
(see Fig.~\ref{f:maps}e).
The population diagram shown in Fig.~\ref{f:popdiag_hnco} uses all but three 
transitions that are clearly detected, plus a few additional lines that are 
more contaminated but for which we have identified and modeled the 
contaminating species. The three transitions of the vibrational ground state 
that we ignore (at 109.50, 109.91, and 110.30~GHz) have an opacity higher than 
4. A fit to the population diagram yields a temperature of $220 \pm 10$~K. 
However, with this temperature, the optically thick lines of the ground state 
would saturate with a brightness temperature that is too low. As a compromise, 
we use a temperature of 240~K. 

The detected lines of all four states of the main isotopologue are relatively 
well reproduced with the same model parameters (Table~\ref{t:coldens}). A few 
issues remain, however. First of all, the peak temperatures of the optically 
thick lines of $\varv=0$ are a bit underestimated, except for the peak 
temperature of the $5_{0,5}$--$4_{0,4}$ transition at 109.91~GHz which is 
overestimated, probably because of spatial filtering of extended emission not 
taken into account in our model \citep[see Fig.~6 of][]{Jones08}. The second 
issue concerns two transitions with an upper energy level 
$E_{\rm up} \sim 650$~K, $34_{1,34}$--$35_{0,35}$ at 
85.37~GHz and $33_{1,33}$--$34_{0,34}$ at 109.96~GHz, which are both overestimated.
The former is located in a frequency range affected by c-C$_3$H$_2$ absorption 
by spiral arm clouds along the line of sight to Sgr~B2, so the discrepancy at 
this frequency may not be a problem. We note, in addition, that the model for
a third transition with $E_{\rm up} \sim 650$~K, $5_{4,1}$--$4_{4,0}$ at 
109.78~GHz, is consistent with the detected signal. Therefore, it is unclear 
why our model overestimates the peak temperature of the transition at 
109.96~GHz.

\subsection{Acetamide (CH$_3$CONH$_2$)}

We use predictions that are based on the measurements and analysis presented in 
\citet{ilyushin2004acetamide} but were recomputed by one of us (VVI) with the 
RAM36 code. We use the partition function calculated in Sect.~\ref{s:spectro}
(Table~\ref{t:qpart}).

With a total of 23 lines clearly detected in its ground state and its 
first and second torsionally excited states toward the hot core Sgr~B2(N2), 
acetamide can be considered as securely detected
(Figs.~\ref{f:spec_ch3conh2_ve0}--\ref{f:spec_ch3conh2_cv}). Fits to the 
integrated intensity maps of three of the detected lines give a median size of 
$\sim$1.0$\arcsec$, with a dispersion of 0.2$\arcsec$
(see Fig.~\ref{f:maps}f). The emission of four 
other detected lines is unresolved, pointing to a smaller size. The maps of 
the remaining detected lines were not fitted because they are to some level
contaminated in their wings, which could bias the size measurements. A fit to 
the population diagram yields a rotational temperature of $226 \pm 33$~K, 
which is not well constrained (Fig.~\ref{f:popdiag_ch3conh2} and 
Table~\ref{t:popfit}). We adopt a temperature of 180~K and a size of 
0.9$\arcsec$ for our LTE model to make the comparison to N-methylformamide 
more straightforward. The detected lines of the ground state 
($\varv_{\rm t} = 0$) and both torsionally excited states ($\varv_{\rm t} = 1$ 
and 2), as well as lines that connect different torsional states 
($\Delta\varv_{\rm t} \neq 0$), are well reproduced with the same model 
parameters (Table~\ref{t:coldens}). One discrepancy can be noticed in 
Fig.~\ref{f:spec_ch3conh2_ve0}: the $\varv_{\rm t}=0$ $25_{20,6}$--$25_{18,7}$ 
transition of the $E$ species at 99.950 GHz does not have a counterpart in the 
observed spectrum (at the 3$\sigma$ level). 
However it was extrapolated from a lower-$J$ fit ($J \leq 20$) and has a 
frequency uncertainty of 200~kHz which means that its frequency could be off 
by several times this number. Indeed, the actual positions of the hyperfine 
components of this line as measured in the laboratory spectrum are 
99951.872~MHz and 99952.331~MHz, both with an uncertainty of 10~kHz. At these
frequencies, strong emission is detected toward Sgr~B2(N2), which reconciles
the spectrum expected for acetamide with the observed spectrum. Therefore,
the discrepancy between the observed and synthetic spectra at 99.950 GHz is 
not a real issue.

\section{Chemical modelling}
\label{s:chemistry}

To investigate the production of the tentatively detected N-methylformamide in 
Sgr B2(N2), we use the chemical kinetics model \textit{MAGICKAL} (Garrod 2013) 
with an expanded gas-grain chemical network. This network is an extension of 
that presented by \citet{Belloche14}, and latterly by \citet{Mueller16a}, and 
includes formation and destruction mechanisms for both CH$_3$NHCHO and the 
related molecule CH$_3$NCO. The model allows for a treatment of the 
fully-coupled gas-phase, 
grain/ice-surface, and ice-mantle chemistry. The physical model follows that 
detailed in previous papers where a cold collapse phase to maximum density 
(n$_{\mathrm{H}}$ = $2 \times 10^{8}$ cm$^{-3}$) and minimum dust-grain 
temperature (8~K) is followed by a warm-up from 8 to 400~K; during this phase, 
the gas and dust temperatures are assumed to be well coupled. The initial 
chemical compositions used in the model follow those of \citet{Garrod13}. The 
reader is referred to the above-mentioned publications for a more detailed 
discussion of the basic physical and chemical model. In the models presented 
here, we use the intermediate warm-up timescale, which generally produces the 
best match between models and observed abundances of other chemical species.
The warm-up model therefore reaches a temperature of 200~K at $2 \times
10^5$~yr, reaching 400~K (and the end of the model run) at $\sim 2.85 \times
10^5$~yr.

The new network concentrates on the grain-surface production of the 
newly-introduced molecules. However, gas-phase destruction mechanisms for both 
molecules (as well as related intermediates) are included in the new network, 
the majority of which are ion-molecule processes or the subsequent 
dissociative recombination with electrons of the resultant molecular ions. 
Ion-molecule reactions are included for the major ionic species C$^+$, He$^+$, 
H$_3$$^+$, H$_3$O$^+$, and HCO$^+$. Estimates for the rates of 
photo-dissociation of new molecules, as caused by cosmic ray-induced and 
(where extinction allows) external UV photons, are also included 
\citep[see][]{Garrod08,Garrod13}. Grain/ice-surface binding 
(desorption) energies for the 
new molecules are estimates based on interpolation/extrapolation of values for 
molecules with similar functional groups, following past publications. 
Binding energy values for CH$_3$NCO and CH$_3$NHCHO are 3575~K and 
6281~K, respectively, based on the formulations [$E_{\mathrm{des}}$(CH$_3$) + 
$E_{\mathrm{des}}$(N) + $E_{\mathrm{des}}$(C) + $E_{\mathrm{des}}$(O)] and 
[$E_{\mathrm{des}}$(NH$_2$CHO) - $E_{\mathrm{des}}$(H) + $E_{\mathrm{des}}$(CH$_3$)].
We are not aware of any experimental determinations of these two quantities 
for appropriate surfaces.

In the new network, grain-surface and ice-mantle formation of CH$_3$NCO occurs 
through a single radical-addition reaction: 
\begin{equation}
\label{RTG-rxn1}
\mathrm{CH}_{3} + \mathrm{OCN} \rightarrow \mathrm{CH}_{3}\mathrm{NCO} 
\end{equation}
Each of the necessary radicals may be formed either through repetitive atomic 
addition or through the photo-dissociation of or chemical H-abstraction from 
either CH$_4$ or HNCO.

To form CH$_3$NHCHO, a reaction involving the addition to CH$_3$NCO of a 
hydrogen atom, followed by another, presents itself as a possible route, i.e.
\begin{eqnarray}
\label{RTG-rxn2}
\mathrm{H} + \mathrm{CH_{3}NCO} \rightarrow \mathrm{CH}_{3}\mathrm{NHCO} \\
\label{RTG-rxn3}
\mathrm{H} + \mathrm{CH}_{3}\mathrm{NHCO}  \rightarrow \mathrm{CH}_{3}\mathrm{NHCHO} 
\end{eqnarray}
The first of these two reactions requires the breaking of a carbon-nitrogen 
double bond, for which no activation energy barrier could be determined from 
the literature. The barrier to the similar reaction of H with HNCO has been 
determined experimentally in the gas phase by \citet{Nguyen96} to be 1390~K, 
although the value for H + CH$_3$NCO could plausibly be higher or lower. The 
expectation, however, is that -- as with most barrier-mediated atomic-H 
reactions on cold grains -- the mechanism would involve the tunneling of the H 
atom through the barrier, introducing further uncertainty into the reaction 
rate. By default, \textit{MAGICKAL} uses a simple rectangular-barrier treatment 
to determine rates for tunneling reactions, typically assuming a uniform 
barrier width of 1 \AA; the assumed height of the energy barrier therefore 
absorbs all other parameters pertaining to the overall reaction rate. The 
ideal activation energy barrier determined for the chemical model may 
therefore not be fully representative of the usual high-temperature value.

Because a broad range of activation energy values could be plausible for 
reaction~(\ref{RTG-rxn2}), we initially assume in the chemical network (models 
M1--M5) that the rate for reaction~(\ref{RTG-rxn2}) is negligible,
introducing non-zero values later.

Radical-addition reactions provide an alternative pathway to the formation of 
N-methylformamide, through the reactions:
\begin{eqnarray}
\label{RTG-rxn4}
\mathrm{CH}_{3} + \mathrm{HNCHO} \rightarrow \mathrm{CH}_{3}\mathrm{NHCHO} \\
\label{RTG-rxn5}
\mathrm{HCO} + \mathrm{HN}\mathrm{CH}_{3} \rightarrow \mathrm{CH}_{3}\mathrm{NHCHO} 
\end{eqnarray}
The larger radicals in each of the above two reactions are produced through 
the addition of NH to either CH$_3$ or HCO. While cosmic ray-induced 
photo-dissociation of CH$_3$NH$_2$ and NH$_2$CHO may also produce the 
necessary radicals, the chemical abstraction of a hydrogen atom by a radical 
from either molecule strongly favours the production of CH$_2$NH$_2$ or 
NH$_2$CO, respectively, rather than the alternative radicals that could play a 
part in forming CH$_3$NHCHO. However, in the case of H-atom abstraction from 
NH$_2$CHO by OH, the model initially considers the barriers to NH$_2$CO and 
NHCHO production to be similar (591 K versus 600 K), following the estimates 
used by Garrod (2013), although neither value has been determined rigorously. 
Consequently, the influence of the production of HNCHO through this mechanism 
is also examined in Sect.~\ref{RTG-mod-res}.

\subsection{Model results}
\label{RTG-mod-res}

Model M1 comprises the initial model in which conversion of CH$_3$NCO to 
CH$_3$NHCHO via consecutive H addition (reactions~\ref{RTG-rxn2} and 
\ref{RTG-rxn3}) is switched off.

\begin{figure}
\centerline{\resizebox{1.0\hsize}{!}{\includegraphics[angle=0]{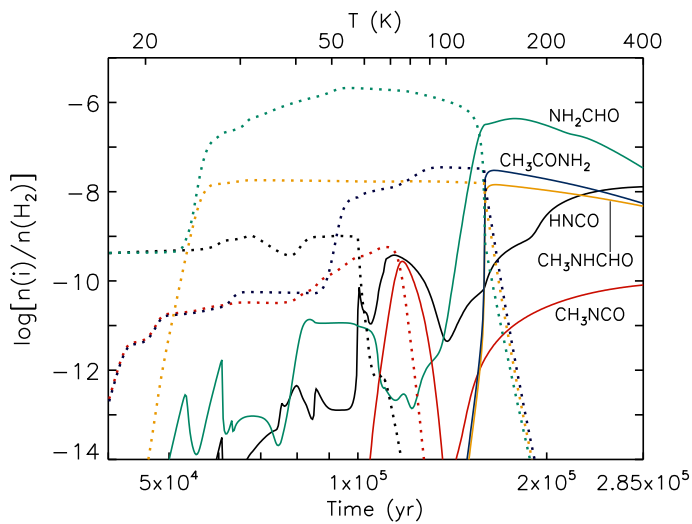}}}
\caption{Calculated abundances of selected chemical model species with respect 
to H$_2$ for model M1, during warm-up from 8~K to 400~K. Solid lines indicate 
gas-phase abundances; dotted lines of the same colour indicate solid-phase 
abundances of the same species.}
\label{RTG-fig1}
\end{figure}

Figure~\ref{RTG-fig1} shows results from model M1. Time-dependent abundances 
for CH$_3$NCO, CH$_3$NHCHO as well as the related species HNCO, NH$_2$CHO and 
CH$_3$CONH$_2$ are shown. CH$_3$NHCHO is seen to be formed on the grains 
(dotted line) in abundance at around 25 K, coincident with a significant 
growth in NH$_2$CHO production. Its formation is dominated by reaction 
(\ref{RTG-rxn4}). CH$_3$NCO production on the grains is much more modest, and 
occurs only through reaction (\ref{RTG-rxn1}). Its gas-phase abundance peaks 
at a relatively low temperature, following its desorption from grains, and 
falls again, although it later begins to rise as the abundant CH$_3$NHCHO is 
photodissociated in the gas phase. CH$_3$CONH$_2$ reaches a peak abundance 
just a little larger than that of CH$_3$NHCHO, forming mainly through the 
addition of CH$_3$ and NH$_2$CO radicals. HNCO is formed on the grains early 
($T<20$~K), via hydrogenation of OCN which itself is formed through the atomic 
addition reactions O + CN and C + NO. The gas-phase abundance of HNCO peaks at 
close to 60~K as it desorbs from the grains, then falls away. It is a 
significant by-product of the destruction of larger molecules at later times 
in the model, and its abundance continues to grow until the final model 
temperature of 400~K is reached.

\begin{table*}
\begin{center}
\caption{Peak gas-phase fractional abundances with respect to H$_2$ and 
temperatures at which the peak abundance values are achieved, for models with 
varying activation energy barriers for the grain-surface reaction 
NH + CO $\rightarrow$ HNCO.}
\label{RTG-tab1}
\renewcommand{\arraystretch}{1.0}
\vspace*{-4ex}
\small
\begin{tabular}[t]{lrrrrrrrrrrrrrrr}
\hline \hline
 && \multicolumn{2}{c}{M1} && \multicolumn{2}{c}{M2} && \multicolumn{2}{c}{M3} && \multicolumn{2}{c}{M4} && \multicolumn{2}{c}{M5} \\
 && \multicolumn{2}{c}{$E_{\mathrm{A}}$=2500 K} && \multicolumn{2}{c}{$E_{\mathrm{A}}$=2000 K} && \multicolumn{2}{c}{$E_{\mathrm{A}}$=1500 K} && \multicolumn{2}{c}{$E_{\mathrm{A}}$=1250 K} && \multicolumn{2}{c}{$E_{\mathrm{A}}$=1000 K}  \\ 
\cline{3-4} \cline{6-7} \cline{9-10} \cline{12-13} \cline{15-16} \\

Species && $n$[i]/$n$[H$_2$] & $T$ (K) && $n$[i]/$n$[H$_2$] & $T$ (K) && $n$[i]/$n$[H$_2$] & $T$ (K) && $n$[i]/$n$[H$_2$] & $T$ (K) && $n$[i]/$n$[H$_2$] & $T$ (K)  \\
\hline\\

CH$_3$NHCHO      && 1.4(-8) & 139  && 1.3(-8) & 139 && 1.6(-9) & 139 && 1.5(-9) & 139 && 1.5(-9) & 139  \\
CH$_3$NCO          && 2.8(-10) & 74    && 1.6(-9) &  74 && 6.9(-9) & 74   && 6.9(-9) & 74  &&  7.0(-9) & 75  \\
NH$_2$CHO          && 4.3(-7) & 160  && 4.4(-7) & 160 && 5.3(-7) & 157 && 5.3(-7) & 157 && 5.4(-7) & 157   \\
HNCO                  && 1.3(-8) & 398  && 1.3(-8) & 398 && 2.2(-8) & 56   && 2.2(-8) & 56  && 2.2(-8) & 56   \\
CH$_3$CONH$_2$ && 3.0(-8) & 138  && 3.0(-8) &  138 && 3.1(-8) & 138 && 3.1(-8) & 138 && 3.2(-8) & 138  \\
\hline
\end{tabular}
\end{center}
\vspace*{-2ex}
\tablefoot{Model M1 uses the value adopted in previous hot-core models. 
$X (Y)$ means $X \times 10^{Y}$.
}
\end{table*}

Table~\ref{RTG-tab1} shows the peak gas-phase abundances of the plotted 
species, along with the temperatures at which those peaks are reached. It may 
be noted that the peak abundance of CH$_3$NHCHO is approximately at parity 
with HNCO, contrary to the observed column densities shown in 
Table~\ref{t:coldens}, where a ratio of $\sim$1:20 is obtained. The amount 
of CH$_3$NHCHO produced in this model
as compared with CH$_3$NCO also appears high, while the abundance of CH$_3$NCO 
is around one order of magnitude too low compared with HNCO. Since the sole 
production mechanism of CH$_3$NCO relies on OCN, we have also considered 
alternative models, in which HNCO -- a possible precursor to OCN -- may be 
more easily formed.

\begin{figure}
\centerline{\resizebox{1.0\hsize}{!}{\includegraphics[angle=0]{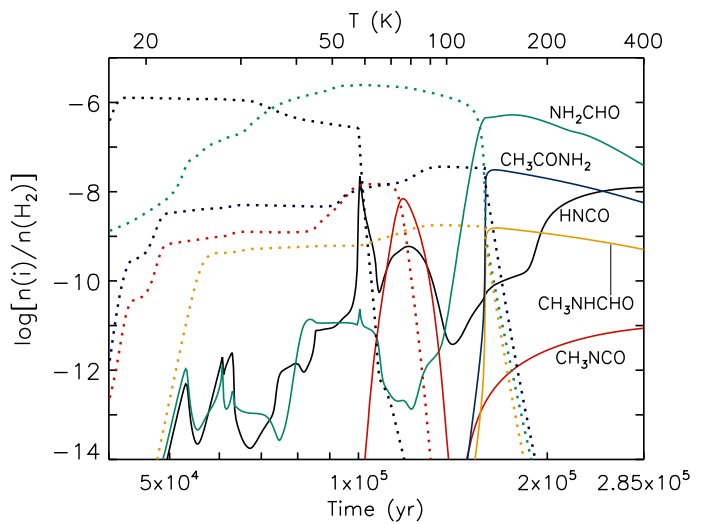}}}
\caption{Same as Fig.~\ref{RTG-fig1} for model M4.}
\label{RTG-fig2}
\end{figure}

Models M2--M5, whose results are also tabulated in 
Table~\ref{RTG-tab1}, allow the grain-surface reaction 
NH + CO $\rightarrow$ HNCO to occur with a lower activation energy barrier 
than the 2500~K first assumed by \citet{Garrod08} -- an estimate based loosely 
on the typically-assumed barrier to the H-addition reaction H + CO 
$\rightarrow$ HCO. These models show two types of behavior, with the threshold 
falling somewhere between models M2 and M3, i.e. an activation energy, 
$E_{\rm A}$, between 2000~K and 1500~K. Below this threshold, the models move 
away from M1-type behavior and instead show significantly increased HNCO 
production on the grains, such that the desorption of HNCO into the gas phase 
produces the peak abundance for this molecule. Figure~\ref{RTG-fig2} plots 
abundances for model M4, for which an activation energy barrier to the NH + CO 
reaction of 1250 K is assumed. The peak abundance of CH$_3$NCO is notably 
increased (see also Table~\ref{RTG-tab1}) as a result of the greater HNCO 
abundance, which contributes to the production of OCN. Conversely, the 
abundance of CH$_3$NHCHO is found to decrease markedly in models M3-M5, as the 
NH + CO reaction becomes competitive with the NH + HCO $\rightarrow$ NHCHO 
reaction at the key temperature ($\sim$25~K) at which CH$_3$ mobility makes 
reaction~(\ref{RTG-rxn4}) important to CH$_3$NHCHO production. Models M3-M5 
all reach within a factor of $\sim$2 of the observed CH$_3$NHCHO/CH$_3$NCO 
abundance ratio.

\begin{table*}
\begin{center}
\caption{Peak gas-phase fractional abundances with respect to H$_2$ and 
temperatures at which the peak abundance values are achieved, for models based 
on model M4, i.e. with an activation energy barrier of 1250~K for the 
grain-surface reaction NH + CO $\rightarrow$ HNCO.}
\label{RTG-tab3}
\renewcommand{\arraystretch}{1.0}
\vspace*{-4ex}
\small
\begin{tabular}[t]{lrrrrrrrrrrrrrrr}
\hline \hline
 && \multicolumn{2}{c}{M4A} && \multicolumn{2}{c}{M4B} && \multicolumn{2}{c}{M4C} && \multicolumn{2}{c}{M4D} && \multicolumn{2}{c}{M4E}  \\
 && \multicolumn{2}{c}{$E_{\mathrm{A}}$=1390 K} && \multicolumn{2}{c}{$E_{\mathrm{A}}$=2000 K} && \multicolumn{2}{c}{$E_{\mathrm{A}}$=2500 K} && \multicolumn{2}{c}{$E_{\mathrm{A}}$=3000 K} && \multicolumn{2}{c}{$E_{\mathrm{A}}$=3500 K}  \\ 
\cline{3-4} \cline{6-7} \cline{9-10} \cline{12-13} \cline{15-16}   \\

Species && $n$[i]/$n$[H$_2$] & $T$ (K) && $n$[i]/$n$[H$_2$] & $T$ (K) && $n$[i]/$n$[H$_2$] & $T$ (K) && $n$[i]/$n$[H$_2$] & $T$ (K) && $n$[i]/$n$[H$_2$] & $T$ (K)  \\
\hline\\

CH$_3$NHCHO     && 1.2(-8) & 139  && 5.3(-9) & 139 && 1.2(-9) & 139 && 6.0(-10) & 139 && 5.1(-10)  & 139  \\
CH$_3$NCO         && 7.0(-11) & 398  && 4.1(-9) & 74   && 6.5(-9) & 74  && 6.9(-9)  & 74 && 6.9(-9)   & 74  \\
NH$_2$CHO         && 5.1(-7) & 157  && 5.1(-7) & 157  && 5.1(-7) & 157 && 5.1(-7) & 157 && 5.1(-7) & 157  \\
HNCO                 && 2.2(-8) & 56    && 2.2(-8) & 56   && 2.2(-8) & 56  && 2.2(-8)  & 56 && 2.2(-8)   & 56  \\
CH$_3$CONH$_2$ && 3.2(-8) & 138  && 3.2(-8) & 138  && 3.2(-8) & 138 && 3.2(-8) & 138 && 3.2(-8) & 138  \\

\hline
\end{tabular}
\end{center}
\vspace*{-2ex}
\tablefoot{The models (M4A--E) vary from model M4 in two ways: (i) the 
OH + NH$_2$CHO $\rightarrow$ NHCHO + H$_2$O reaction pathway has been switched 
off; (ii) the reaction H + CH$_3$NCO $\rightarrow$ CH$_3$NHCO is switched on, 
with an activation energy barrier, $E_{\rm A}$, as indicated. Model M4A uses 
the value assumed for the H + HNCO $\rightarrow$ NH$_2$CO reaction 
\citep[1390~K,][]{Nguyen96}, while the other models use larger values. $X (Y)$ 
means $X \times 10^{Y}$.
}
\end{table*}

To investigate the importance of H-addition to CH$_3$NCO, we adjust the 
conditions assumed in model M4, taking a selection of activation energy 
barriers for reaction~(\ref{RTG-rxn2}) to give a non-zero reaction rate.
M4A assumes the same value as the reaction H + HNCO, while models M4B--E 
increase this value incrementally (see Table~\ref{RTG-tab3}). We also 
switch off the H-abstraction reaction OH + NH$_2$CHO 
$\rightarrow$ NHCHO, to test its influence on CH$_3$NHCHO production. 

Model M4A demonstrates an extreme degree of conversion of CH$_3$NCO to 
CH$_3$NHCHO that is not borne out by the observations, while the somewhat 
higher barrier to hydrogenation of model M4B improves the match to the 
detected CH$_3$NHCHO/CH$_3$NCO ratio (see also 
Table~\ref{RTG-tab5}), albeit with a value greater than unity. Model M4C, as 
with M4, shows a modest dominance of CH$_3$NCO over CH$_3$NHCHO, in line with 
observations, and is similar to M4 in the quality of its overall match with 
observations. At the higher activation energies used in models M4D and 
M4E, the importance of reaction~(\ref{RTG-rxn2}) diminishes. Here, the 
influence of the removal of the OH + NH$_2$CHO $\rightarrow$ NHCHO reaction 
becomes apparent, via comparison with model M4 (Table~\ref{RTG-tab1}); the 
abundance of CH$_3$NHCHO falls by two-thirds. For each of models M4A--E, the 
behavior of HNCO, NH$_2$CHO, and CH$_3$CONH$_2$ are little affected versus 
model M4.

Note that particularly in the cases of CH$_3$NCO and CH$_3$CONH$_2$, the 
temperatures at which peak abundances are attained are somewhat lower than the 
rotational temperatures obtained from the spectroscopic model fits to the 
observational data. This may be due to the imprecision of either or both of 
the binding energy estimates and the spectroscopic fits. The observed HNCO 
temperature is best represented by the late-time/high-temperature peak found 
in the models, rather than the brief gas-phase spike at around 60~K. The 
precise temperature at which the high-temperature peak is reached is dependent 
on the destruction rates and warm-up timescale assumed in the model.
In the case of CH$_3$NCO, the larger spatial extent of this molecule (1.2''),
as derived from the observations, compared with that of CH$_3$NHCHO (0.9'') 
is consistent with the idea that the former is released from grains at lower 
temperatures than the latter.

\begin{table*}
\begin{center}
\caption{Relative abundances for a selection of chemical species, from the initial model (M1) and the three other most successful models, and from the observations toward Sgr B2(N2)}
\label{RTG-tab5}
\renewcommand{\arraystretch}{1.0}
\vspace*{-2ex}
\small
\begin{tabular}[t]{lrrrrrrrr}
\hline \hline
Species & M1 & M4 & M4B & M4C & Observations \\
\hline\\
CH$_3$NHCHO / HNCO           &  1.1   & 0.070   &  0.24 &  0.055 & 0.050  \\
CH$_3$NCO / HNCO             &  0.021 &  0.32   &  0.19 &  0.30  & 0.11   \\
NH$_2$CHO / HNCO             & 34     &   24    & 23    & 23     & 1.8    \\
CH$_3$CONH$_2$ / HNCO        &  2.3   &   1.4   &  1.5  &  1.5   & 0.070  \\

\\

CH$_3$NHCHO / CH$_3$NCO      &  52    &   0.22  &  1.3  &  0.18  & 0.45  \\
CH$_3$NHCHO / CH$_3$CONH$_2$ &   0.48 &   0.050 &  0.17 &  0.037 & 0.71  \\

\hline
\end{tabular}
\end{center}
\vspace*{-2ex}
\end{table*}

Table~\ref{RTG-tab5} shows ratios of the abundances of the main molecules with 
that of HNCO, as well as the CH$_3$NHCHO/CH$_3$NCO and 
CH$_3$NHCHO/CH$_3$CONH$_2$ ratios. Values are shown for 
the initial model M1, as well as representative models from 
Tables~\ref{RTG-tab1} and \ref{RTG-tab3} that show a good match with observed 
values. Models M4 and M4C both show similarly good agreement with observed 
ratios involving CH$_3$NCO and CH$_3$NHCHO, while M4B is significantly further 
from the observed CH$_3$NHCHO/HNCO ratio. However, all four models produce an 
overabundance of NH$_2$CHO and CH$_3$CONH$_2$ by more 
than an order of magnitude. Because acetamide, CH$_3$CONH$_2$, is 
predominantly formed as a result of H-abstraction from formamide, followed by 
methyl-group addition, its overabundance is related to that of NH$_2$CHO. The 
majority of the latter molecule is formed via the reaction NH$_2$ + H$_2$CO 
$\rightarrow$ NH$_2$CHO + H, which we include on the grains as well as in the 
gas phase, assuming the activation energy barrier determined by 
\citet{Barone15} of 26.9~K. However, \citet{Song16} find a 
substantially higher barrier that would render the reaction rate negligible 
(as determined for a gas-phase interaction). In order to test this 
possibility, we remove the NH$_2$+H$_2$CO mechanism both in the gas phase and 
on grains. This reduces NH$_2$CHO production by around an order of magnitude 
and brings the peak NH$_2$CHO:HNCO ratio achieved in the models to a very 
good match with the
observed value (1.2 versus 1.8). The removal of either the gas-phase or 
grain-surface mechanism alone is not sufficient to reduce NH$_2$CHO abundance, 
as both contribute significantly  in the present implementation. However, even 
the better match to observed NH$_2$CHO is not sufficient to bring down 
acetamide abundances to appropriate levels, as the alternative NH$_2$ + 
CH$_3$CO formation mechanism is also important to its formation. The 
production of acetamide in hot cores merits further detailed study.

While models M4 and M4C show similar agreement with observations in spite of 
the different dominant chemical pathways involved for CH$_3$NHCHO production, 
the fact that NH$_2$CO is a more likely product of the reaction between OH and 
NH$_2$CHO than NHCHO (judging by typical barriers to H-abstraction from an 
amino versus a carbonyl group) makes model M4C more plausible in this respect.
The observed CH$_3$NHCHO/CH$_3$NCO ratio is bracketed by the values obtained
with models M4B and M4C, indicating indeed that a mechanism of 
direct hydrogenation from CH$_3$NCO to CH$_3$NHCHO is capable of reproducing 
observations of these molecules. However, given an appropriate activation 
energy barrier, both reactions~(\ref{RTG-rxn2}) and (\ref{RTG-rxn4}) may 
produce sufficient quantities of CH$_3$NHCHO to agree reasonably with 
observations. 

In order to achieve observed abundances of CH$_3$NCO, a barrier 
to the reaction NH + CO $\rightarrow$ HNCO of no more than $\sim$1500 K is 
required, allowing this to become the dominant mechanism by which HNCO is 
formed on grains (although its abundance in the gas phase is still dominated 
by its formation as a by-product of the destruction of larger species). Under 
such conditions, a barrier to the hydrogenation of CH$_3$NCO may be estimated 
to be around 2000~K to best reproduce observed abundances of N-methylformamide.

\section{Discussion}
\label{s:discussion}

The spectroscopic results obtained in Sect.~\ref{s:spectro} represent a 
significant improvement in the characterization of the spectrum of 
N-methylformamide. First of all, we have substantially expanded the frequency 
coverage. Second, the rotational lines belonging to the first and second 
excited torsional states of N-methylformamide have been assigned and fitted 
for the first time. Finally, the resulting best model is capable of 
reproducing the assigned data set within experimental error, 
including the lines from previous studies that were excluded from the fits due 
to large observed-minus-calculated values \citep{kawashima2010dynamical}.

All the chemical models computed in the course of this work predict a 
substantial abundance of CH$_3$NHCHO compared to CH$_3$NCO. The smallest 
ratios were produced by models M4D and M4E, with [CH$_3$NHCHO]/[CH$_3$NCO] 
$\sim$ 0.07--0.09, but all other models have [CH$_3$NHCHO]/[CH$_3$NCO] 
{\lower.5ex\hbox{$\buildrel > \over \sim$}} 0.2. Two of the chemical models 
provide values that bracket the observed ratio, thus a fine tuning of the
barrier against hydrogenation of CH$_3$NCO would allow the observed ratio to 
be achieved. The chemical modelling thus gives some additional support to the 
tentative interstellar detection of CH$_3$NHCHO.

N-methylformamide, CH$_3$NHCHO, is a structural isomer of acetamide, 
CH$_3$CONH$_2$. As mentioned in Sect.~\ref{s:introduction}, CH$_3$NHCHO is the 
second most stable C$_2$H$_5$NO isomer, CH$_3$CONH$_2$ being the most stable one
\citep[][]{Lattelais10}. In Sect.~\ref{s:obsresults}, we found that, provided 
its detection is confirmed, N-methylformamide is slightly less abundant
than acetamide in Sgr~B2(N2), which at first sight appears to be in agreement 
with the minimum energy principle initially put forward by 
\citet{Lattelais09}. This principle states that the most abundant isomer of a 
given generic chemical formula should be the most stable one thermodynamically. 
\citet{Lattelais09} found a correlation between the observed abundance ratios 
of isomers of several generic chemical formulae as a function of their 
zero-point energy difference. According to this relation, N-methylformamide 
should be $\sim$3.5 times less abundant than acetamide while it is only a 
factor 1.4 less abundant in Sgr~B2(N2). The discrepancy is only slightly 
larger than a factor of two, but it tends to suggest that N-methylformamide 
does not 
follow this correlation closely. In addition, the range of kinetic parameters 
explored in the chemical models presented here produce variations in the 
abundance of CH$_3$NHCHO of more than one order of magnitude. In the case of 
acetamide, while the models all produce an excess over the abundance of 
N-methylformamide, variations in other model parameters within a plausible 
range can produce variations in CH$_3$CONH$_2$ abundance that are comparable 
with those of CH$_3$NHCHO. The availablility of key precursor radicals at the 
optimal temperature for diffusion is one of the key influences on the 
production of such molecules, and one that is unlikely to be controlled purely 
by the thermodynamic properties of the products. Therefore, the abundance 
ratio of the two isomers has most likely an origin based on kinetics over a 
simple thermal equilibrium. A similar conclusion was obtained by 
\citet{Loomis15} and \citet{Loison16} based on observations of the isomers of 
C$_3$H$_2$O, whose abundance ratios are not consistent with the minimum energy 
principle.

A further test would be to measure the abundance of the next C$_2$H$_5$NO 
isomer, acetimidic acid, CH$_3$C(OH)NH  
which has a slightly higher zero-point energy than 
N-methylformamide \citep[][]{Lattelais10}. This would require laboratory 
measurements to characterize the rotational spectrum of this molecule and 
produce spectroscopic predictions suitable for an astronomical search. The
dipole moment of this molecule, however, is more than a factor of 2 smaller 
than the ones of acetamide and N-methylformamide \citep[][]{Lattelais09},
making its detection more challenging.

\section{Conclusions}
\label{s:conclusions}

The rotational spectrum of the \textit{trans} conformer of N-methylformamide  
was studied in the laboratory in the frequency range from 45 to 630 GHz using 
two different spectrometers in Kharkiv and Lille. The new data provides 
significant expansion both in frequency range (from 118 GHz to 630 GHz) and 
quantum number coverage (from $J=11$ to $J = 62$), including the first 
assignment of the rotational spectra of the first and second excited torsional 
states of N-methylformamide. The final data set contains 12456 $A$- 
and $E$-type transitions in the ground, first, and second excited torsional 
states of the \textit{trans} conformer. Our theoretical model fits the 
available data with a weighted root-mean-square deviation of 0.84, i.e. within 
experimental error. The obtained results provide a firm basis for reliable 
predictions of the N-methylformamide spectrum in the millimeter and 
submillimeter wavelength range for the needs of radio astronomy.

Using the spectroscopic predictions obtained here, we report the first 
tentative interstellar detection of N-methylformamide. The main results of 
this study can be summarized as follows:
\begin{enumerate}
 \item Five transitions of N-methylformamide are coincident with 
spectral lines detected toward the hot molecular core Sgr~B2(N2). These lines
are not contaminated by other species, and their intensities are 
well reproduced by our LTE model of N-methylformamide. This suggests that the 
molecule may be present in this source.
 \item We derive a column density of $\sim$$1 \times 10^{17}$~cm$^{-2}$ for
N-methylformamide. The molecule is more than one order of magnitude less 
abundant than formamide, twice less abundant than methyl isocyanate, and only 
slightly less abundant than acetamide.
 \item Our gas-grain chemical kinetics model is able to reproduce the 
abundance ratio of N-methylformamide to methyl isocyanate using kinetic
parameters within a plausible range, supporting the tentative detection of the 
former.
 \item The chemical models indicate that the 
efficient formation of HNCO via NH + CO on grains is a necessary step in the 
achievement of the observed gas-phase abundance of CH$_3$NCO.
 \item Production of CH$_3$NHCHO may plausibly occur on grains either through 
the direct addition of functional-group radicals or through the hydrogenation 
of CH$_3$NCO.
 \item We also report the tentative detection of the $^{15}$N isotopologue of
formamide toward Sgr~B2(N2) with a $^{14}$N/$^{15}$N isotopic ratio of 300.
\end{enumerate}
Provided the detection of N-methylformamide is confirmed, the only slight 
underabundance of this molecule compared to its more stable structural isomer 
and the sensitivity of the model abundances to the variations 
of the model parameters suggest that the formation of these two molecules is 
controlled by kinetics rather than thermal equilibrium. The interstellar 
detection of the next stable isomer of the C$_2$H$_5$NO family, CH$_3$C(OH)NH, 
may therefore become possible once its rotational spectrum will have been 
measured in the laboratory.

\begin{acknowledgements}
This paper makes use of the following ALMA data: 
ADS/JAO.ALMA\#2011.0.00017.S, ADS/JAO.ALMA\#2012.1.00012.S. 
ALMA is a partnership of ESO (representing its member states), NSF (USA), and 
NINS (Japan), together with NRC (Canada), NSC and ASIAA (Taiwan), and KASI 
(Republic of Korea), in cooperation with the Republic of Chile. The Joint ALMA 
Observatory is operated by ESO, AUI/NRAO, and NAOJ. The interferometric data 
are available in the ALMA archive at https://almascience.eso.org/aq/. 
Part of this work was done within the Ukrainian-French CNRS-PICS \# 6051 
project.
This work has been in part supported by the Deutsche Forschungsgemeinschaft 
(DFG) through the collaborative research grant SFB 956 ``Conditions and Impact 
of Star Formation'', project area B3. 
Our research benefited from NASA's Astrophysics Data System (ADS).
\end{acknowledgements}

\onecolumn
\begin{appendix}
\label{Appendix}
\section{Complementary figures}

\begin{figure*}
\centerline{\resizebox{0.82\hsize}{!}{\includegraphics[angle=0]{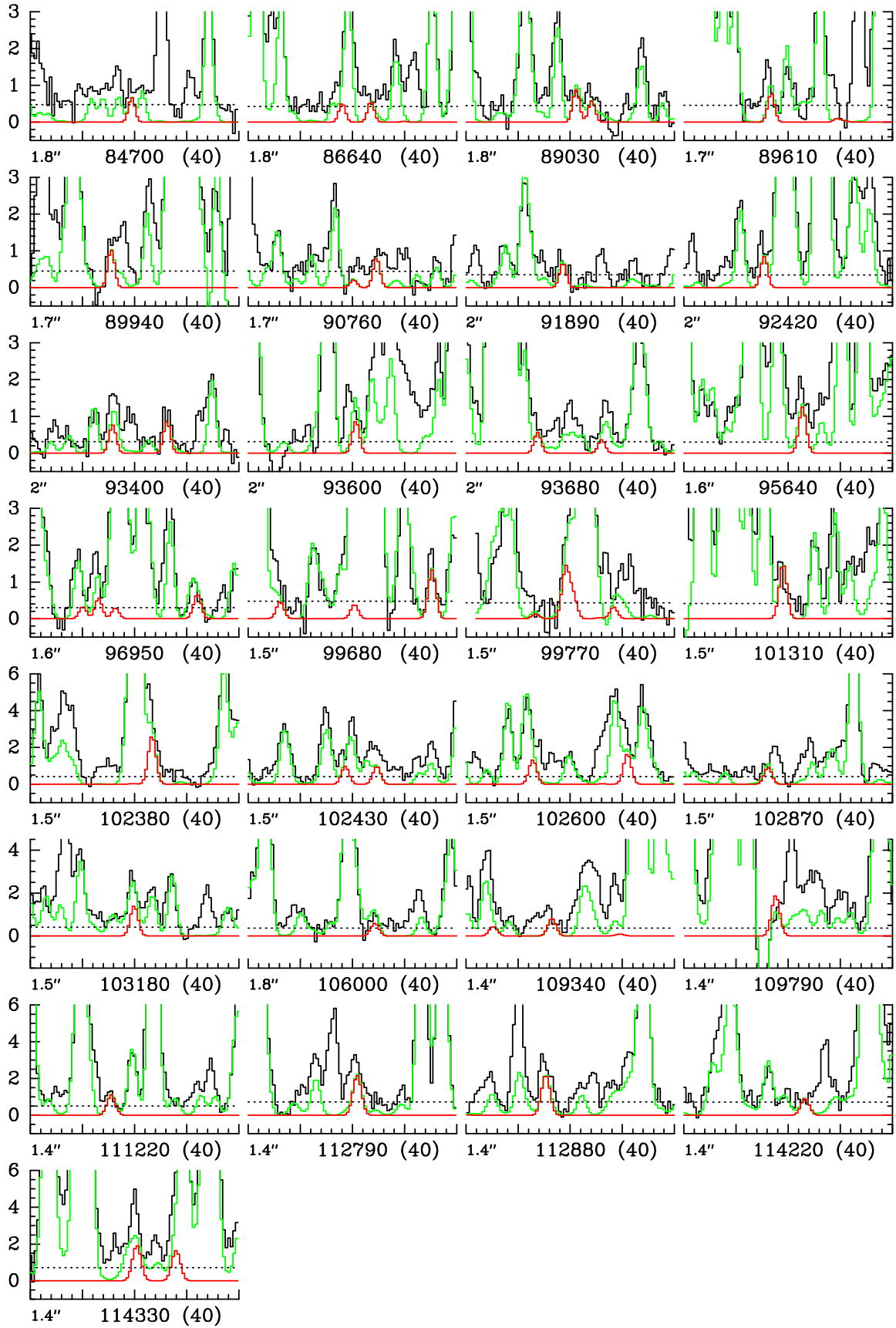}}}
\caption{Transitions of CH$_3$NHCHO, $\varv_{\rm t} = 0$ covered by our ALMA 
survey. The best-fit LTE synthetic spectrum of CH$_3$NHCHO, $\varv_{\rm t} = 0$ 
is displayed in red and overlaid on the observed spectrum of Sgr~B2(N2) shown 
in black. The green synthetic spectrum contains the contributions of all 
molecules identified in our survey so far, including the species shown in red. 
The central frequency and width are indicated in MHz below each panel. The 
angular resolution (HPBW) is also indicated. The y-axis is labeled in 
brightness temperature units (K). The dotted line indicates the $3\sigma$ 
noise level.
}
\label{f:spec_ch3nhcho_ve0}
\end{figure*}

\begin{figure*}
\centerline{\resizebox{0.82\hsize}{!}{\includegraphics[angle=0]{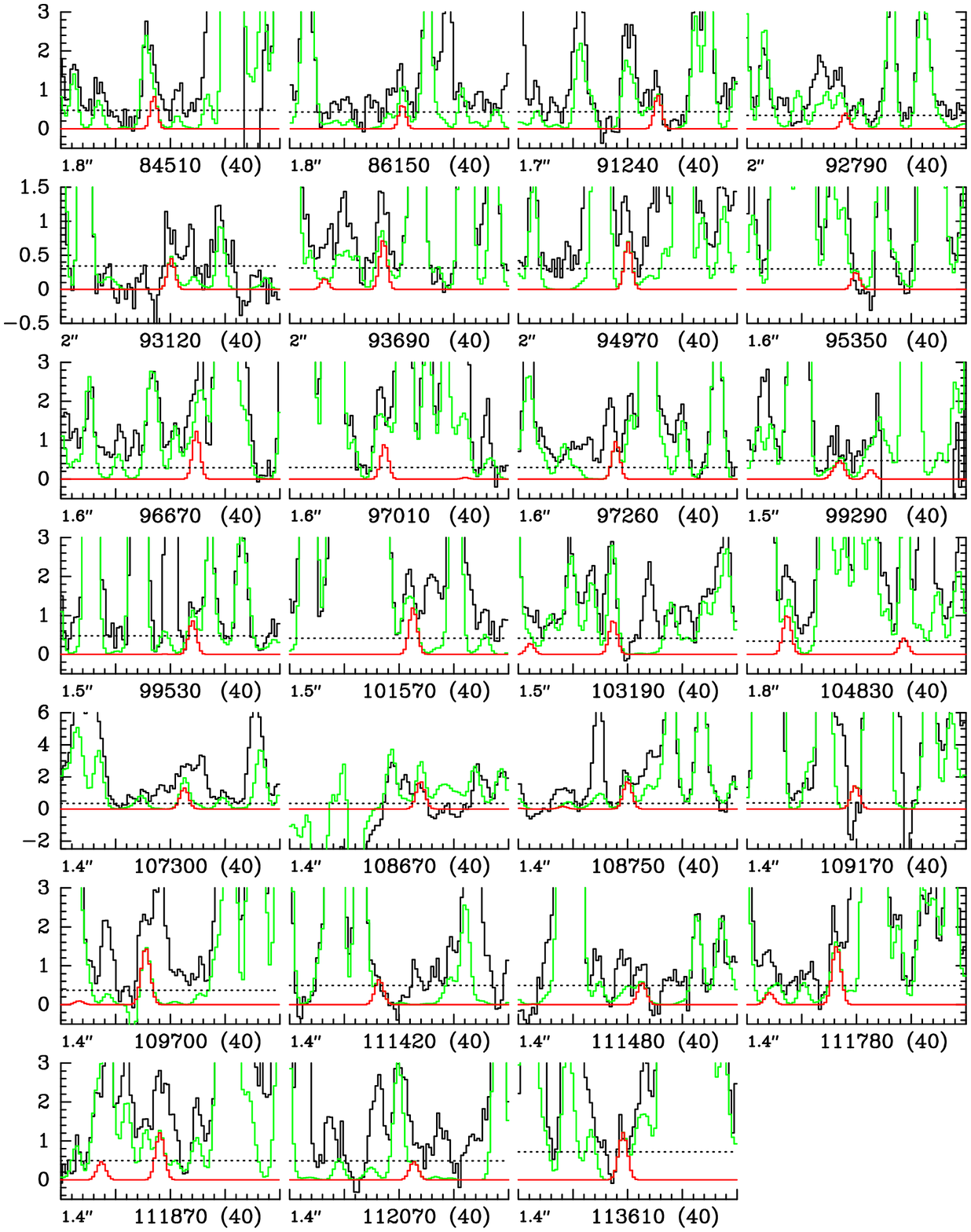}}}
\caption{Same as Fig.~\ref{f:spec_ch3nhcho_ve0} for CH$_3$NHCHO, $\varv_{\rm t} = 1$.
}
\label{f:spec_ch3nhcho_ve1}
\end{figure*}

\begin{figure*}
\centerline{\resizebox{0.82\hsize}{!}{\includegraphics[angle=0]{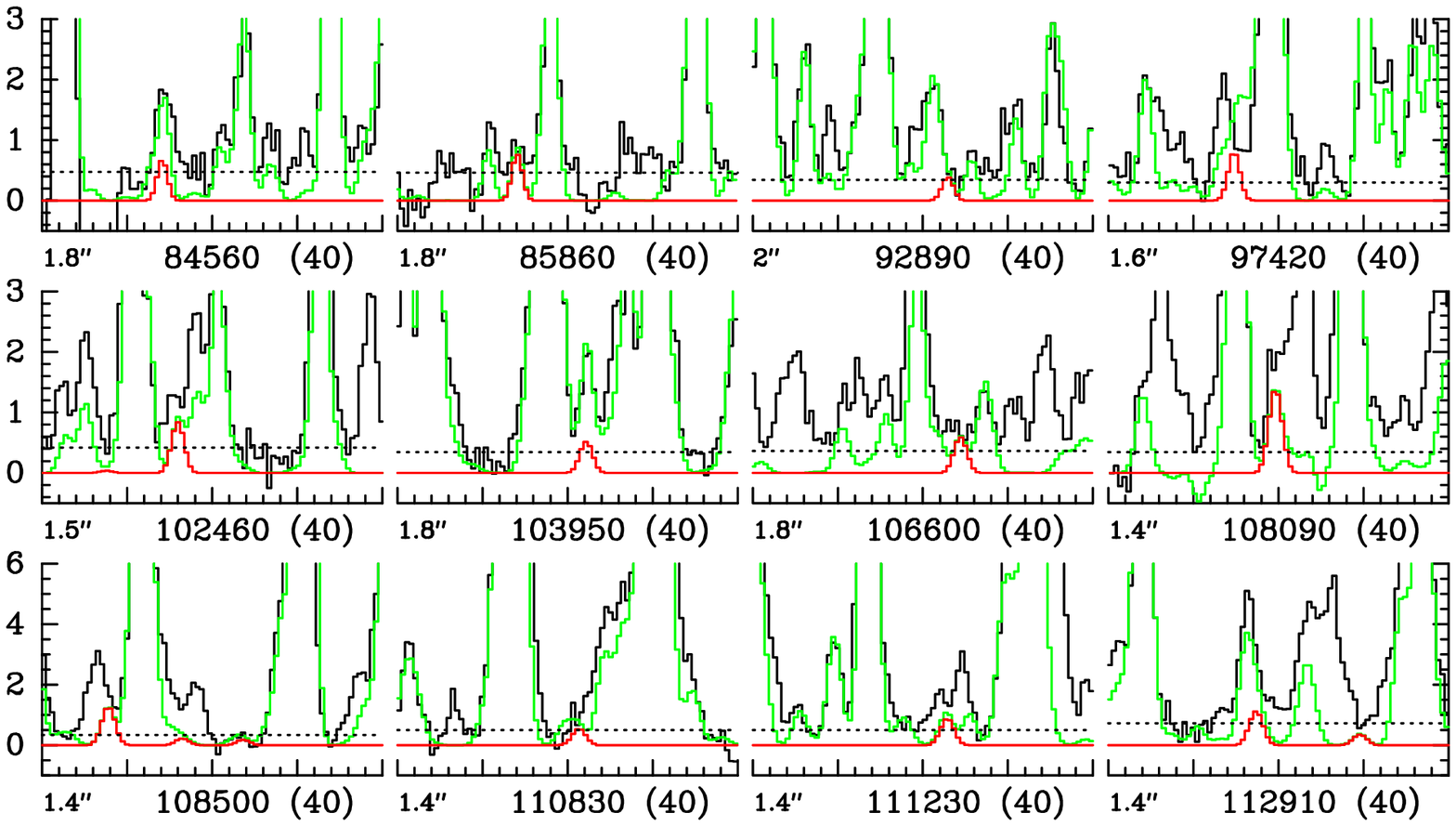}}}
\caption{Same as Fig.~\ref{f:spec_ch3nhcho_ve0} for CH$_3$NHCHO, $\varv_{\rm t} = 2$.
}
\label{f:spec_ch3nhcho_ve2}
\end{figure*}

Figures~\ref{f:spec_ch3nhcho_ve0}--\ref{f:spec_ch3conh2_cv} show the
transitions of CH$_3$NHCHO, CH$_3$NCO, NH$_2$CHO, HNCO, CH$_3$CONH$_2$, and 
some of their isotopologues or vibrationally excited
states that are covered by the EMoCA survey and contribute significantly to 
the signal detected toward Sgr~B2(N2). 
Figures~\ref{f:popdiag_ch3nco}--\ref{f:popdiag_ch3conh2} show the population 
diagrams of these molecules except the one of CH$_3$NHCHO which is shown in 
Fig.~\ref{f:popdiag_ch3nhcho}.

\clearpage

\clearpage

\clearpage
\begin{figure*}
\centerline{\resizebox{0.82\hsize}{!}{\includegraphics[angle=0]{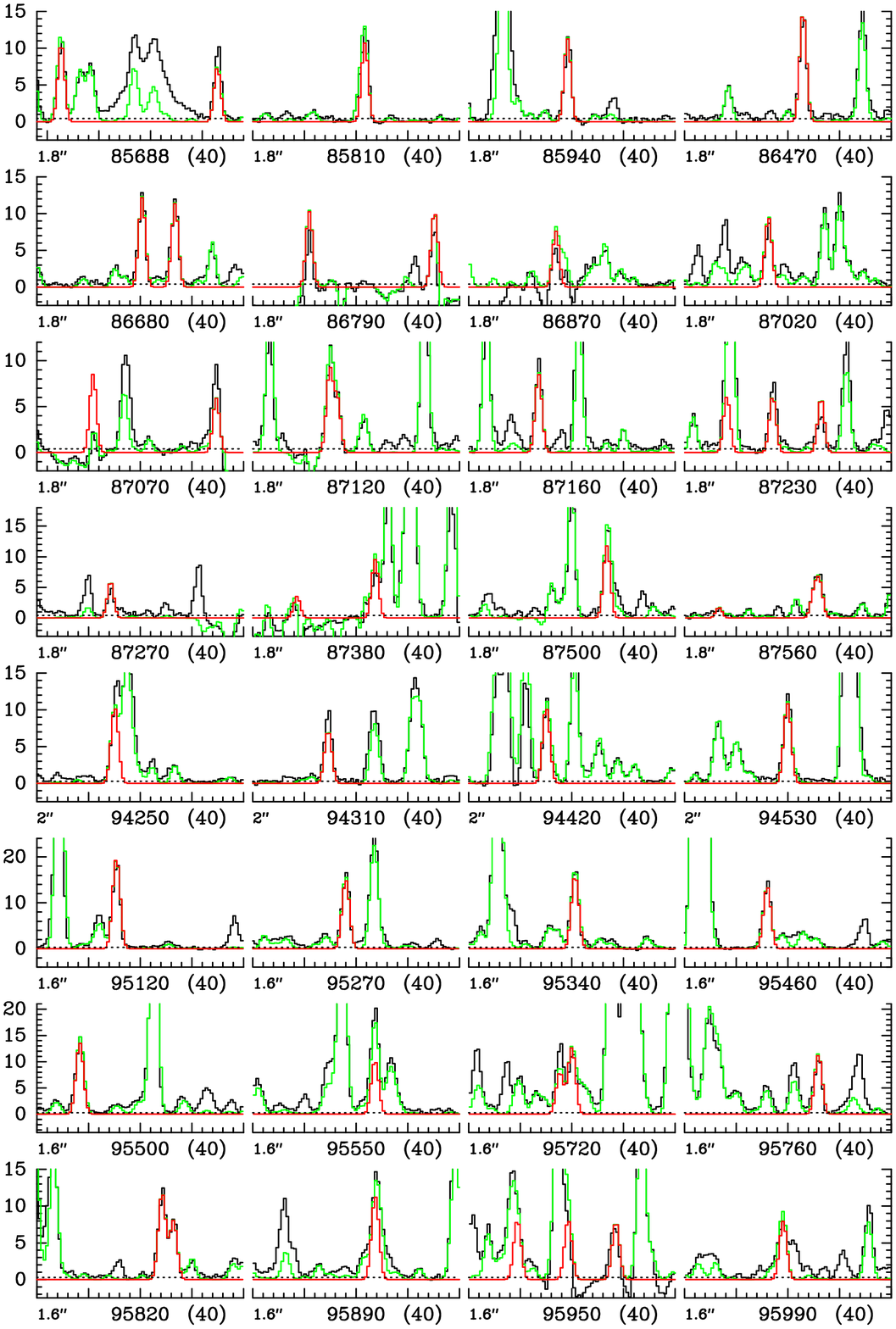}}}
\caption{Same as Fig.~\ref{f:spec_ch3nhcho_ve0} for CH$_3$NCO, $\varv_{\rm b} = 0$.
}
\label{f:spec_ch3nco_ve0}
\end{figure*}

\clearpage
\begin{figure*}
\addtocounter{figure}{-1}
\centerline{\resizebox{0.82\hsize}{!}{\includegraphics[angle=0]{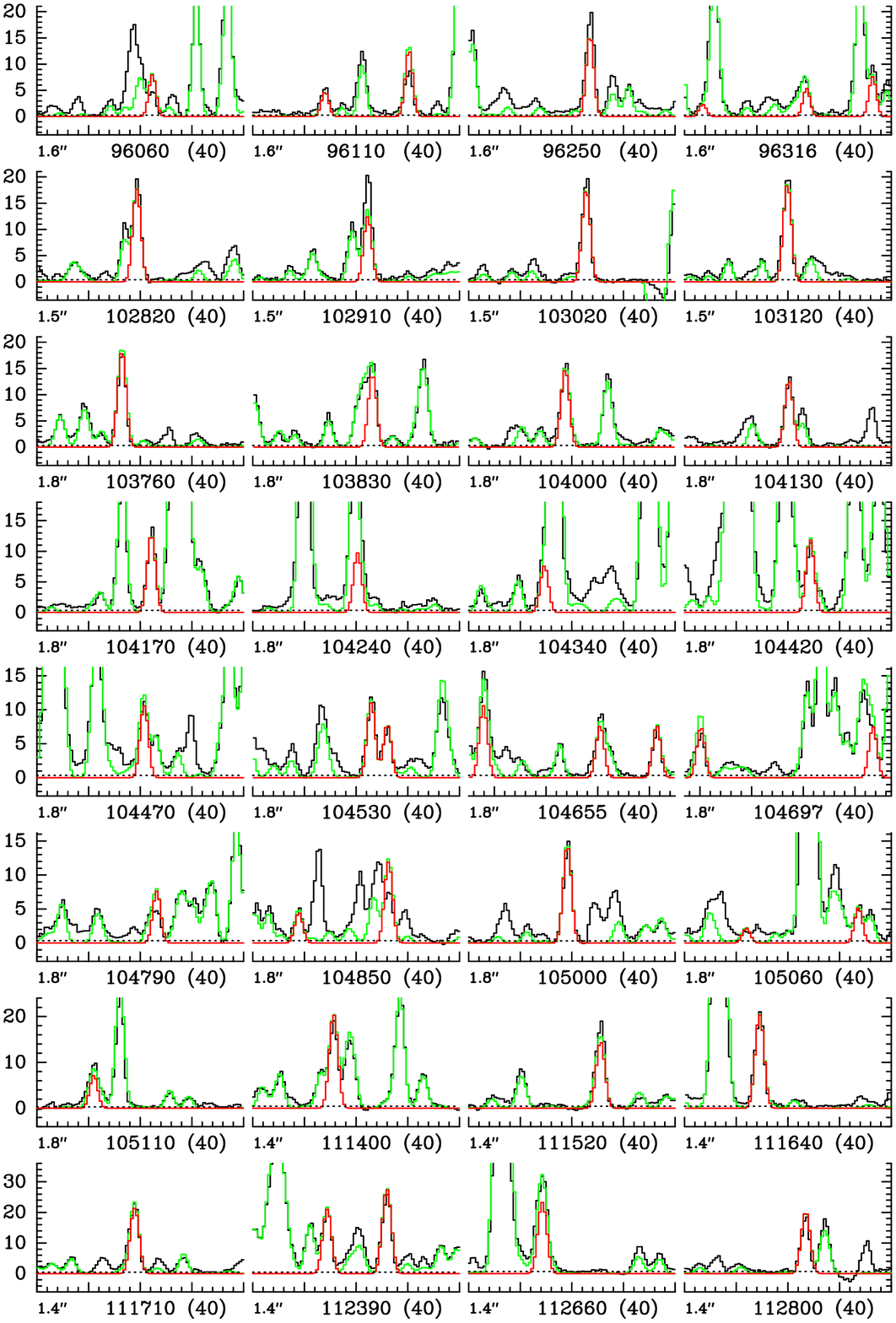}}}
\caption{continued.}
\end{figure*}

\clearpage
\begin{figure*}
\addtocounter{figure}{-1}
\centerline{\resizebox{0.82\hsize}{!}{\includegraphics[angle=0]{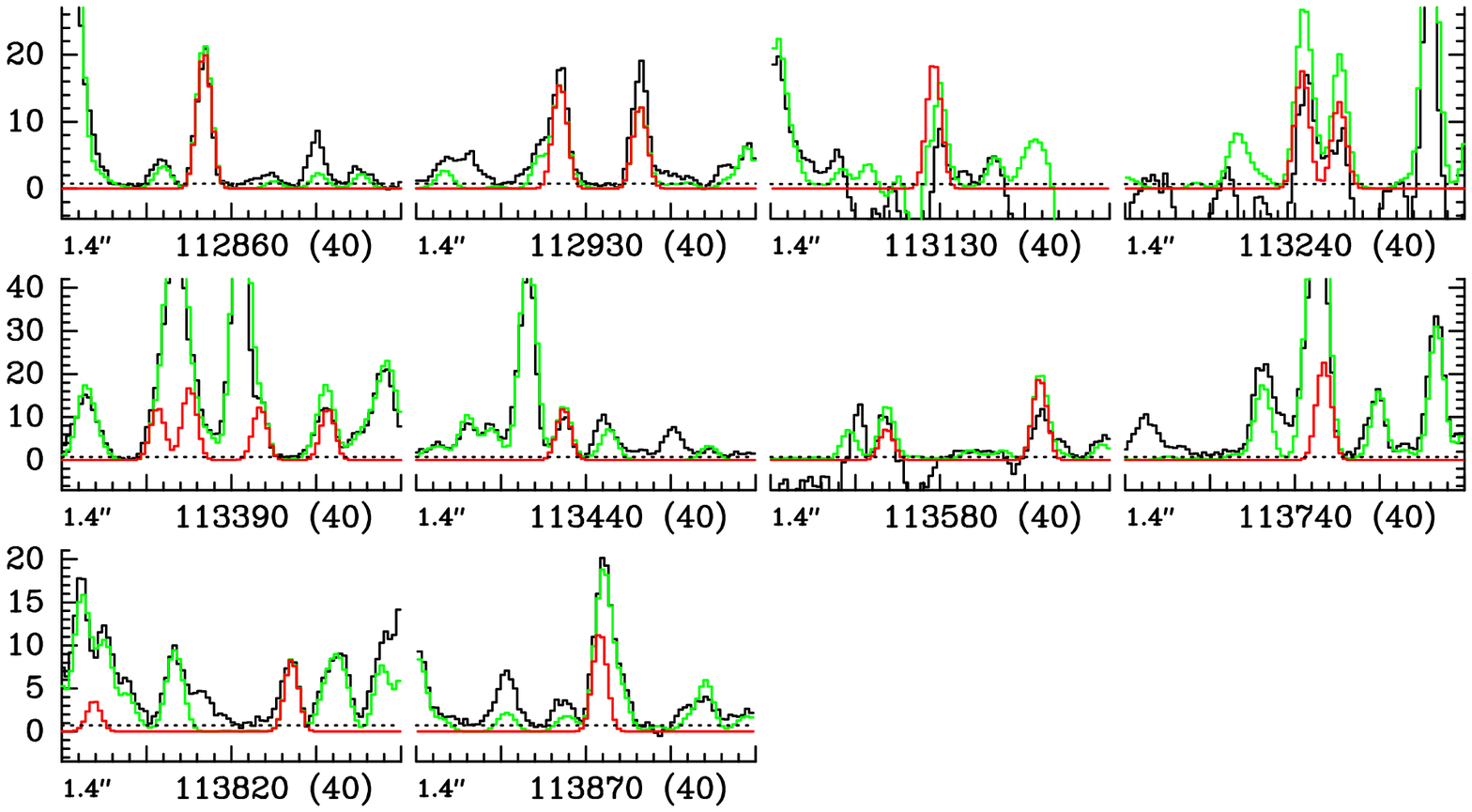}}}
\caption{continued.}
\end{figure*}

\begin{figure*}
\centerline{\resizebox{0.82\hsize}{!}{\includegraphics[angle=0]{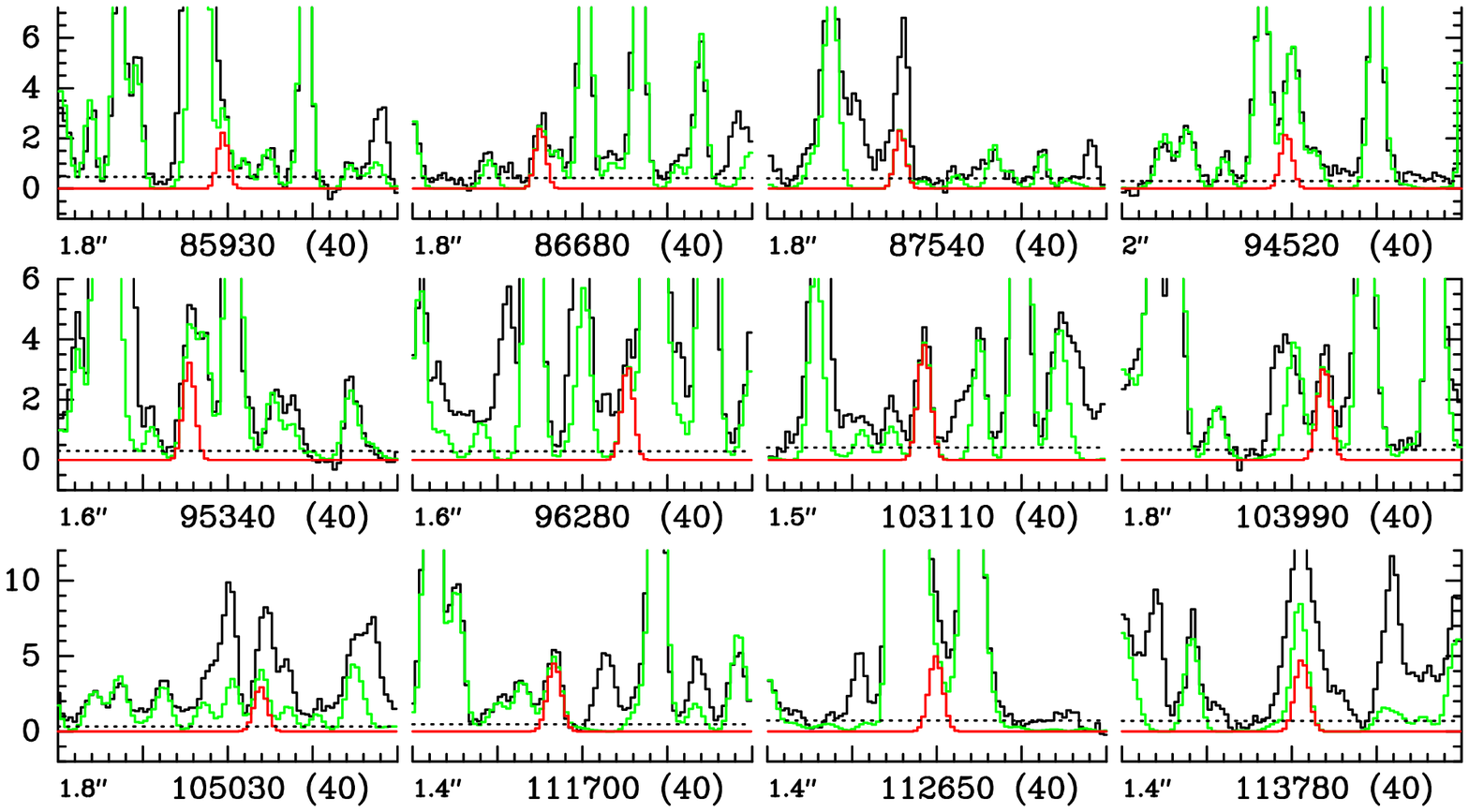}}}
\caption{Same as Fig.~\ref{f:spec_ch3nhcho_ve0} for CH$_3$NCO, $\varv_{\rm b} = 1$.
}
\label{f:spec_ch3nco_ve1}
\end{figure*}

\clearpage
\begin{figure*}
\centerline{\resizebox{0.82\hsize}{!}{\includegraphics[angle=0]{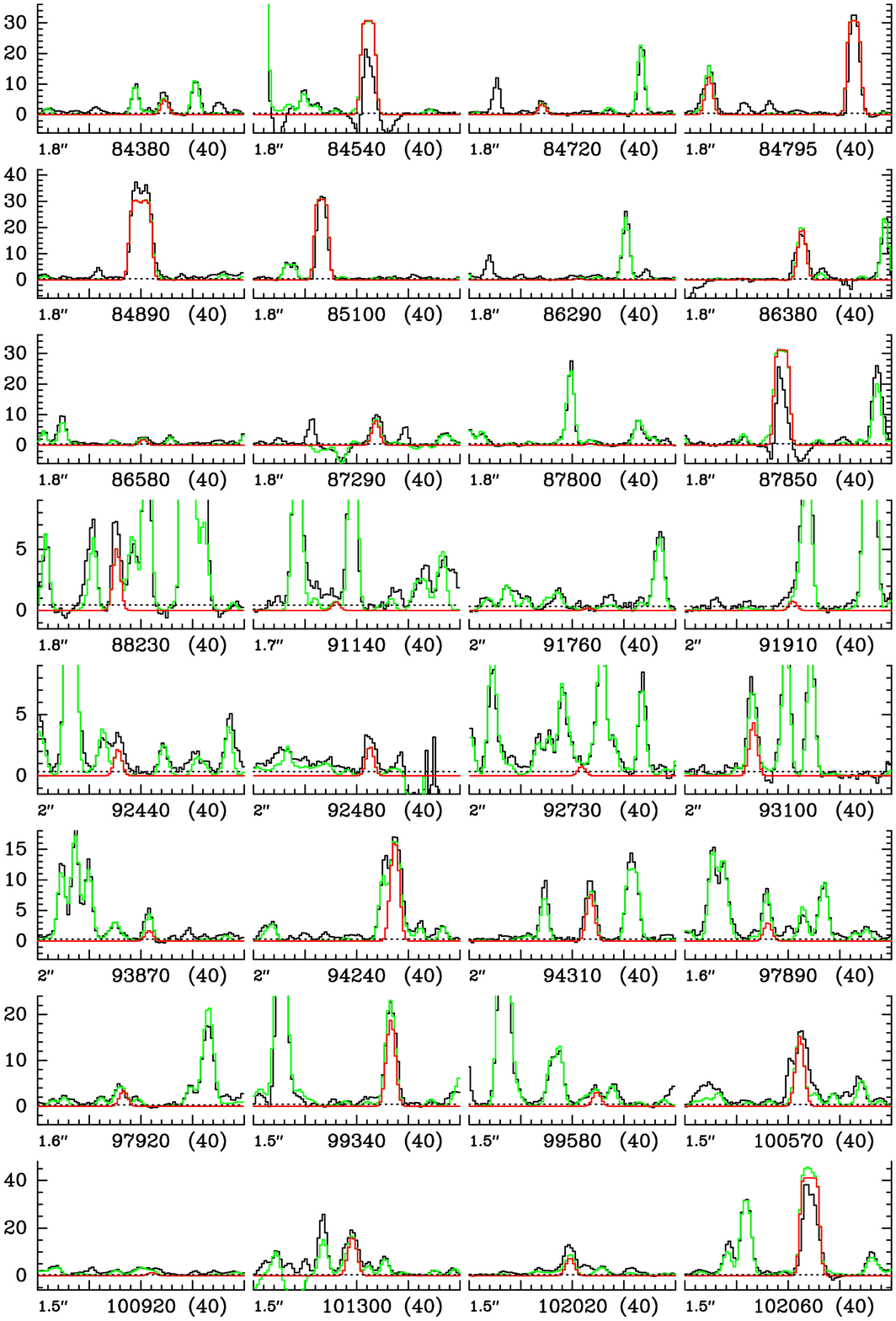}}}
\caption{Same as Fig.~\ref{f:spec_ch3nhcho_ve0} for NH$_2$CHO, $\varv = 0$.
}
\label{f:spec_nh2cho_ve0}
\end{figure*}

\clearpage
\begin{figure*}
\addtocounter{figure}{-1}
\centerline{\resizebox{0.82\hsize}{!}{\includegraphics[angle=0]{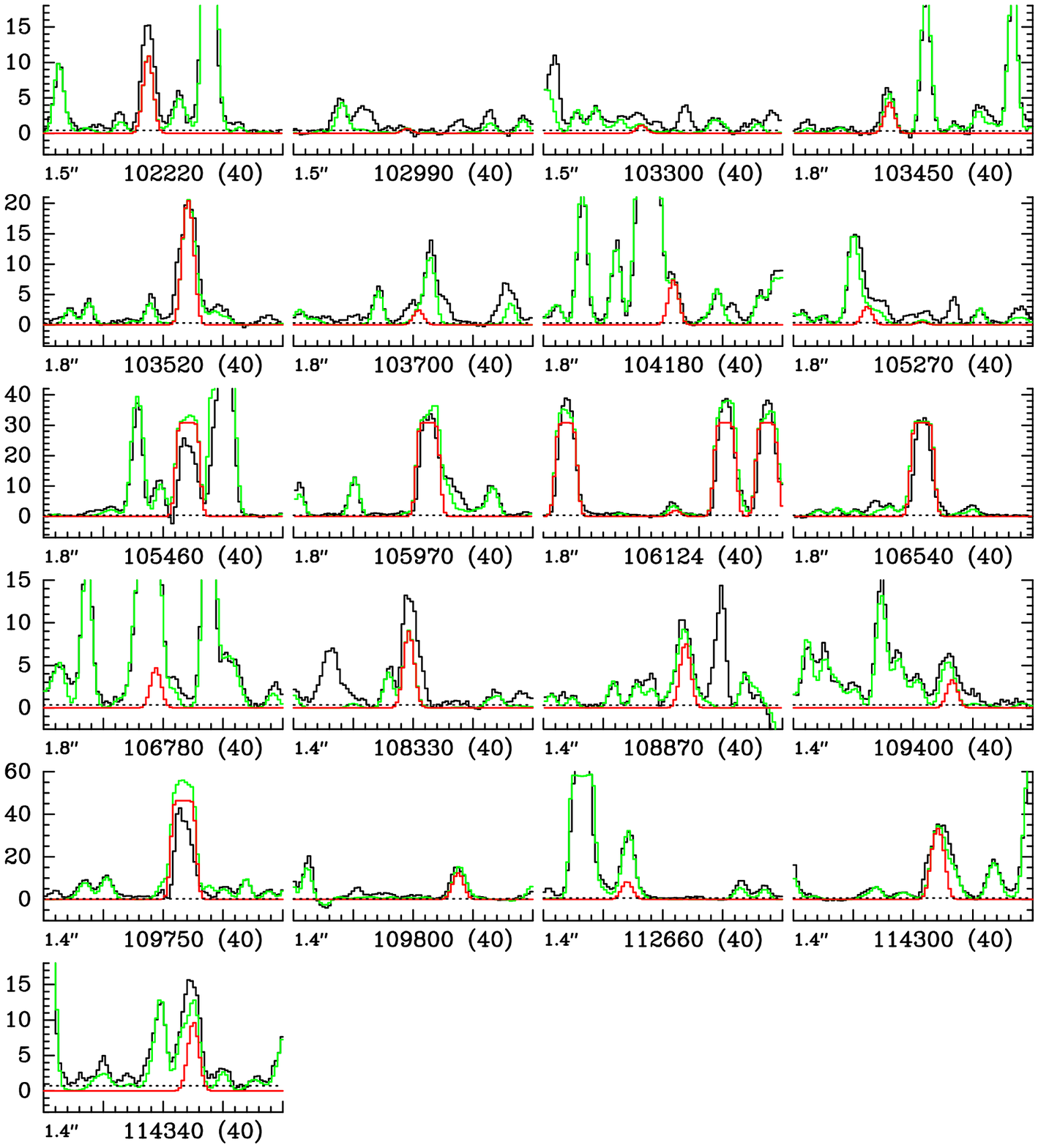}}}
\caption{continued.}
\end{figure*}

\clearpage
\begin{figure*}
\centerline{\resizebox{0.82\hsize}{!}{\includegraphics[angle=0]{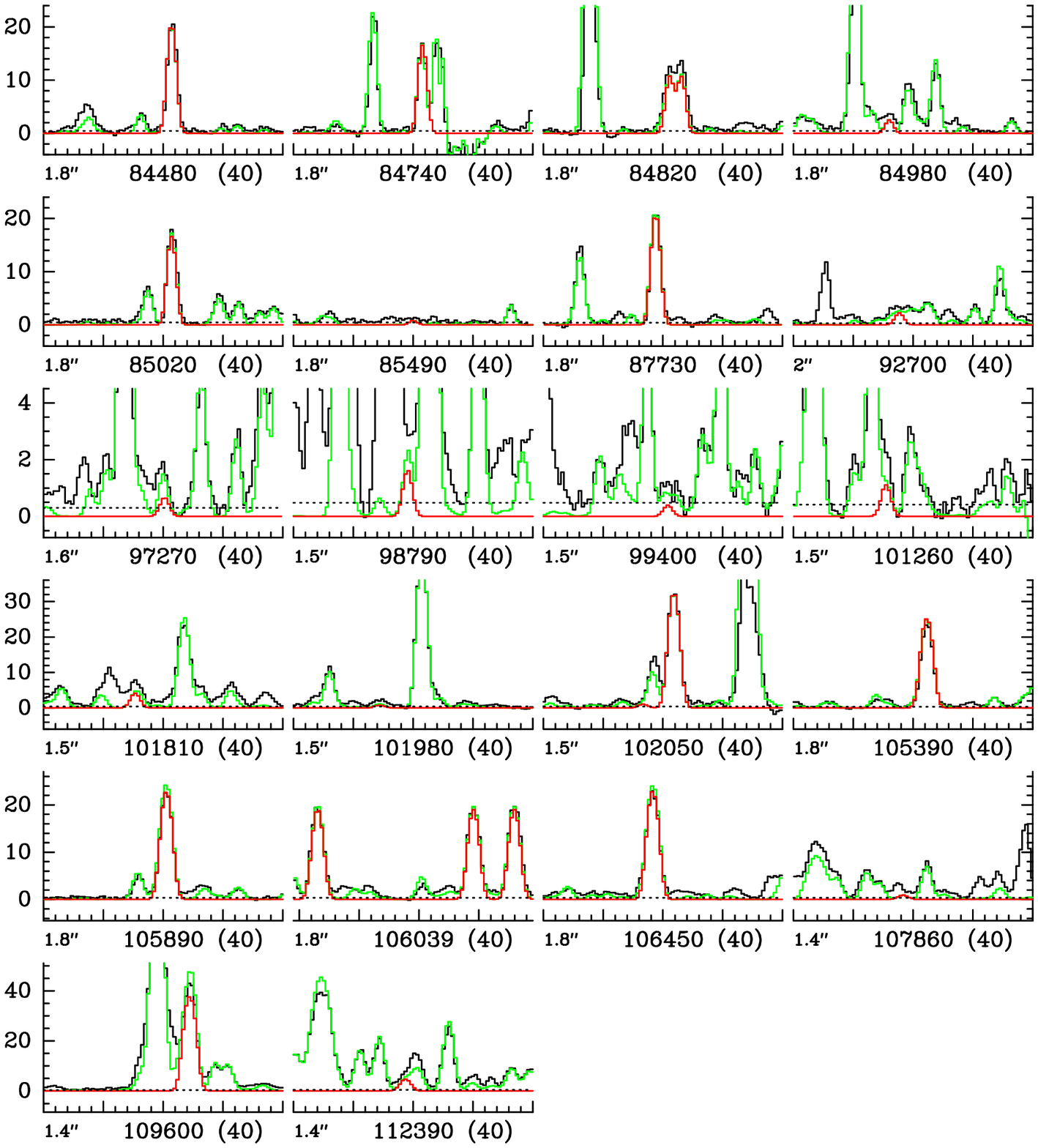}}}
\caption{Same as Fig.~\ref{f:spec_ch3nhcho_ve0} for NH$_2$CHO, $\varv_{12} = 1$.
}
\label{f:spec_nh2cho_v12e1}
\end{figure*}

\clearpage
\begin{figure*}
\centerline{\resizebox{0.82\hsize}{!}{\includegraphics[angle=0]{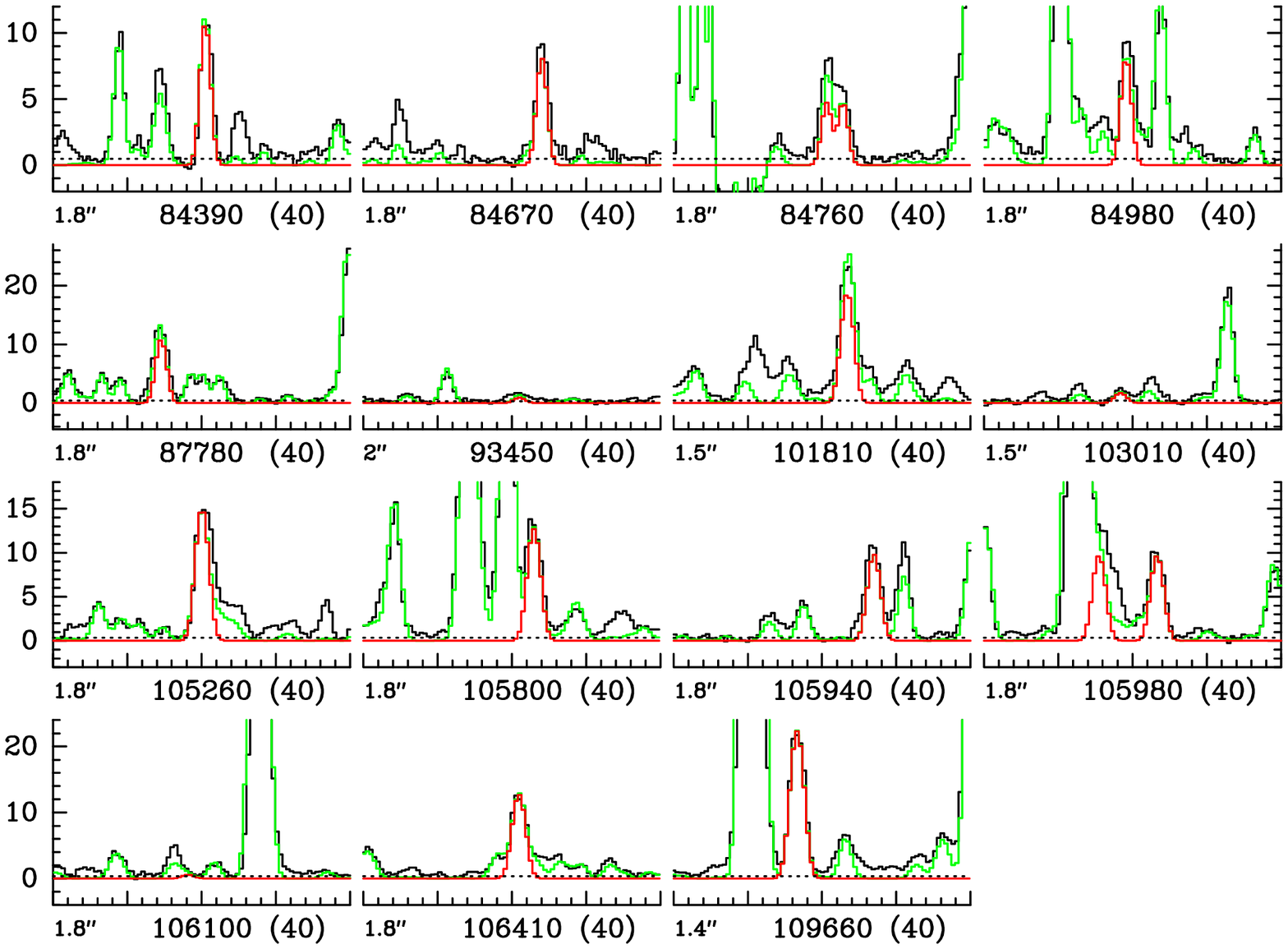}}}
\caption{Same as Fig.~\ref{f:spec_ch3nhcho_ve0} for NH$_2$$^{13}$CHO,
$\varv = 0$.}
\label{f:spec_nh2cho_13c_ve0}
\end{figure*}

\begin{figure*}
\centerline{\resizebox{0.82\hsize}{!}{\includegraphics[angle=0]{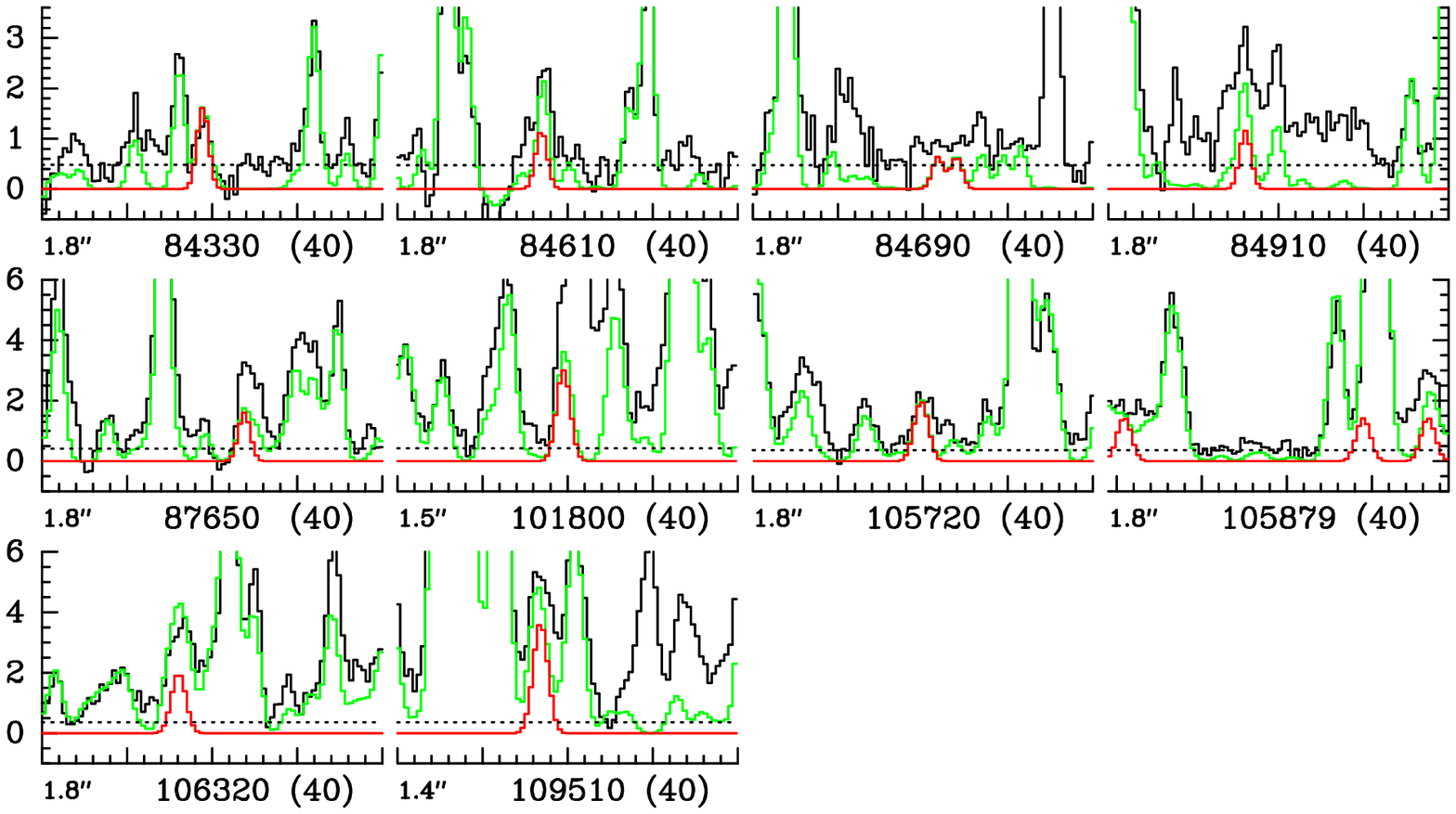}}}
\caption{Same as Fig.~\ref{f:spec_ch3nhcho_ve0} for NH$_2$$^{13}$CHO,
$\varv_{12} = 1$.}
\label{f:spec_nh2cho_13c_v12e1}
\end{figure*}

\clearpage
\begin{figure*}
\centerline{\resizebox{0.82\hsize}{!}{\includegraphics[angle=0]{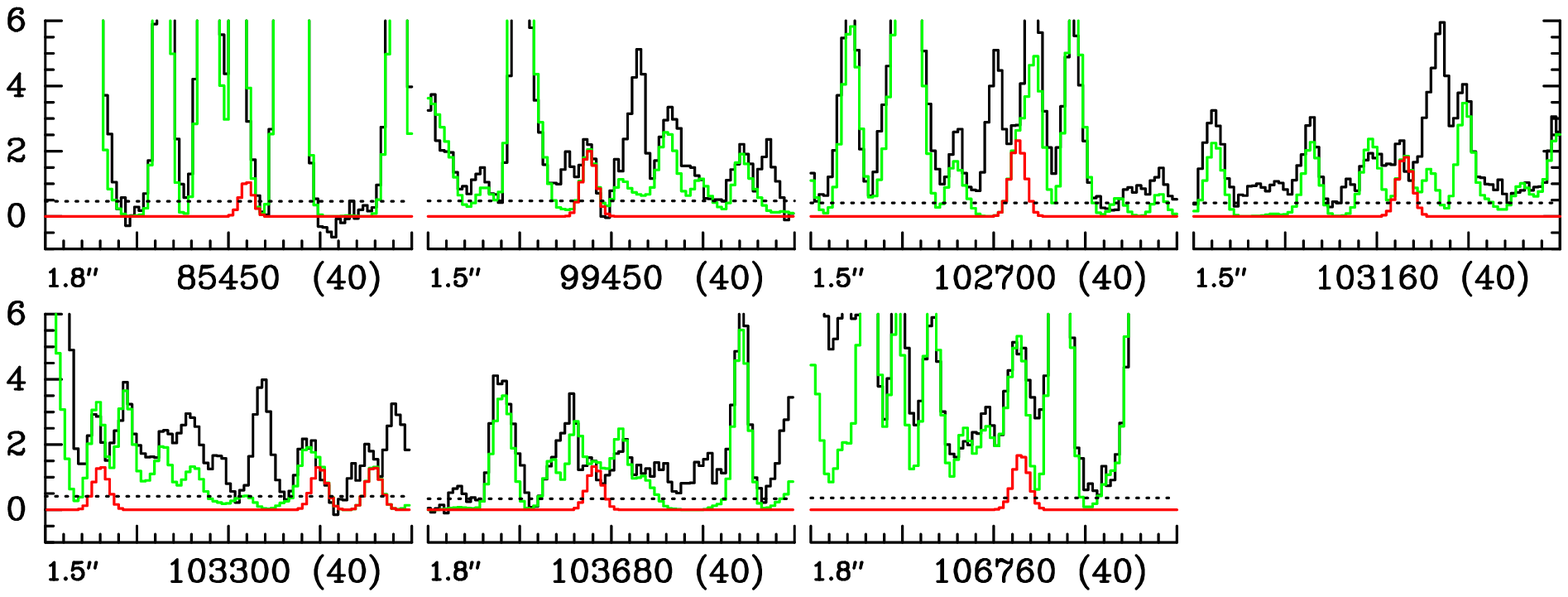}}}
\caption{Same as Fig.~\ref{f:spec_ch3nhcho_ve0} for $^{15}$NH$_2$CHO,
$\varv = 0$.}
\label{f:spec_nh2cho_15n_ve0}
\end{figure*}

\begin{figure*}
\centerline{\resizebox{0.82\hsize}{!}{\includegraphics[angle=0]{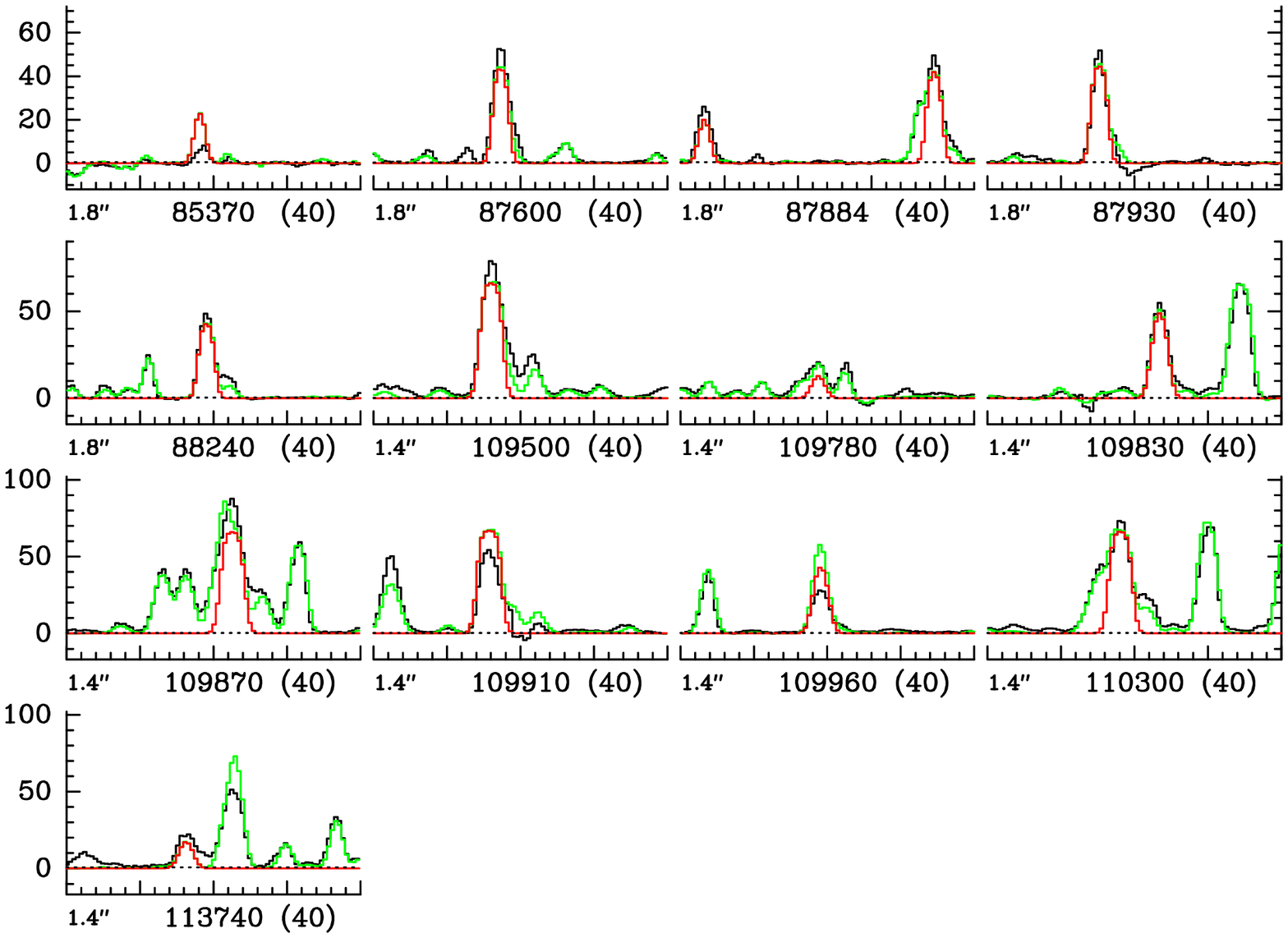}}}
\caption{Same as Fig.~\ref{f:spec_ch3nhcho_ve0} for HNCO, $\varv = 0$.
}
\label{f:spec_hnco_ve0}
\end{figure*}

\clearpage
\begin{figure*}
\centerline{\resizebox{0.82\hsize}{!}{\includegraphics[angle=0]{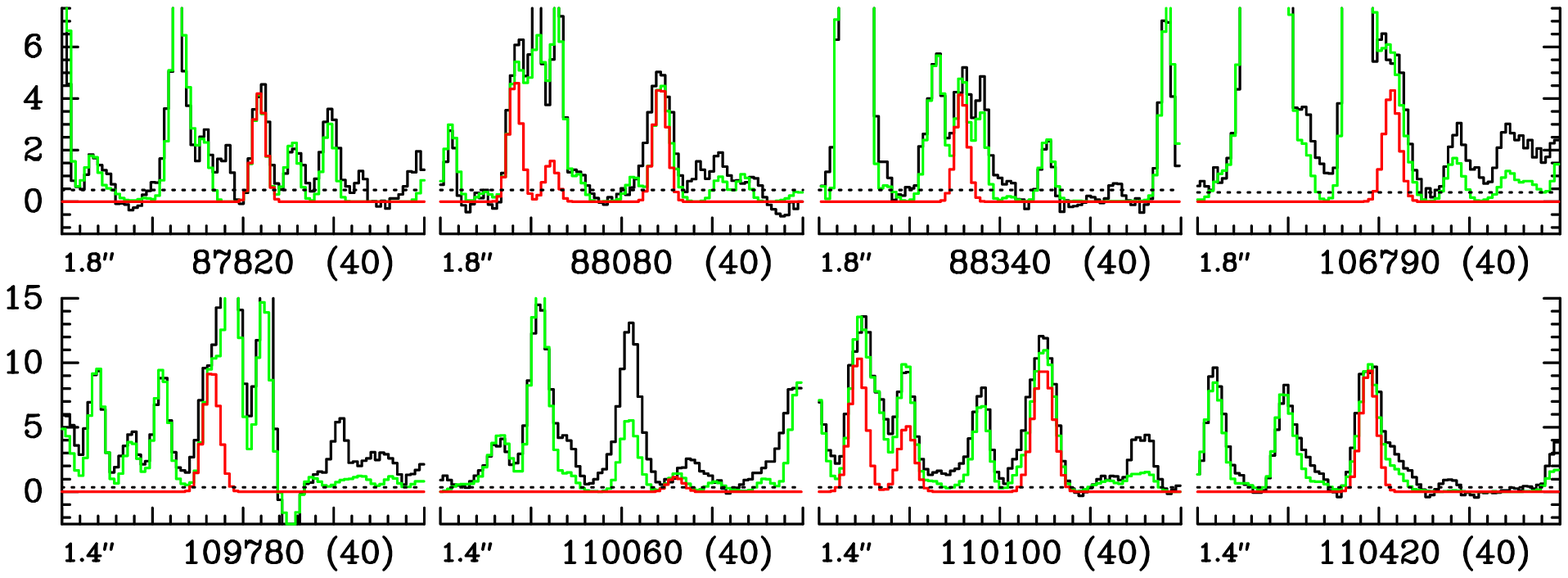}}}
\caption{Same as Fig.~\ref{f:spec_ch3nhcho_ve0} for HNCO, $\varv_5 = 1$.
}
\label{f:spec_hnco_v5e1}
\end{figure*}

\begin{figure*}
\centerline{\resizebox{0.82\hsize}{!}{\includegraphics[angle=0]{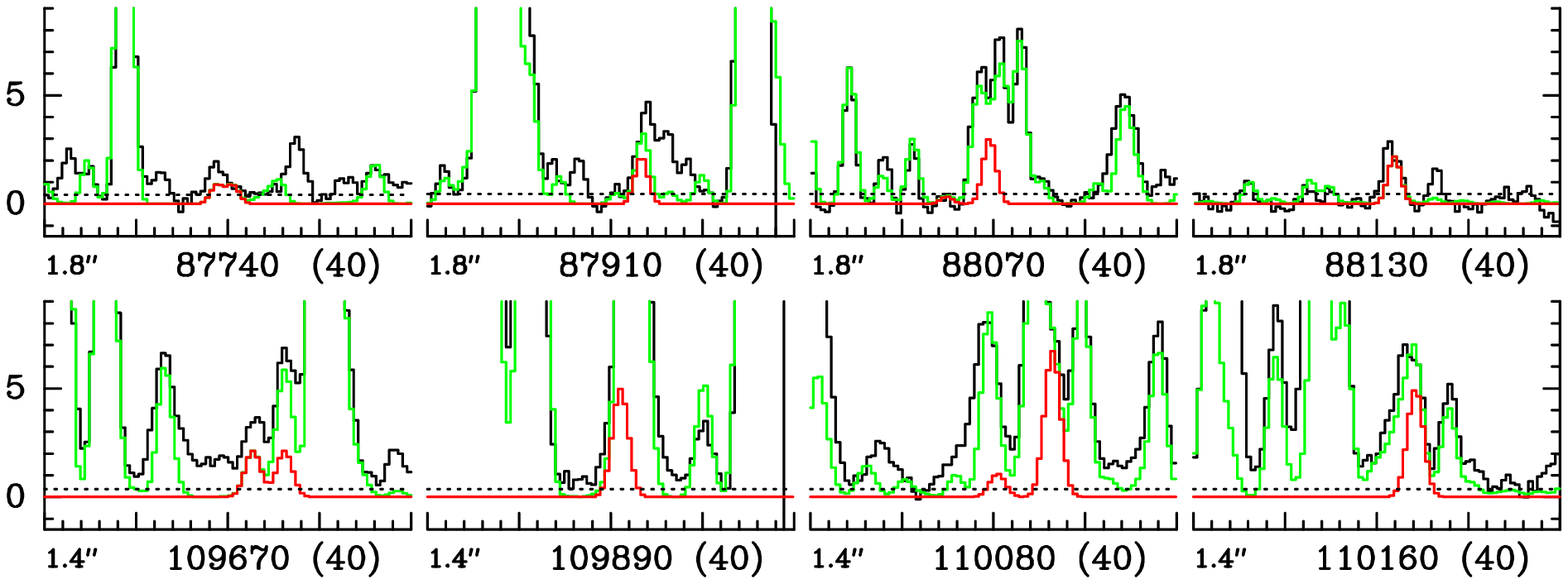}}}
\caption{Same as Fig.~\ref{f:spec_ch3nhcho_ve0} for HNCO, $\varv_6 = 1$.
}
\label{f:spec_hnco_v6e1}
\end{figure*}

\begin{figure*}
\centerline{\resizebox{0.82\hsize}{!}{\includegraphics[angle=0]{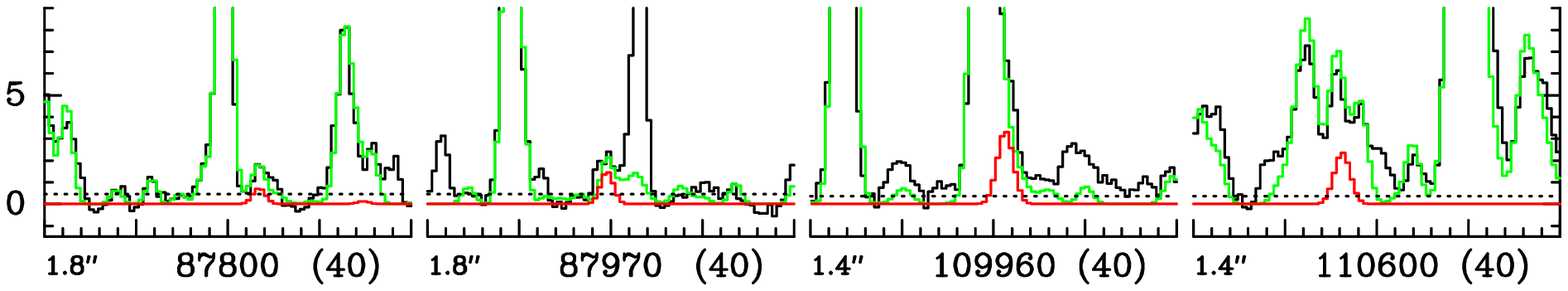}}}
\caption{Same as Fig.~\ref{f:spec_ch3nhcho_ve0} for HNCO, $\varv_4 = 1$.
}
\label{f:spec_hnco_v4e1}
\end{figure*}

\clearpage
\begin{figure*}
\centerline{\resizebox{0.82\hsize}{!}{\includegraphics[angle=0]{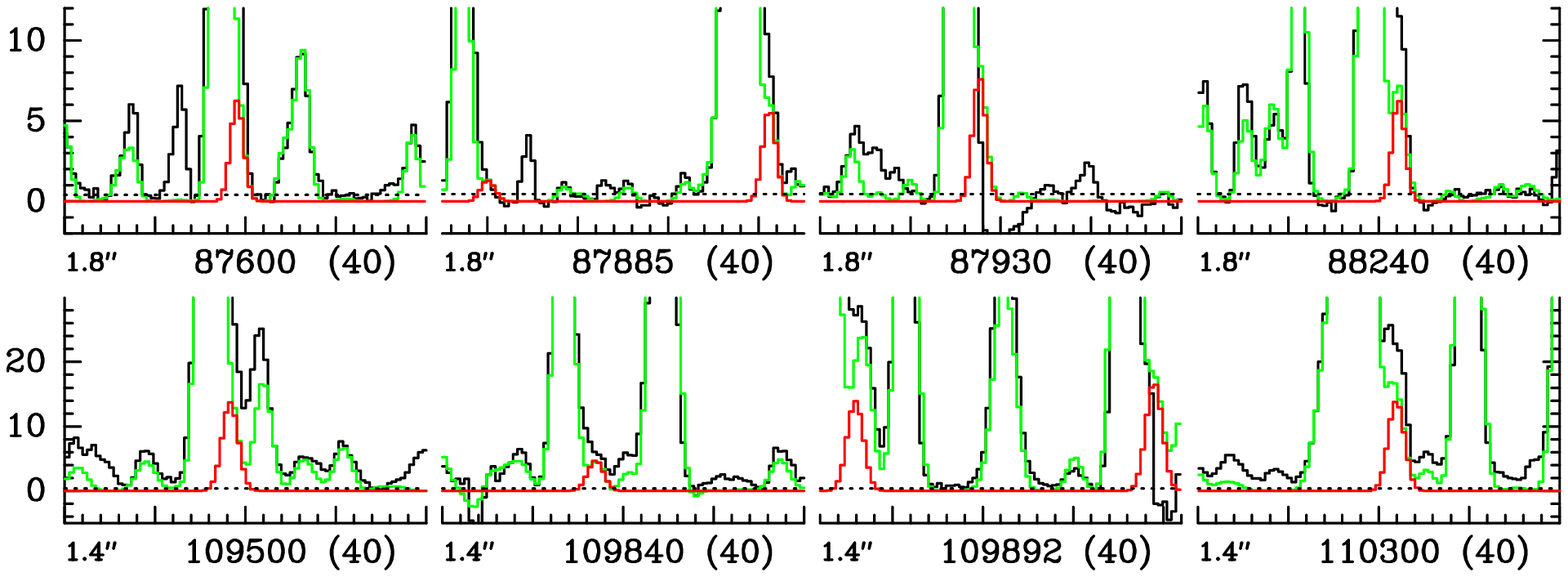}}}
\caption{Same as Fig.~\ref{f:spec_ch3nhcho_ve0} for HN$^{13}$CO, 
$\varv = 0$.
}
\label{f:spec_hnco_13c_ve0}
\end{figure*}

\clearpage
\begin{figure*}
\centerline{\resizebox{0.82\hsize}{!}{\includegraphics[angle=0]{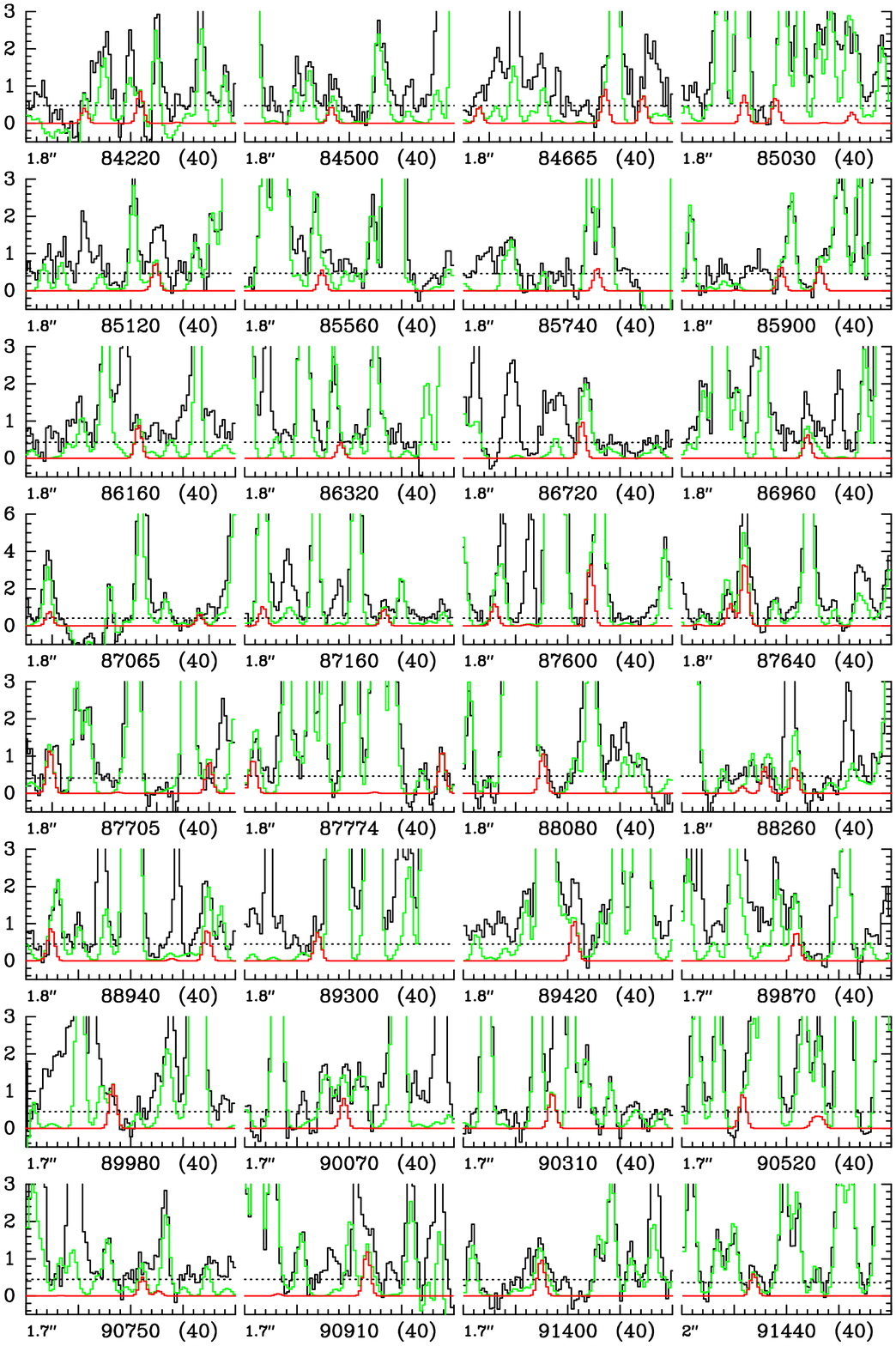}}}
\caption{Same as Fig.~\ref{f:spec_ch3nhcho_ve0} for CH$_3$CONH$_2$, 
$\varv_{\rm t} = 0$.
}
\label{f:spec_ch3conh2_ve0}
\end{figure*}

\clearpage
\begin{figure*}
\addtocounter{figure}{-1}
\centerline{\resizebox{0.82\hsize}{!}{\includegraphics[angle=0]{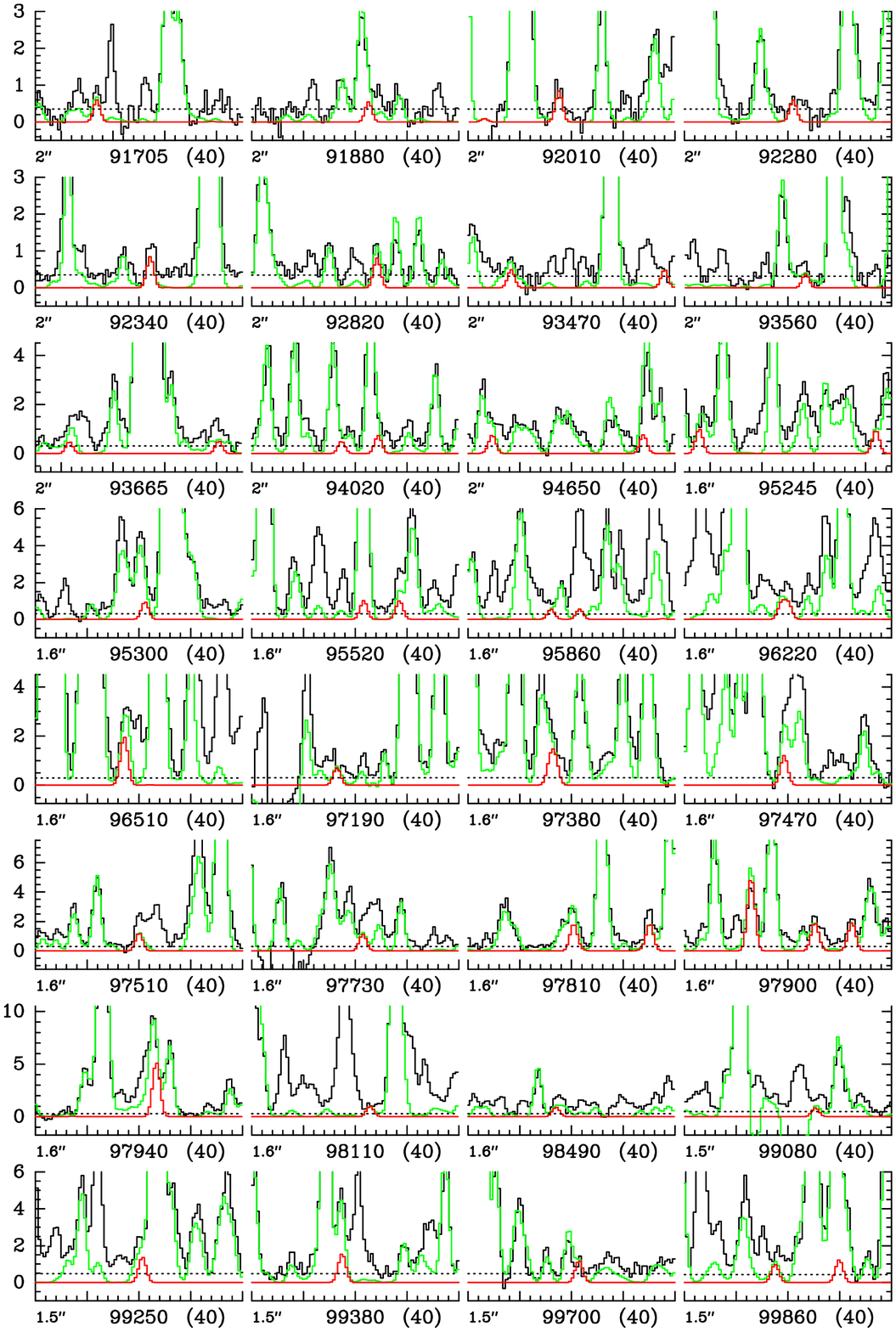}}}
\caption{continued.}
\end{figure*}

\clearpage
\begin{figure*}
\addtocounter{figure}{-1}
\centerline{\resizebox{0.82\hsize}{!}{\includegraphics[angle=0]{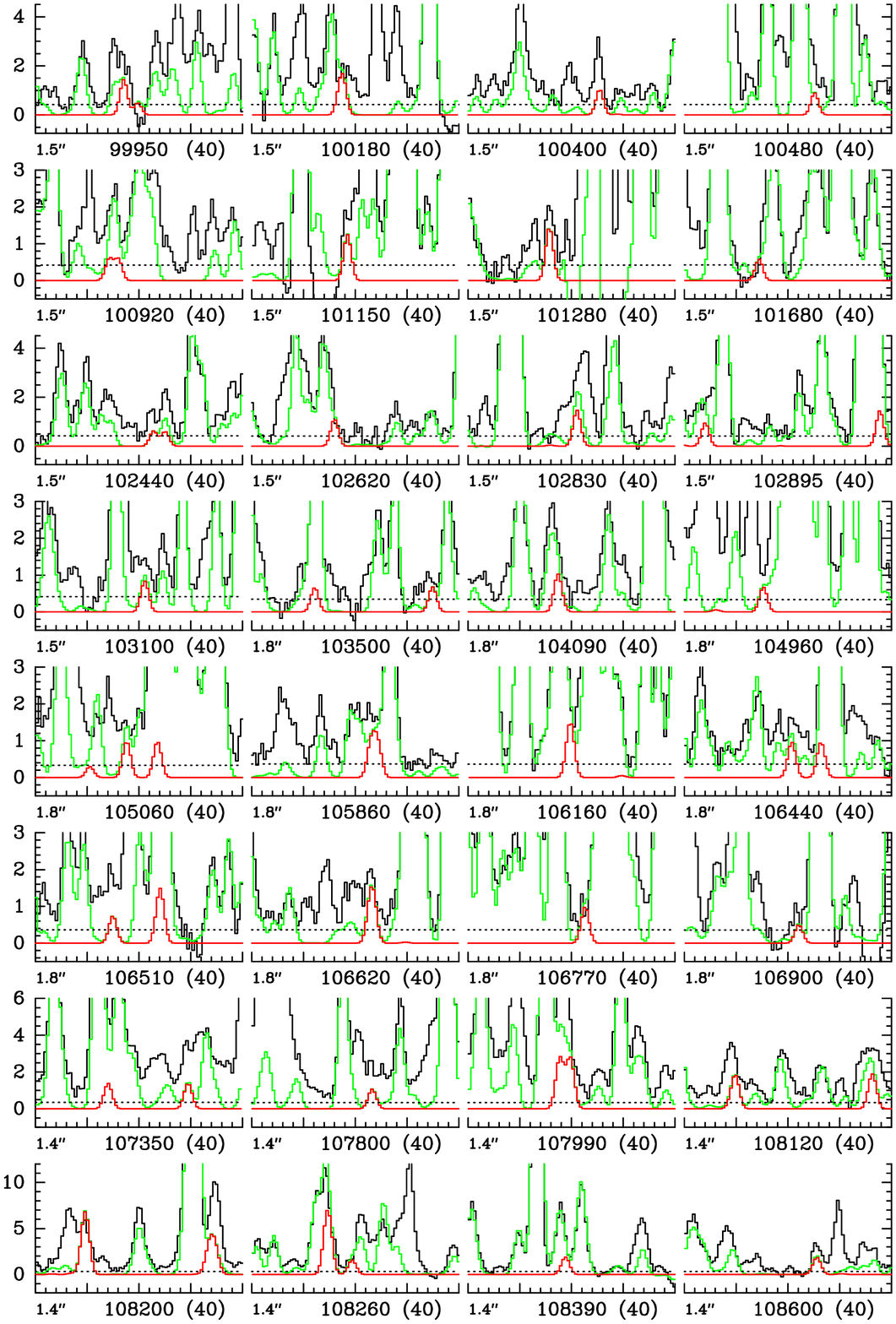}}}
\caption{continued.}
\end{figure*}

\clearpage
\begin{figure*}
\addtocounter{figure}{-1}
\centerline{\resizebox{0.82\hsize}{!}{\includegraphics[angle=0]{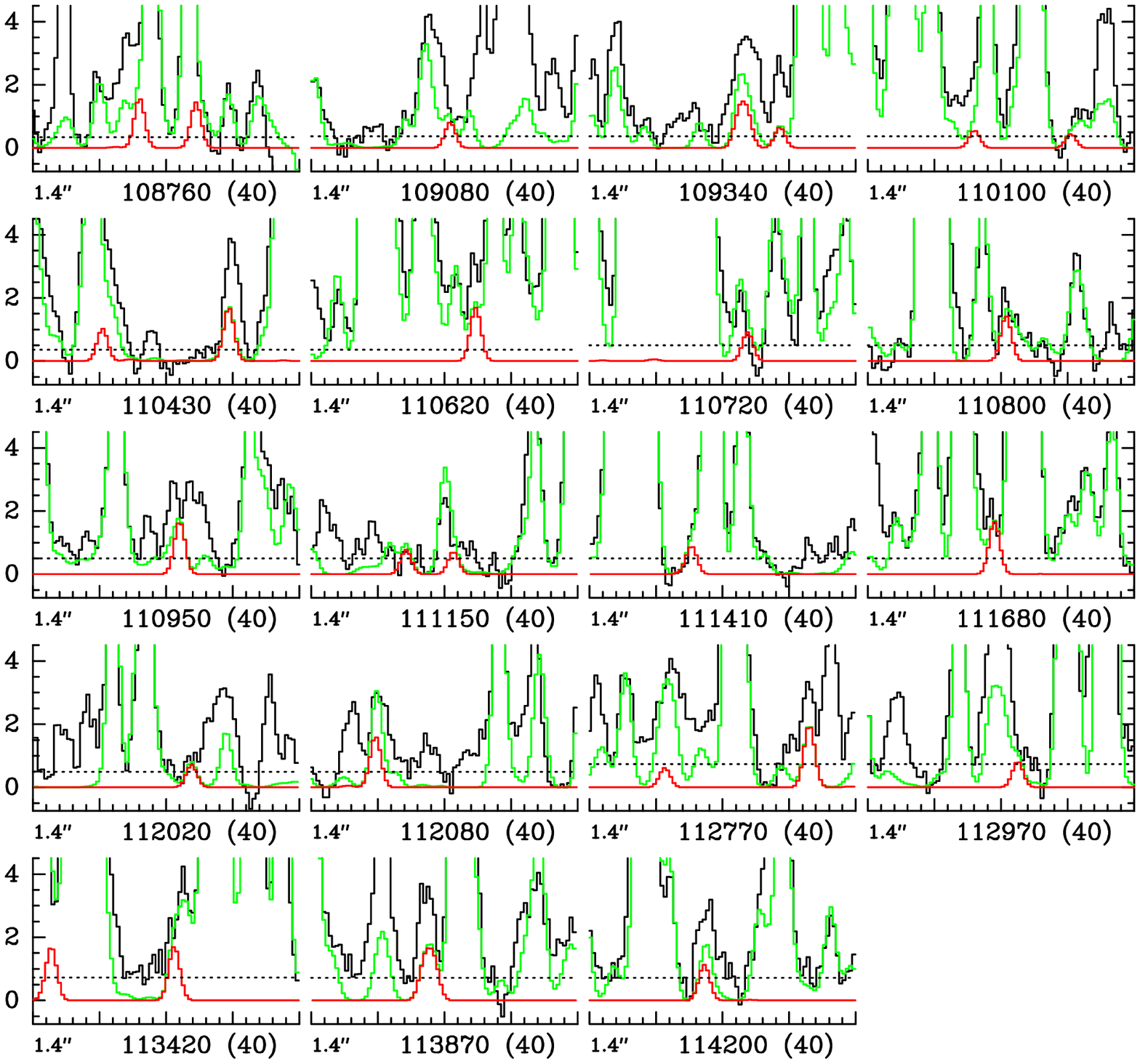}}}
\caption{continued.}
\end{figure*}

\clearpage
\begin{figure*}
\centerline{\resizebox{0.82\hsize}{!}{\includegraphics[angle=0]{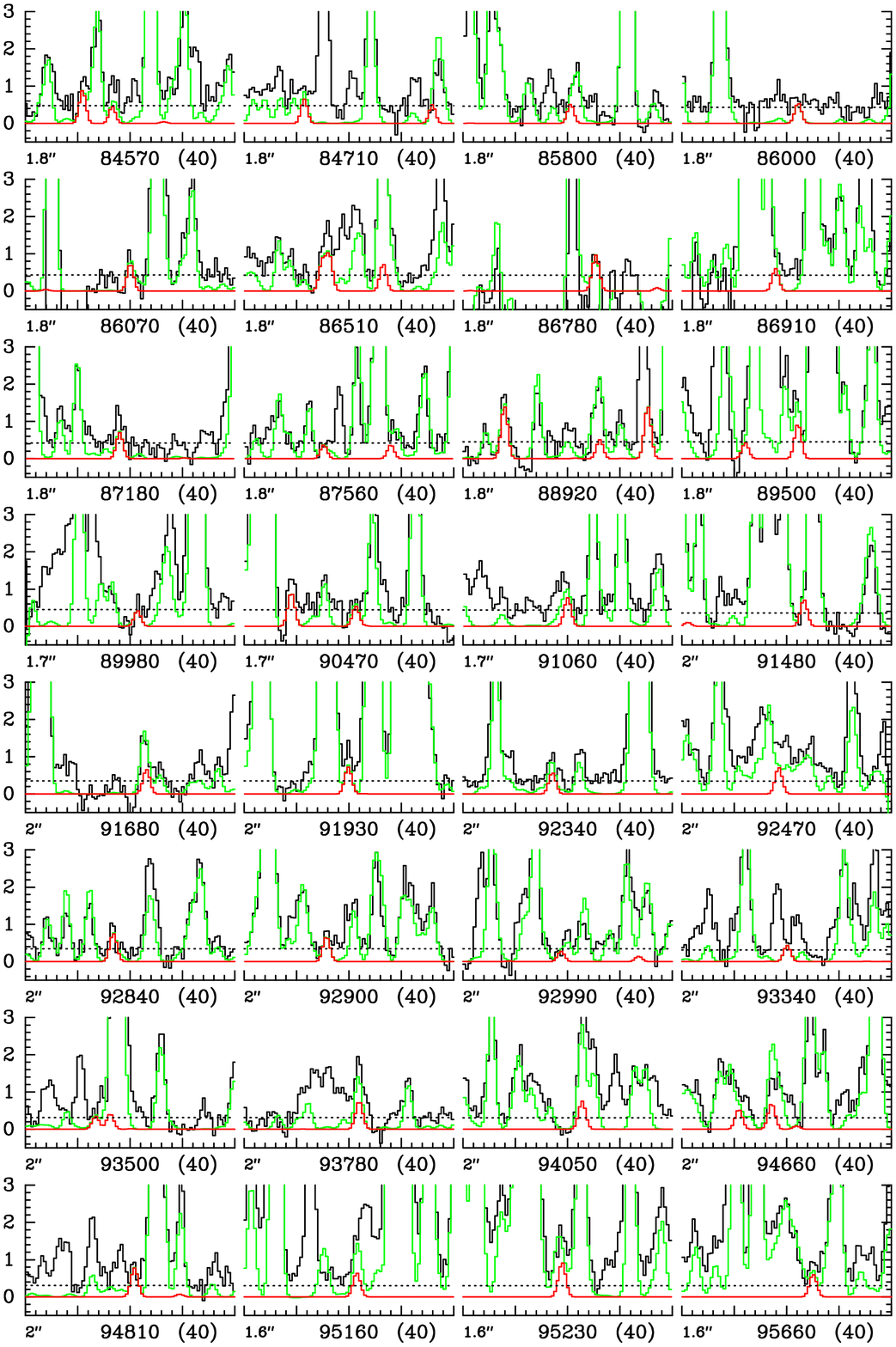}}}
\caption{Same as Fig.~\ref{f:spec_ch3nhcho_ve0} for CH$_3$CONH$_2$, 
$\varv_{\rm t} = 1$.
}
\label{f:spec_ch3conh2_ve1}
\end{figure*}

\clearpage
\begin{figure*}
\addtocounter{figure}{-1}
\centerline{\resizebox{0.82\hsize}{!}{\includegraphics[angle=0]{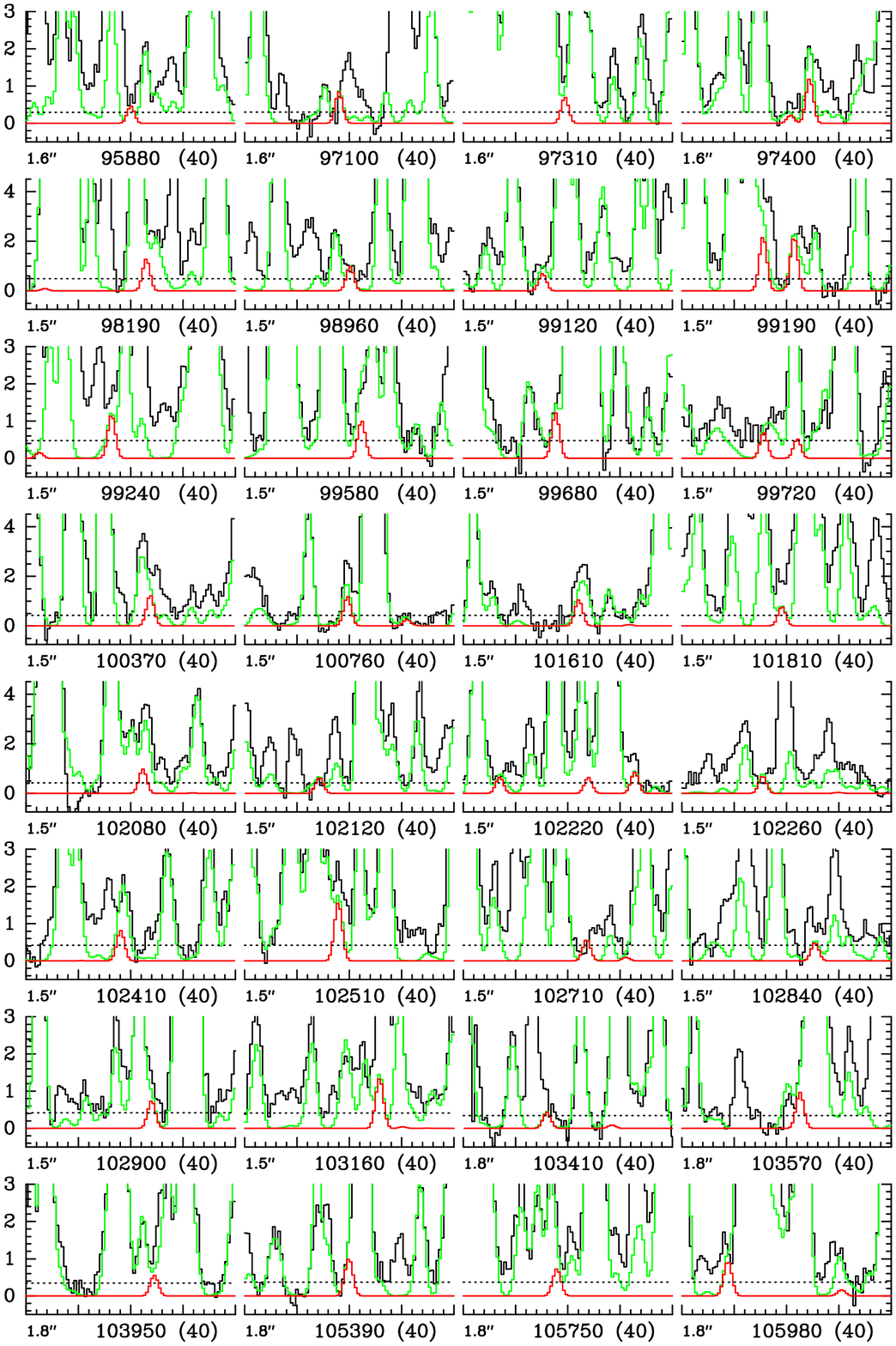}}}
\caption{continued.}
\end{figure*}

\clearpage
\begin{figure*}
\addtocounter{figure}{-1}
\centerline{\resizebox{0.82\hsize}{!}{\includegraphics[angle=0]{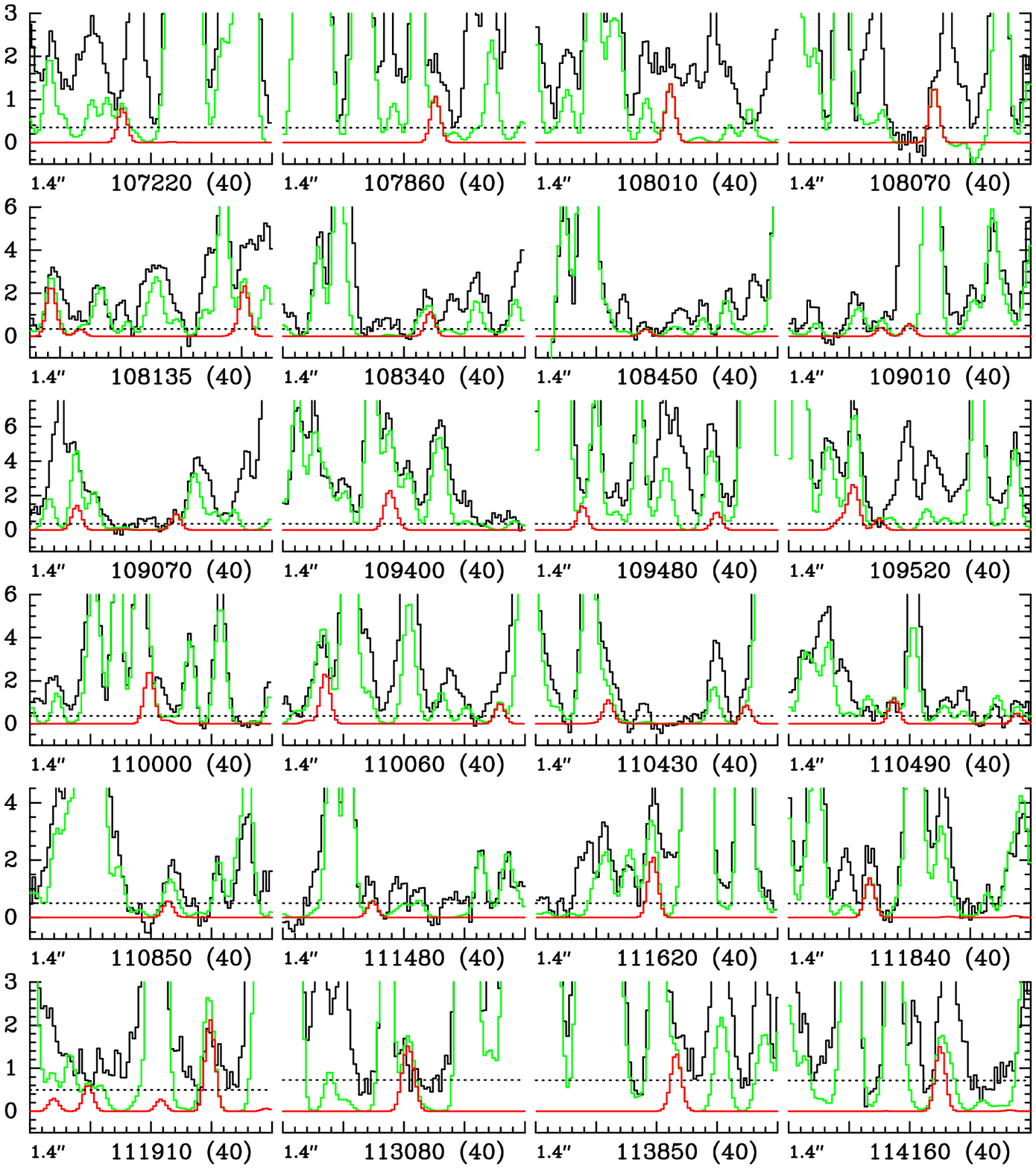}}}
\caption{continued.}
\end{figure*}

\clearpage
\begin{figure*}
\centerline{\resizebox{0.82\hsize}{!}{\includegraphics[angle=0]{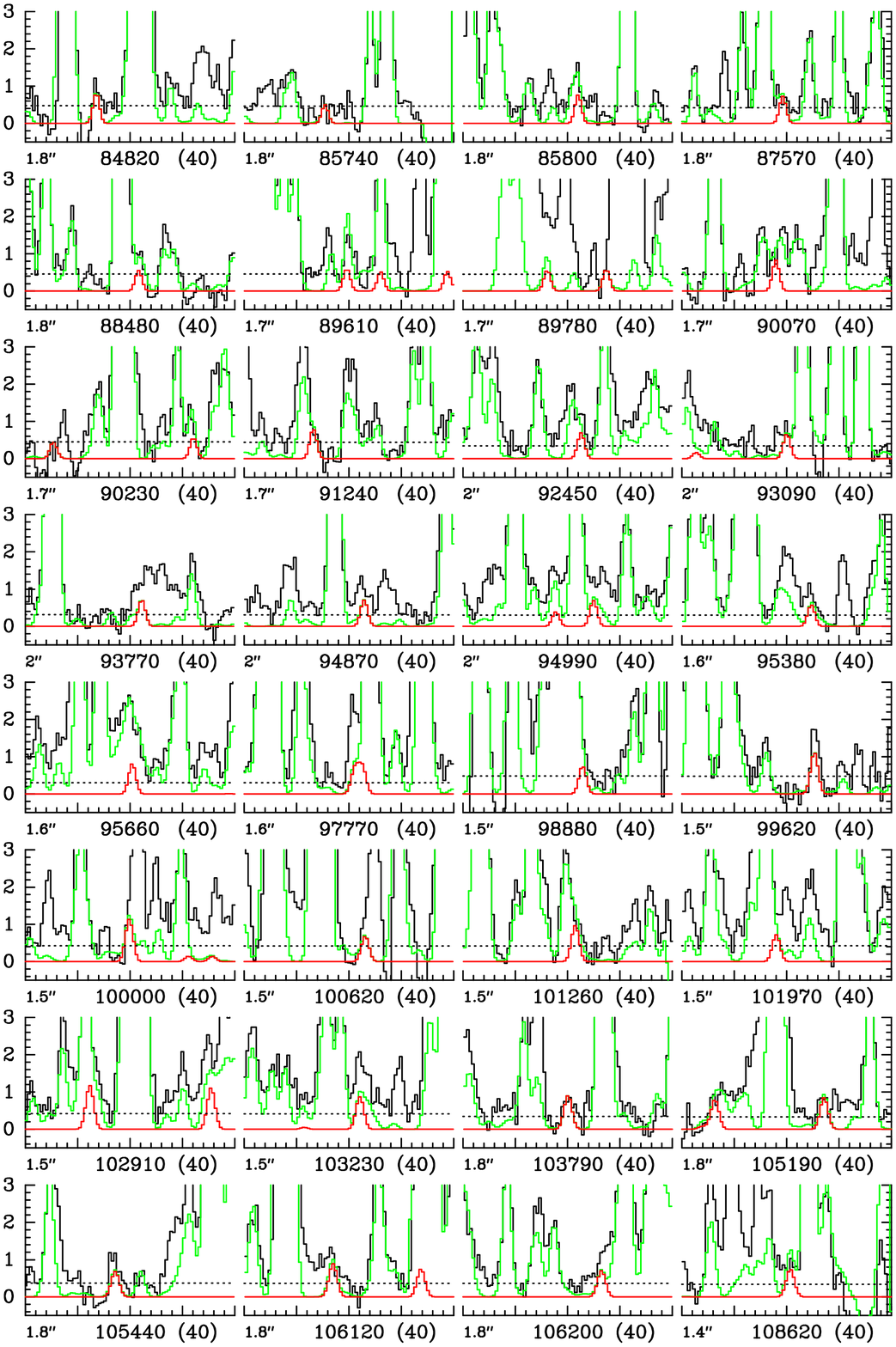}}}
\caption{Same as Fig.~\ref{f:spec_ch3nhcho_ve0} for CH$_3$CONH$_2$, 
$\varv_{\rm t} = 2$.
}
\label{f:spec_ch3conh2_ve2}
\end{figure*}

\clearpage
\begin{figure*}
\addtocounter{figure}{-1}
\centerline{\resizebox{0.82\hsize}{!}{\includegraphics[angle=0]{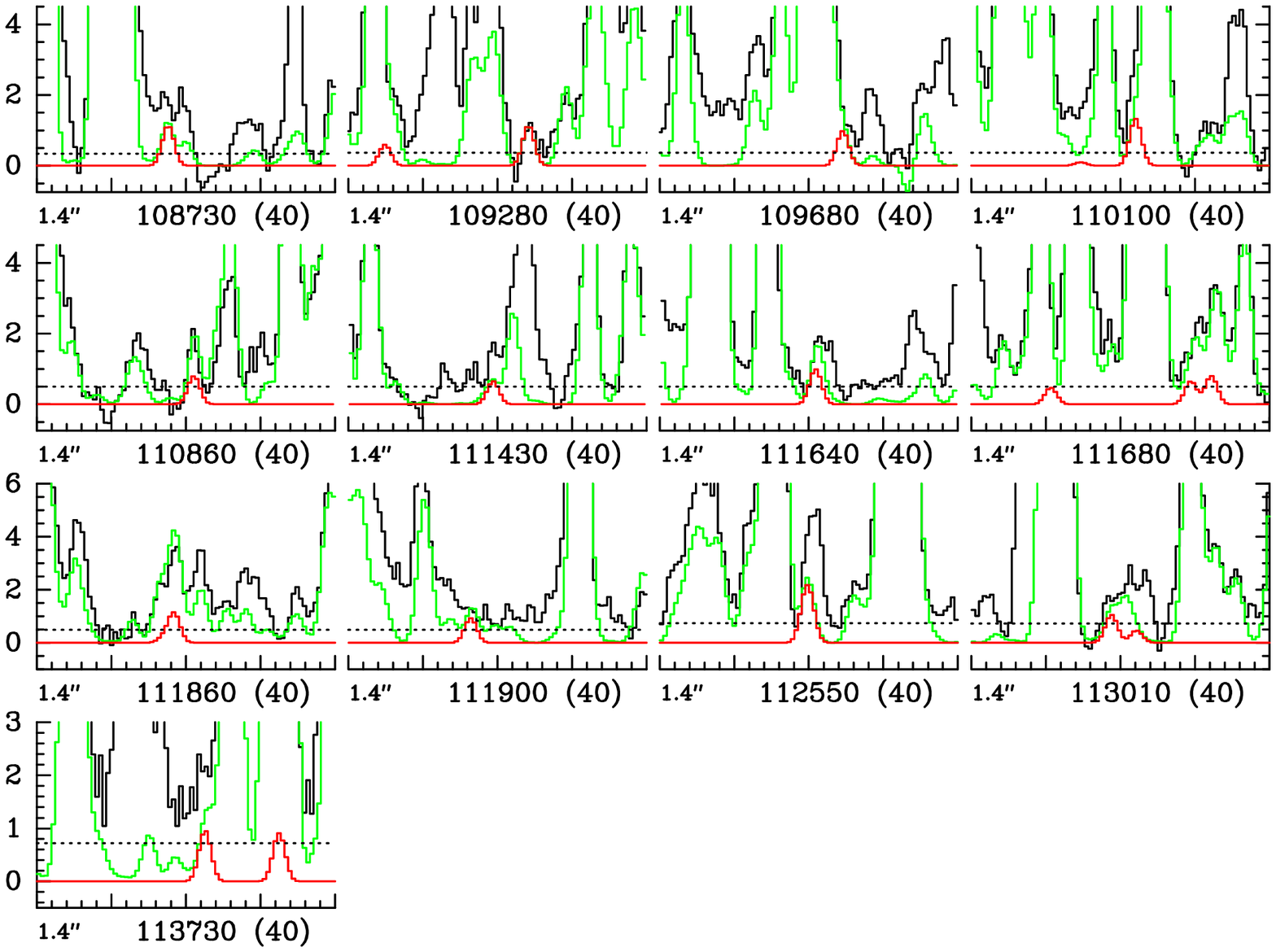}}}
\caption{continued.}
\end{figure*}

\clearpage
\begin{figure*}
\centerline{\resizebox{0.82\hsize}{!}{\includegraphics[angle=0]{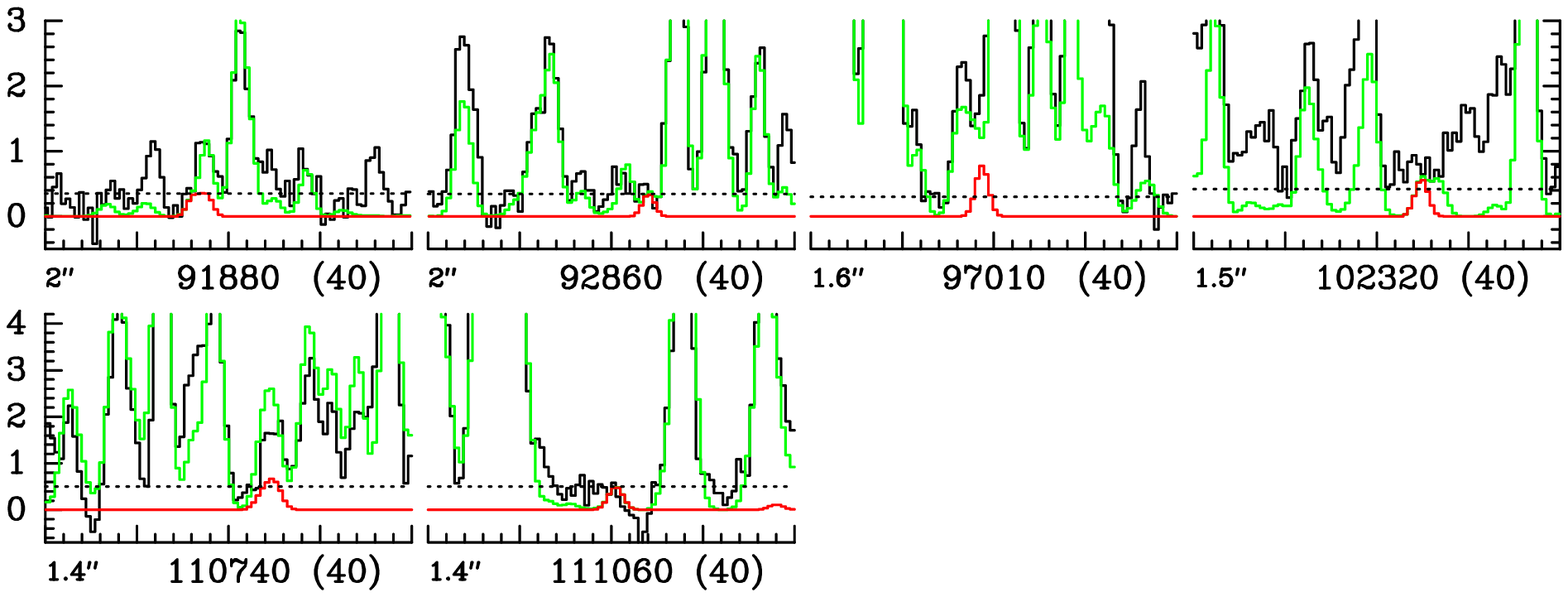}}}
\caption{Same as Fig.~\ref{f:spec_ch3nhcho_ve0} for CH$_3$CONH$_2$,
$\Delta\varv_{\rm t} \neq 0$.
}
\label{f:spec_ch3conh2_cv}
\end{figure*}

\begin{figure}
\centerline{\resizebox{0.50\hsize}{!}{\includegraphics[angle=0]{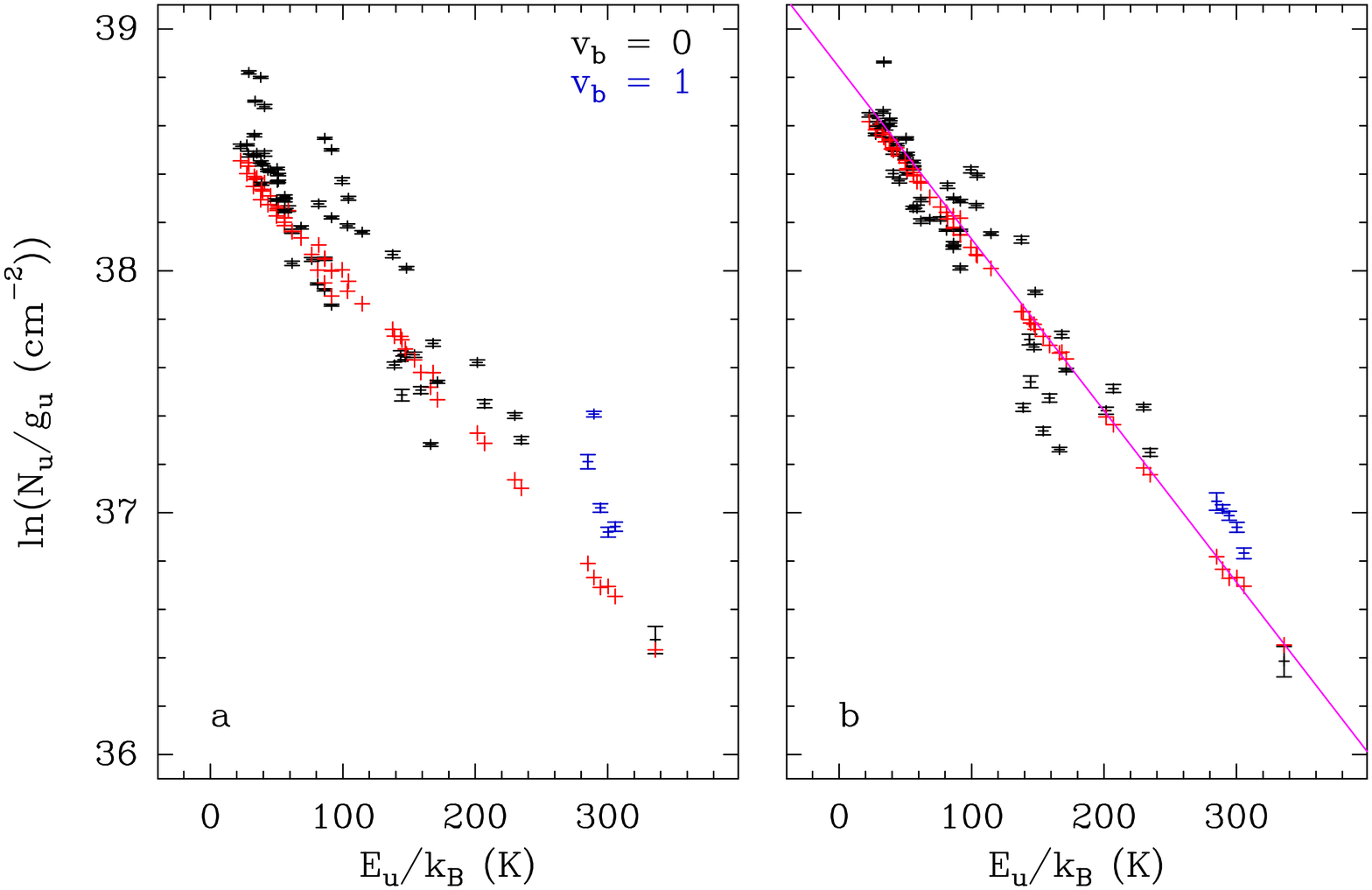}}}
\caption{Same as Fig.~\ref{f:popdiag_ch3nhcho} for CH$_3$NCO.
}
\label{f:popdiag_ch3nco}
\end{figure}

\begin{figure}
\centerline{\resizebox{0.50\hsize}{!}{\includegraphics[angle=0]{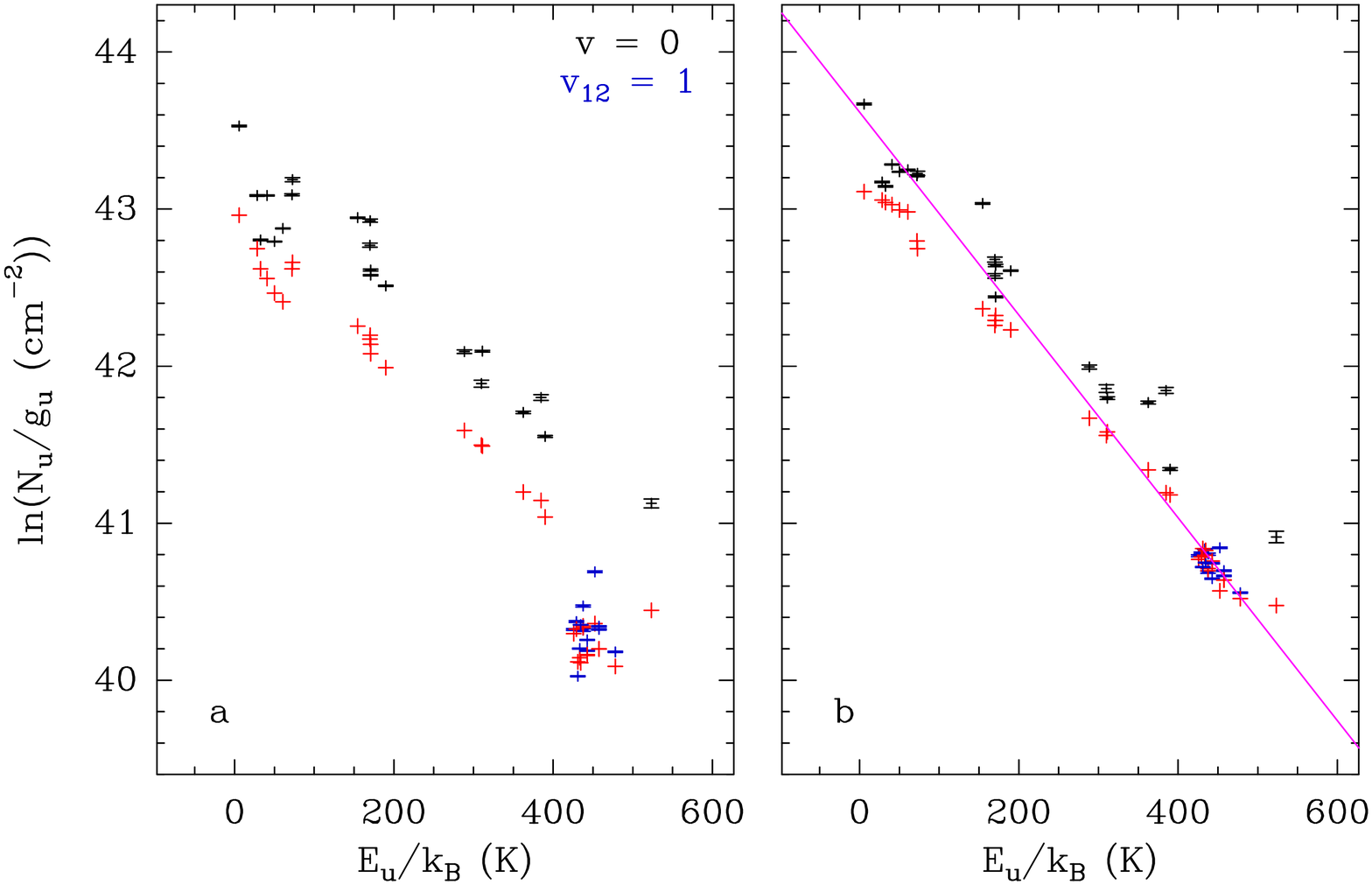}}}
\caption{Same as Fig.~\ref{f:popdiag_ch3nhcho} for NH$_2$CHO.
}
\label{f:popdiag_nh2cho}
\end{figure}

\begin{figure}
\centerline{\resizebox{0.50\hsize}{!}{\includegraphics[angle=0]{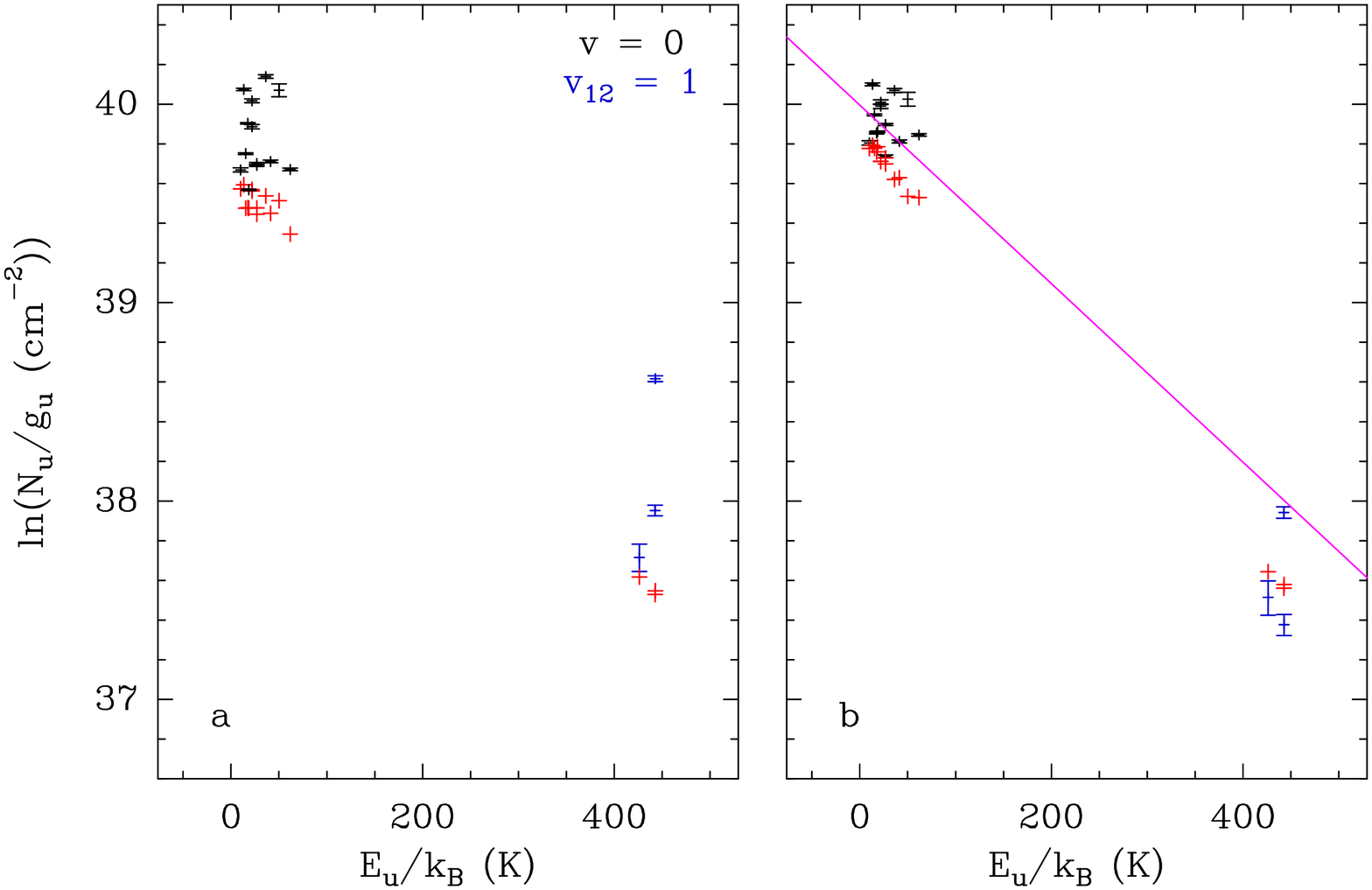}}}
\caption{Same as Fig.~\ref{f:popdiag_ch3nhcho} for NH$_2$$^{13}$CHO.
}
\label{f:popdiag_nh2cho_13c}
\end{figure}

\begin{figure}
\centerline{\resizebox{0.50\hsize}{!}{\includegraphics[angle=0]{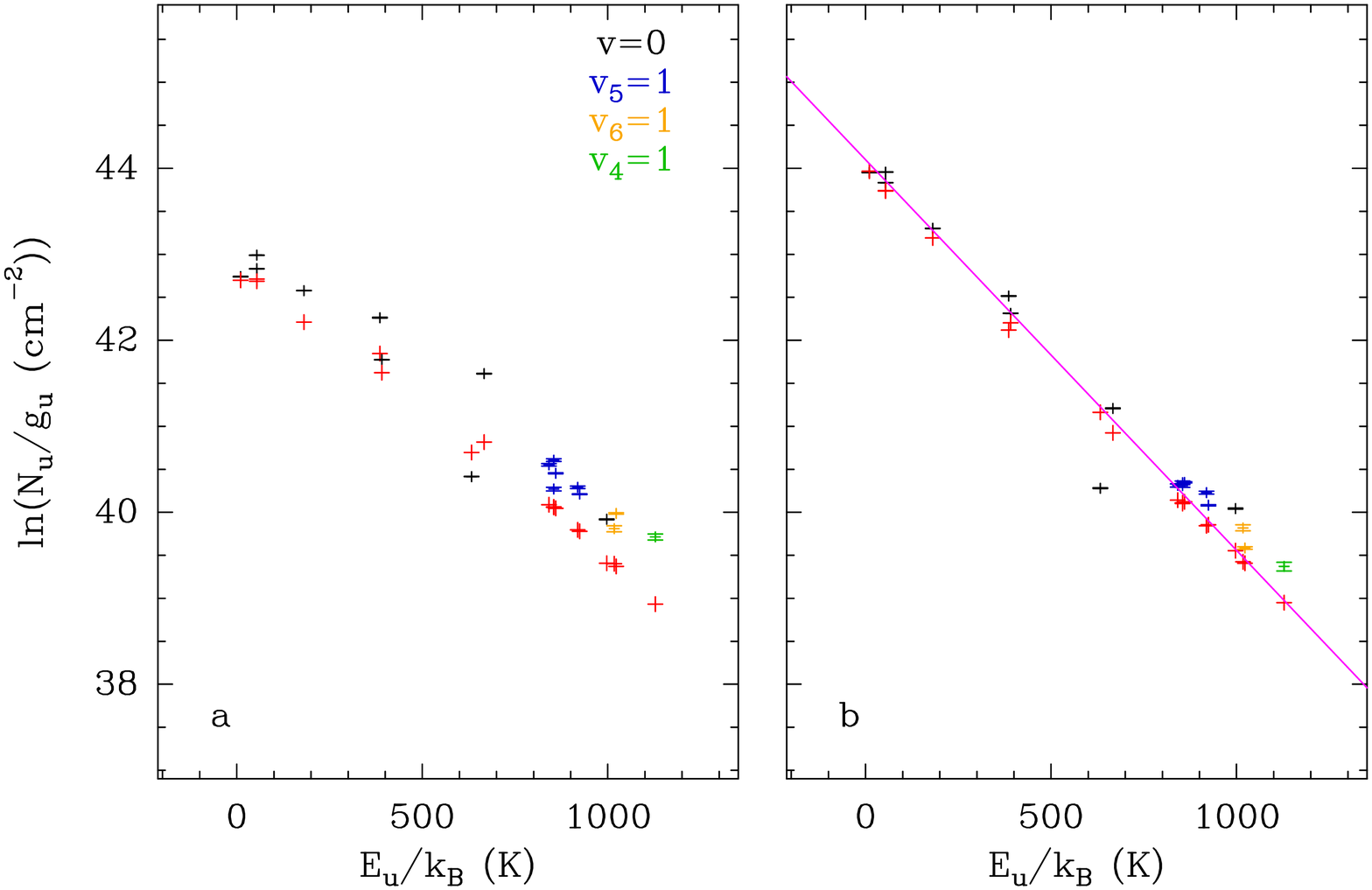}}}
\caption{Same as Fig.~\ref{f:popdiag_ch3nhcho} for HNCO.
}
\label{f:popdiag_hnco}
\end{figure}

\begin{figure}
\centerline{\resizebox{0.50\hsize}{!}{\includegraphics[angle=0]{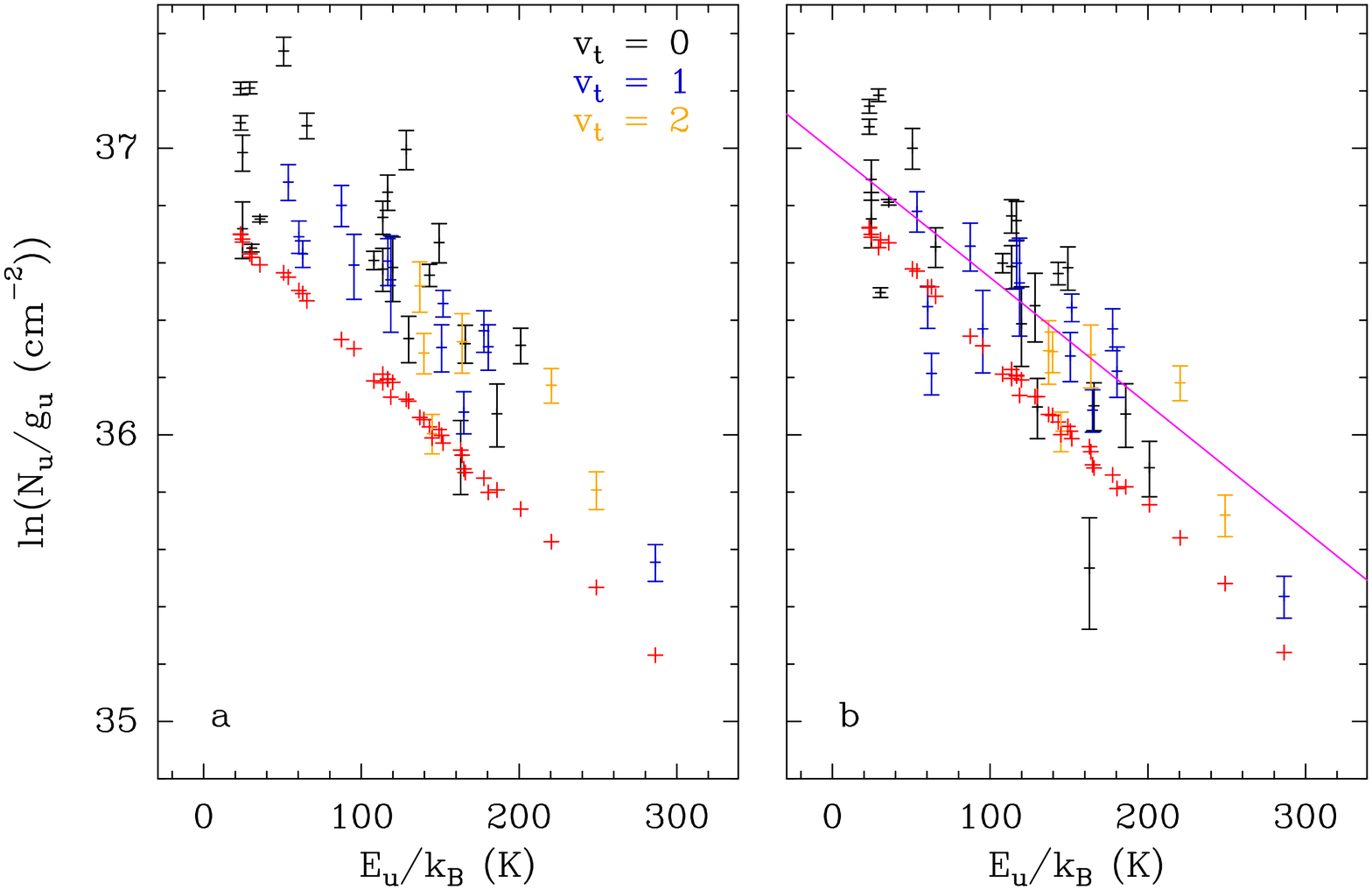}}}
\caption{Same as Fig.~\ref{f:popdiag_ch3nhcho} for CH$_3$CONH$_2$.
}
\label{f:popdiag_ch3conh2}
\end{figure}

\clearpage

\section{Complementary tables}

Table~\ref{t:Scompar} provides the direct comparison between low-order 
spectral parameters determined in this study and the corresponding parameters 
from \citet{kawashima2010dynamical}.

Tables~\ref{t:list_ch3nhcho_ve0}--\ref{t:list_ch3nhcho_ve2} list the 
transitions that are used to build the population diagram of CH$_3$NHCHO 
(Fig.~\ref{f:popdiag_ch3nhcho}). The transitions that we consider as clearly
detected individually, i.e. that are above the $3\sigma$ level and do 
not suffer too much from contamination by other species, are marked with a star.
Because there are only five such transitions, the assignment of the
corresponding five detected lines to CH$_3$NHCHO is only tentative (see 
Sect.~\ref{s:obs_ch3nhcho}).

\begin{table*}
\centering
\caption{Comparison of the main spectroscopic parameters with previous results.}
\label{t:Scompar}	
\begin{tabular}{ccc}
\hline 
\hline 
 Parameter                  & This study     & \citet{kawashima2010dynamical} \\
\hline 
$V_3$ (cm$^{-1}$)	        & 51.7199088(90) & 	53.91824(11) \\
$I_\alpha$ (u$\text{\AA}^2$)	                & 3.1904202\tablefootmark{a}	 & 	3.1542735    \\
$\lambda_a$		            & 0.91962767	 & 	0.90814582   \\
$\rho_a$		                & 0.0744683		 & 0.0768384     \\
$A$ (MHz)		            & 19246.36(12)	 & 	19455.9905   \\
$B$ (MHz)		            & 6376.416(72)	 & 	6297.7305    \\
$C$ (MHz)		            & 4933.8717(42)	 & 4905.1546\tablefootmark{b}    \\
$\chi_{aa}$ (MHz)	        & 	2.1013(27)	 & 	2.115 (22)   \\
$\chi_{bb}-\chi_{cc}$ (MHz)	& 	5.9666(61)	 & 	5.952 (28)   \\
$\chi_{ab}$ (MHz)	        & 	0.2630(23)	 & 	0.265 (17)   \\ 
\hline
\end{tabular} 
\tablefoot{
\tablefoottext{a}{Recalculated from $\rho$ value.}
\tablefoottext{b}{Average of $A$- and $E$-state values from Table 2 of \citet{kawashima2010dynamical}.}
}
\end{table*}

\begin{table*}
 \begin{center}
 \caption{
 Selection of lines of CH$_3$NHCHO $\varv_{\rm t} = 0$ covered by the EMoCA survey of Sgr B2(N2).
}
 \label{t:list_ch3nhcho_ve0}
 \vspace*{-2.0ex}
 \begin{tabular}{lrcccrccrrcrr}
 \hline\hline
 \multicolumn{1}{c}{Transition\tablefootmark{a}} & \multicolumn{1}{c}{Frequency} & \multicolumn{1}{c}{Unc.\tablefootmark{b}} & \multicolumn{1}{c}{$E_{\rm up}$\tablefootmark{c}} & \multicolumn{1}{c}{$g_{\rm up}$\tablefootmark{d}} & \multicolumn{1}{c}{$A_{\rm ul}$\tablefootmark{e}} & \multicolumn{1}{c}{$\sigma$\tablefootmark{f}} & \multicolumn{1}{c}{$\tau_{\rm peak}$\tablefootmark{g}} & \multicolumn{2}{c}{Frequency range\tablefootmark{h}} & \multicolumn{1}{c}{$I_{\rm obs}$\tablefootmark{i}} & \multicolumn{1}{c}{$I_{\rm mod}$\tablefootmark{j}} & \multicolumn{1}{c}{$I_{\rm all}$\tablefootmark{k}} \\ 
  & \multicolumn{1}{c}{\scriptsize (MHz)} & \multicolumn{1}{c}{\scriptsize (kHz)} &  \multicolumn{1}{c}{\scriptsize (K)} & & \multicolumn{1}{c}{\scriptsize ($10^{-5}$ s$^{-1}$)} & \multicolumn{1}{c}{\scriptsize (mK)} & & \multicolumn{1}{c}{\scriptsize (MHz)} & \multicolumn{1}{c}{\scriptsize (MHz)} & \multicolumn{1}{c}{\scriptsize (K km s$^{-1}$)} & \multicolumn{2}{c}{\scriptsize (K km s$^{-1}$)} \\ 
 \hline
8$_{2,6}$ -- 7$_{2,5}$ &   91888.760 &   1 &   27 & 17 &  3.1 &  117 &  0.024 &  91887.2 &  91890.1 &   6.2(5)$^\star$ &   3.6 &   4.0 \\ 
9$_{0,9}$ -- 8$_{0,8}$ &   93406.380 &   1 &   23 & 19 &  3.6 &  104 &  0.032 &  93404.9 &  93407.8 &   7.7(4)$^\star$ &   4.5 &   4.8 \\ 
9$_{2,8}$ -- 8$_{2,7}$ &   99695.472 &   1 &   27 & 19 &  4.2 &  162 &  0.033 &  99693.6 &  99697.0 &  10.5(6)$^\star$ &   7.3 &   7.5 \\ 
9$_{1,8}$ -- 8$_{1,7}$ &  101309.201 &   1 &   30 & 19 &  4.7 &  133 &  0.034 & 101307.6 & 101310.6 &  11.1(5) &   7.5 &   7.7 \\ 
10$_{6,4}$ -- 9$_{6,3}$ &  111215.743 &   1 &   59 & 21 &  4.1 &  166 &  0.024 & 111213.9 & 111217.3 &   7.9(6)$^\star$ &   5.8 &   6.2 \\ 
11$_{0,11}$ -- 1$_{0,1}$ &  112875.483 &   1 &   37 & 23 &  6.6 &  242 &  0.044 & 112873.3 & 112877.3 &  22.6(9) &  11.7 &  12.8 \\ 
 \hline
 \end{tabular}
 \end{center}
 \vspace*{-2.5ex}
 \tablefoot{
 \tablefoottext{a}{Quantum numbers of the upper and lower levels.}
 \tablefoottext{b}{Frequency uncertainty.}
 \tablefoottext{c}{Upper level energy.}
 \tablefoottext{d}{Upper level degeneracy.}
 \tablefoottext{e}{Einstein coefficient for spontaneous emission.}
 \tablefoottext{f}{Measured rms noise level.}
 \tablefoottext{g}{Peak opacity of the synthetic line.}
 \tablefoottext{h}{Frequency range over which the emission was integrated.}
 \tablefoottext{i}{Integrated intensity of the observed spectrum in brightness temperature scale. The statistical standard deviation is given in parentheses in unit of the last digit. Values marked with a star indicate the lines that suffer little contamination and are thus unambiguously detected.}
 \tablefoottext{j}{Integrated intensity of the synthetic spectrum of CH$_3$NHCHO $\varv_{\rm t} = 0$.}
 \tablefoottext{k}{Integrated intensity of the model that contains the contribution of all identified molecules, including CH$_3$NHCHO $\varv_{\rm t} = 0$.}
 }
 \end{table*}

\begin{table*}
 \begin{center}
 \caption{
 Selection of lines of CH$_3$NHCHO $\varv_{\rm t} = 1$ covered by the EMoCA survey of Sgr B2(N2).
}
 \label{t:list_ch3nhcho_ve1}
 \vspace*{-2.0ex}
 \begin{tabular}{lrcccrccrrcrr}
 \hline\hline
 \multicolumn{1}{c}{Transition\tablefootmark{a}} & \multicolumn{1}{c}{Frequency} & \multicolumn{1}{c}{Unc.\tablefootmark{b}} & \multicolumn{1}{c}{$E_{\rm up}$\tablefootmark{c}} & \multicolumn{1}{c}{$g_{\rm up}$\tablefootmark{d}} & \multicolumn{1}{c}{$A_{\rm ul}$\tablefootmark{e}} & \multicolumn{1}{c}{$\sigma$\tablefootmark{f}} & \multicolumn{1}{c}{$\tau_{\rm peak}$\tablefootmark{g}} & \multicolumn{2}{c}{Frequency range\tablefootmark{h}} & \multicolumn{1}{c}{$I_{\rm obs}$\tablefootmark{i}} & \multicolumn{1}{c}{$I_{\rm mod}$\tablefootmark{j}} & \multicolumn{1}{c}{$I_{\rm all}$\tablefootmark{k}} \\ 
  & \multicolumn{1}{c}{\scriptsize (MHz)} & \multicolumn{1}{c}{\scriptsize (kHz)} &  \multicolumn{1}{c}{\scriptsize (K)} & & \multicolumn{1}{c}{\scriptsize ($10^{-5}$ s$^{-1}$)} & \multicolumn{1}{c}{\scriptsize (mK)} & & \multicolumn{1}{c}{\scriptsize (MHz)} & \multicolumn{1}{c}{\scriptsize (MHz)} & \multicolumn{1}{c}{\scriptsize (K km s$^{-1}$)} & \multicolumn{2}{c}{\scriptsize (K km s$^{-1}$)} \\ 
 \hline
9$_{1,9}$ -- 8$_{1,8}$ &   91245.678 &   1 &   71 & 19 &  3.2 &  149 &  0.023 &  91244.4 &  91247.3 &   7.6(6) &   4.6 &   5.1 \\ 
10$_{5,6}$ -- 9$_{5,5}$ &   99534.218 &   1 &  114 & 21 &  3.9 &  162 &  0.021 &  99532.4 &  99535.3 &   7.9(6) &   4.6 &   6.8 \\ 
11$_{1,11}$ -- 1$_{0,11}$ &  108750.182 &   1 &  106 & 23 &  7.2 &  115 &  0.036 & 108748.1 & 108751.5 &  15.6(4) &   9.2 &  11.9 \\ 
11$_{1,11}$ -- 1$_{0,11}$ &  111776.580 &   1 &   81 & 23 &  5.9 &  166 &  0.033 & 111774.6 & 111778.0 &  14.5(6) &   8.0 &   9.2 \\ 
10$_{3,8}$ -- 9$_{3,7}$ &  113609.358 &   1 &   86 & 21 &  5.3 &  242 &  0.025 & 113607.5 & 113610.9 &   8.7(8)$^\star$ &   6.4 &   7.1 \\ 
 \hline
 \end{tabular}
 \end{center}
 \vspace*{-2.5ex}
 \tablefoot{
 Same as Table~\ref{t:list_ch3nhcho_ve0} but for CH$_3$NHCHO $\varv_{\rm t} = 1$.
 }
 \end{table*}

\begin{table*}
 \begin{center}
 \caption{
 Selection of lines of CH$_3$NHCHO $\varv_{\rm t} = 2$ covered by the EMoCA survey of Sgr B2(N2).
}
 \label{t:list_ch3nhcho_ve2}
 \vspace*{-2.0ex}
 \begin{tabular}{lrcccrccrrcrr}
 \hline\hline
 \multicolumn{1}{c}{Transition\tablefootmark{a}} & \multicolumn{1}{c}{Frequency} & \multicolumn{1}{c}{Unc.\tablefootmark{b}} & \multicolumn{1}{c}{$E_{\rm up}$\tablefootmark{c}} & \multicolumn{1}{c}{$g_{\rm up}$\tablefootmark{d}} & \multicolumn{1}{c}{$A_{\rm ul}$\tablefootmark{e}} & \multicolumn{1}{c}{$\sigma$\tablefootmark{f}} & \multicolumn{1}{c}{$\tau_{\rm peak}$\tablefootmark{g}} & \multicolumn{2}{c}{Frequency range\tablefootmark{h}} & \multicolumn{1}{c}{$I_{\rm obs}$\tablefootmark{i}} & \multicolumn{1}{c}{$I_{\rm mod}$\tablefootmark{j}} & \multicolumn{1}{c}{$I_{\rm all}$\tablefootmark{k}} \\ 
  & \multicolumn{1}{c}{\scriptsize (MHz)} & \multicolumn{1}{c}{\scriptsize (kHz)} &  \multicolumn{1}{c}{\scriptsize (K)} & & \multicolumn{1}{c}{\scriptsize ($10^{-5}$ s$^{-1}$)} & \multicolumn{1}{c}{\scriptsize (mK)} & & \multicolumn{1}{c}{\scriptsize (MHz)} & \multicolumn{1}{c}{\scriptsize (MHz)} & \multicolumn{1}{c}{\scriptsize (K km s$^{-1}$)} & \multicolumn{2}{c}{\scriptsize (K km s$^{-1}$)} \\ 
 \hline
10$_{0,10}$ -- 9$_{0,9}$ &   85854.246 &   3 &  147 & 21 &  4.0 &  158 &  0.023 &  85853.0 &  85855.4 &   6.9(6) &   4.1 &   4.7 \\ 
 \hline
 \end{tabular}
 \end{center}
 \vspace*{-2.5ex}
 \tablefoot{
 Same as Table~\ref{t:list_ch3nhcho_ve0} but for CH$_3$NHCHO $\varv_{\rm t} = 2$.
 }
 \end{table*}

\end{appendix}


\begin{thebibliography}{}

\bibitem[{Alekseev {et~al.}(2012)Alekseev, Motiyenko, \&
  Margul{\`e}s}]{alekseev2012millimeter}
Alekseev, E., Motiyenko, R., \& Margul{\`e}s, L. 2012, Radio Physics and Radio
  Astronomy, 3

\bibitem[Barone et al.(2015)]{Barone15} Barone, V., Latouche, C., Skouteris, 
D., et al.\ 2015, \mnras, 453, L31

\bibitem[Belloche et al.(2013)]{Belloche13} Belloche, A., M{\"u}ller, 
H.~S.~P., Menten, K.~M., Schilke, P., \& Comito, C.\ 2013, \aap, 559, A47 

\bibitem[Belloche et al.(2014)]{Belloche14} Belloche, A., Garrod, R.~T., 
M{\"u}ller, H.~S.~P., \& Menten, K.~M.\ 2014, Science, 345, 1584

\bibitem[Belloche et al.(2016)]{Belloche16} 
Belloche, A., M{\"u}ller, H.~S.~P., Garrod, R.~T., \& Menten, K.~M. 
2016, \aap, 587, A91

\bibitem[Cernicharo et al.(2016)]{Cernicharo16} Cernicharo, J., Kisiel, Z., 
Tercero, B., et al.\ 2016, \aap, 587, L4 

\bibitem[{Fantoni \& Caminati(1996)}]{fantoni1996rotational}
Fantoni, A. \& Caminati, W. 1996, J. Chem. Soc., Faraday Trans., 92, 343

\bibitem[{Fantoni {et~al.}(2002)Fantoni, Caminati, Hartwig, \&
  Stahl}]{fantoni2002very}
Fantoni, A.~C., Caminati, W., Hartwig, H., \& Stahl, W. 2002, J. Mol. Struc.,
  612, 305

\bibitem[Garrod et al.(2008)]{Garrod08} Garrod, R.~T., Widicus Weaver, S.~L., 
\& Herbst, E.\ 2008, \apj, 682, 283-302

\bibitem[Garrod(2013)]{Garrod13} Garrod, R.~T.\ 2013, \apj, 765, 60

\bibitem[Goesmann et al.(2015)]{Goesmann15} Goesmann, F., Rosenbauer, H., 
Bredeh{\"o}ft, J.~H., et al.\ 2015, Science, 349,  aab0689

\bibitem[Gordy \& Cook(1984)]{gordycook84} Gordy, W., Cook, R.~L. 
1984, Microwave Molecular Spectra, Techniques of Chemistry, Vol. XVIII (New York: Wiley)

\bibitem[Halfen et al.(2011)]{Halfen11} Halfen, D.~T., Ilyushin, V., \& 
Ziurys, L.~M.\ 2011, \apj, 743, 60 

\bibitem[Halfen et al.(2015)]{Halfen15} Halfen, D.~T., Ilyushin, V.~V., \& 
Ziurys, L.~M.\ 2015, \apjl, 812, L5 

\bibitem[Hirota {et~al.}(2010)Hirota, Kawashima, Usami \& Seto]{hirota2010acetamide}
  Hirota, E., Kawashima, Y., Usami, T., Seto, K. 2010, J. Mol. Spec., 260, 30

\bibitem[Hocking et al.(1975)]{Hocking75} Hocking, W.~H., Gerry, M.~C.~L., \& 
Winnewisser, G.\ 1975, Can. J. Phys., 53, 1869 

\bibitem[{Hollis {et~al.}(2006)Hollis, Lovas, Remijan, Jewell, Ilyushin, \&
  Kleiner}]{hollis2006detection}
Hollis, J., Lovas, F.~J., Remijan, A.~J., {et~al.} 2006, \apjl, 643, L25

\bibitem[Hougen {et~al.}(1994)Hougen, Kleiner, \& Godefroid]{hougen1994ram}
  Hougen, J.~T., Kleiner, I., Godefroid, M. 1994, J. Mol. Spec., 163, 559

\bibitem[{Ilyushin {et~al.}(2004)Ilyushin, Alekseev, Dyubko, Kleiner, \&
  Hougen}]{ilyushin2004acetamide}
 Ilyushin, V.~V., Alekseev, E.~A., Dyubko, S.~F., Kleiner, I., Hougen, J.~T.
  2004, J. Mol. Spec., 227, 115

\bibitem[{Ilyushin {et~al.}(2013)Ilyushin, Endres, Lewen, Schlemmer, \&
  Drouin}]{ilyushin2013submillimeter}
Ilyushin, V.~V., Endres, C.~P., Lewen, F., Schlemmer, S., \& Drouin, B.~J.
  2013, J. Mol. Spec., 290, 31

\bibitem[{Ilyushin {et~al.}(2010)Ilyushin, Kisiel, Pszcz{\'o}lkowski,
  M{\"a}der, \& Hougen}]{ilyushin2010new}
Ilyushin, V.~V., Kisiel, Z., Pszcz{\'o}lkowski, L., M{\"a}der, H., \& Hougen,
  J.~T. 2010, J. Mol. Spec., 259, 26

\bibitem[Jones et al.(2008)]{Jones08} Jones, P.~A., Burton, M.~G., Cunningham, 
M.~R., et al.\ 2008, \mnras, 386, 117 

\bibitem[{Kaiser {et~al.}(2013)Kaiser, Stockton, Kim, Jensen, \&
  Mathies}]{kaiser2013formation}
Kaiser, R., Stockton, A., Kim, Y., Jensen, E., \& Mathies, R. 2013, \apj, 765,
  111

\bibitem[{Kawashima {et~al.}(2010)Kawashima, Usami, Suenram, Golubiatnikov, \&
  Hirota}]{kawashima2010dynamical}
Kawashima, Y., Usami, T., Suenram, R.~D., Golubiatnikov, G.~Y., \& Hirota, E.
  2010, J. Mol. Spec., 263, 11

\bibitem[{Kleiner (2010)}]{kleinerreview2010} 
Kleiner, I.  2010, J. Mol. Spec., 260, 1

\bibitem[Koput(1986)]{Koput86} Koput, J.\ 1986, J. Mol. Spec., 115, 131 

\bibitem[Kryvda et al.(2009)]{Kryvda09} Kryvda, A.~V., Gerasimov, V.~G., 
Dyubko, S.~F., Alekseev, E.~A., \& Motiyenko, R.~A.\ 2009, J. Mol. Spec., 254, 
28 

\bibitem[Kutzelnigg \& Mecke(1962)]{kutz62} Kutzelnigg, W., Mecke, R. 1962,
   Spectrochim. Acta, 18, 549 

\bibitem[Kydd \& Dunham(1980)]{kydd80} Kydd, R.~A., Dunham, A.~R.~C. 1980,
   J. Mol. Struct., 69, 79 

\bibitem[Lapinov et al.(2007)]{Lapinov07} Lapinov, A.~V., Golubiatnikov, 
G.~Y., Markov, V.~N., \& Guarnieri, A.\ 2007, Astron. Lett., 33, 121 

\bibitem[Lattelais et al.(2009)]{Lattelais09} Lattelais, M., Pauzat, F., 
Ellinger, Y., \& Ceccarelli, C.\ 2009, \apjl, 696, L133 

\bibitem[Lattelais et al.(2010)]{Lattelais10}
Lattelais, M., Pauzat, F., Ellinger, Y., \& Ceccarelli, C. 2010, \aap, 519, A30

\bibitem [{Lin \& Swalen (1959)}]{linswalen1959}
Lin, C.~C., \& Swalen, J.~D. 1959, Rev. Mod. Phys., 31, 841

\bibitem[Loison et al.(2016)]{Loison16} Loison, J.-C., Ag{\'u}ndez, M., 
Marcelino, N., et al.\ 2016, \mnras, 456, 4101

\bibitem[Loomis et al.(2015)]{Loomis15} Loomis, R.~A., McGuire, B.~A., 
Shingledecker, C., et al.\ 2015, \apj, 799, 34 

\bibitem[Margul{\`e}s et al.(2016)]{Margules16} Margul{\`e}s, L., 
Belloche, A., M{\"u}ller, H.~S.~P., et al.\ 2016, \aap, 590, A93

\bibitem[Motiyenko et al.(2012)]{motiyenko2012rotational}
Motiyenko, R., Tercero, B., Cernicharo, J., \& Margul{\`e}s, L. 2012, \aap,
  548, A71

\bibitem[M{\"u}ller et al.(2005)]{Mueller05}
M{\"u}ller, H.~S.~P., Schl{\"o}der, F., Stutzki, J., \& Winnewisser, G.
2005, J. Mol. Struct., 742, 215

\bibitem[M{\"u}ller et al.(2016a)]{Mueller16a} M{\"u}ller, H.~S.~P., Belloche, 
A., Xu, L.-H., et al.\ 2016a, \aap, 587, A92

\bibitem[M{\"u}ller et al.(2016b)]{Mueller16b} M{\"u}ller, H.~S.~P., 
Walters, A., Wehres, N., et al.\ 2016b, \aap, 595, A87

\bibitem[Nguyen et al.(1996)]{Nguyen96} Nguyen, M.~T., Sengupta, D., 
Vereecken, L., Peeters, J. \& Vanquickenborne, L.~G.\ 1996, J. Phys. Chem., 
100, 1615

\bibitem[Niedenhoff et al.(1996)]{Niedenhoff96} Niedenhoff, M., Yamada, 
K.~M.~T., \& Winnewisser, G.\ 1996, J. Mol. Spec., 176, 342 

\bibitem[{Rubin {et~al.}(1971)Rubin, Swenson~Jr, Benson, Tigelaar, \&
  Flygare}]{rubin1971microwave}
Rubin, R., Swenson~Jr, G., Benson, R., Tigelaar, H., \& Flygare, W. 1971, \apj,
  169, L39

\bibitem[Song \& K{\"a}stner(2016)]{Song16} Song, L., \& K{\"a}stner, 
J.\ 2016, Phys. Chem. Chem. Phys., 18, 29278

\bibitem[Stubgaard(1978)]{Stubgaard78} Stubgaard, M. 1978, Ph.D. Thesis, 
K{\o}benhavns Universitet, Denmark

\bibitem[{Xu {et~al.}(2008)Xu, Fisher, Lees, Shi, Hougen, Pearson, Drouin,
  Blake, \& Braakman}]{Xu2008305}
Xu, L.-H., Fisher, J., Lees, R., {et~al.} 2008, J. Mol. Spec., 251, 305

\bibitem[Yamada \& Winnewisser(1977)]{Yamada77a} Yamada, K., \& Winnewisser, 
M.\ 1977, J. Mol. Spec., 68, 307 

\bibitem[Yamada(1977)]{Yamada77b} Yamada, K.\ 1977, J. Mol. Spec., 68, 423 

\bibitem[{Zakharenko {et~al.}(2015)Zakharenko, Motiyenko, Margul{\`e}s, \&
  Huet}]{zakharenko2015terahertz}
Zakharenko, O., Motiyenko, R.~A., Margul{\`e}s, L., \& Huet, T.~R. 2015, J.
  Mol. Spec., 317, 41

\end{thebibliography}
\end{document}